\documentclass[a4paper,12pt]{article}
\usepackage{graphicx}
\graphicspath{ {figures/} }

\usepackage{adjustbox}
\usepackage{multirow}

\usepackage{scicite}

\usepackage{todonotes}

\usepackage{times}

\usepackage{subcaption}
\usepackage{amsmath} 

\newenvironment{sciabstract}{%
\begin{quote} \bf}
{\end{quote}}

\title{Machine learning dismantling and early-warning signals of disintegration in complex systems}

\author
{
Marco Grassia$^{1}$,
Manlio De Domenico$^{2\ast}$,
Giuseppe Mangioni$^{1\ast}$\\
\\
\normalsize{$^{1}$Dip. Ingegneria Elettrica, Elettronica e Informatica - Universit\`{a} degli Studi di Catania - Italy}\\
\normalsize{$^{2}$CoMuNe Lab, Fondazione Bruno Kessler, Via Sommarive 18, 38123 Povo (TN), Italy}\\
\\
\normalsize{$^\ast$To whom correspondence should be addressed;}\\
\normalsize{E-mail: giuseppe.mangioni@dieei.unict.it; mdedomenico@fbk.eu.}
}

\date{}

\usepackage{xpatch}

\newcounter{mybibstartvalue}
\xpatchcmd{\thebibliography}{%
  \usecounter{enumiv}%
}{%
  \usecounter{enumiv}%
  \setcounter{enumiv}{\value{mybibstartvalue}}%
}{}{}

\begin{document}


\baselineskip24pt

\maketitle


\begin{sciabstract}
From physics to engineering, biology and social science, natural and artificial systems are characterized by interconnected topologies whose features -- e.g., heterogeneous connectivity, mesoscale organization, hierarchy -- affect their robustness to external perturbations, such as targeted attacks to their units. Identifying the minimal set of units to attack to disintegrate a complex network, i.e. network dismantling, is a computationally challenging (NP-hard) problem which is usually attacked with heuristics. Here, we show that a machine trained to dismantle relatively small systems is able to identify higher-order topological patterns, allowing to disintegrate large-scale social, infrastructural and technological networks more efficiently than human-based heuristics. Remarkably, the machine assesses the probability that next attacks will disintegrate the system, providing a quantitative method to quantify systemic risk and detect early-warning signals of system's collapse. This demonstrates that machine-assisted analysis can be effectively used for policy and decision making to better quantify the fragility of complex systems and their response to shocks.
\end{sciabstract}

\section*{Introduction}
Several empirical systems consist of nonlinearly interacting units, whose structure and dynamics can be suitably represented by complex networks~\cite{boccaletti2006complex}. Heterogeneous connectivity~\cite{barabasi1999emergence}, mesoscale~\cite{newman2012communities,fortunato2010community}, higher-order~\cite{benson2016higher,lambiotte2019networks} and hierarchical~\cite{clauset2008hierarchical} organization, efficiency in information exchange~\cite{watts1998collective} and multiplexity~\cite{de2013mathematical,kivela2014multilayer,boccaletti2014structure,de2016physics}, are distinctive features of biological molecules within the cell~\cite{guimera2005functional}, connectomes~\cite{bassett2017network}, mutualistic interactions among species~\cite{suweis2013emergence}, urban ~\cite{barthelemy2019statistical}, trade~\cite{alves2019} and social~\cite{lazer2009computational,johnson2016new,centola2018experimental} systems.

However, the structure of complex networks can dramatically affect its proper functioning, with crucial effects on collective behavior and phenomena such as synchronization in populations of coupled oscillators~\cite{arenas2008synchronization}, the spreading of infectious diseases~\cite{pastor2015epidemic,matamalas2018effective} and cascade failures~\cite{yang2017small}, the emergence of misinformation~\cite{vosoughi2018spread,stella2018bots} and hate~\cite{johnson2019hidden} in socio-technical systems or the emergence of social conventions~\cite{baronchelli2018emergence}. While heterogeneous connectivity is known to make such complex networks more sensitive to shocks and other perturbations occurring to hubs~\cite{albert2000error}, a clear understanding of the topological factors -- and their interplay --  responsible for a system's vulnerability still remains elusive. For this reason, the identification of the minimum set of units to target for driving a system towards its collapse -- a procedure known as network dismantling -- attracted increasing attention~\cite{kitsak2010identification,morone2015influence,morone2016collective,braunstein2016network,ren2019generalized} for practical applications and their implications for policy making. Dismantling is efficient if such a set is small and, simultaneously, the system quickly breaks down into smaller isolated clusters.
The problem is, however, NP-hard and while percolation theory provides the tools to understand large-scale transitions as units are randomly disconnected~\cite{buldyrev2010catastrophic,bashan2013extreme,radicchi2015percolation,osat2017optimal}, a general theory of network dismantling is missing and applications mostly rely on approximated theories or heuristics.

\begin{figure}[htbp!]
	\centering
   \includegraphics[width=\textwidth]{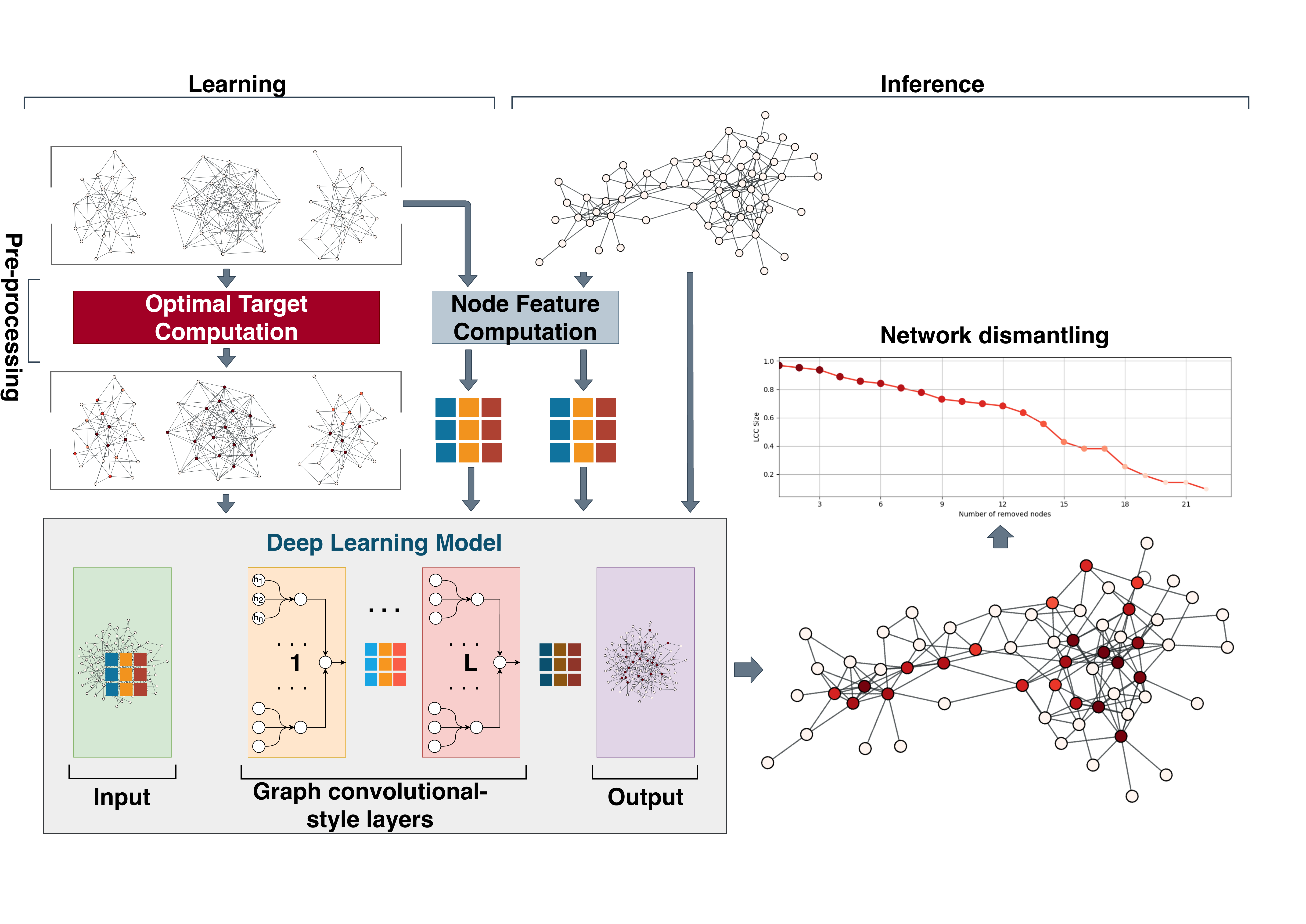}
   \caption{\textbf{Training a machine to learn complex topological patterns for network dismantling.} To build our training data, we generate and dismantle small networks optimally and compute the node features.
	 After the model is trained, it can be fed the target network (again, with its nodes' features) and it will assign each node $n$ a value $p_n$, the probability that it belongs to the (sub-)optimal dismantling set. Nodes are then ranked and removed until the dismantling target is reached.
	 }%
   \label{f:ml_flow}
\end{figure}

Here, we develop a computationally efficient framework -- named GDM and conceptually described in Figure~\ref{f:ml_flow} -- based on machine learning, to provide a scalable solution, tackle the dismantling challenge, and gain new insights about the latent features of the topological organization of complex networks.
Specifically, we employ graph convolutional-style layers, overcoming the limitations of classic (Euclidean) deep learning and operate on graph-structured data. These layers, inspired by the convolutional layers that empower most of the deep learning models nowadays, aggregate the features of each node with the ones found in its neighborhood by means of a learned non-trivial function, producing high-level node features.
While the machine is trained on identifying the critical point from dismantling of relatively small systems -- that can be easily and optimally dismantled -- we show that it exhibits remarkable inductive capabilities, being able to generalize to previously unseen nodes and way larger networks after the learning phase.






\section*{Results}
%
%
%

\subsection*{Model architecture}
The machine learning framework proposed here consists of a (geometric) deep learning model, composed of graph convolutional-style layers and a regressor (a multilayer perceptron), that is trained to predict attack strategies on small synthetic networks -- that can be easily and optimally dismantled -- and then used to dismantle large networks, for which the optimal solution cannot be found in reasonable time.
To give an insight, the graph convolutional-style layers aggregate the features of each node with the ones found in its neighborhood by means of a learned non-trivial function, as they are inspired by the convolutional layers that empower most of the (Euclidean) deep learning models nowadays.
More practically, the (higher-order) node features are propagated by the neural network when many layers are stacked: deeper the architecture, i.e., the more convolutional layers, the farther the features propagate, capturing the importance of the neighborhood of each node.
Specifically, we stack a variable number of state-of-the-art layers, namely \emph{Graph Attention Networks (GAT)}~\cite{gat_networks}, that are based on the self-attention mechanism (also known as intra-attention) which was shown to improve the performance in natural language processing tasks~\cite{attentionisall}.
These layers are able to handle the whole neighborhood of nodes without any sampling, which is one of the major limitations of other popular convolutional-style layers (e.g., \emph{GraphSage}~\cite{hamilton2017inductive}), and also to assign a relative importance factor to the features of each neighboring node that depends on the node itself thanks to the attention mechanism.

Such detailed model takes as input one network at a time plus the features of its nodes and returns a scalar value $p_n$ between zero and one for every node $n$.
During the dismantling of a network, nodes are sorted and removed (if they belong to the LCC) in descending order of $p_n$ until the target is reached.
\subsection*{Dismantling real-world systems}

In our experiments, we dismantle empirical complex systems of high societal or strategic relevance, our main goal being to learn an efficient attack strategy. To validate the goodness of such a strategy, we compare against state-of-the-art dismantling methods, such as \emph{Generalized Network Dismantling (GND)}~\cite{ren2019generalized}, \emph{Explosive Immunization (EI)}~\cite{PhysRevLett.117.208301}, \emph{CoreHD}~\cite{Zdeborov__2016}, \emph{Min-Sum (MS)}~\cite{braunstein2016network} and \emph{Collective Influence} (CI)~\cite{morone2016collective}, using node degree, $k$--core value and local clustering coefficient as node features.

We refer the reader to the Supplementary Materials for a detailed description of our models, for additional discussion and experiments (also on large real-world and synthetic networks, plus results in table form), and also for an extensive list of the real-world test networks, that include  biological, social, infrastructure, communication, trophic and technological ones.

To quantify the goodness of each method in dismantling the network, we consider the \emph{Area Under the Curve}~(AUC) encoding changes in the \emph{Largest Connected Component}~(LCC) size across the attacks. The LCC size is commonly used in the literature to quantify the robustness of a network, because systems need the existence of a giant cluster to work properly.
The AUC indicator\footnote{We compute the AUC value by integrating the $LCC(x) / |N|$ values using Simpson's rule.} has the advantage of accounting for how quickly, overall, the LCC is disintegrated: the lower the area under the curve, the more efficient is the network dismantling.

\begin{figure}[htbp!]
	\centering
	\begin{subfigure}{0.33\textwidth}
		\centering
		\vspace{0.3cm}
		\includegraphics[width=\textwidth]{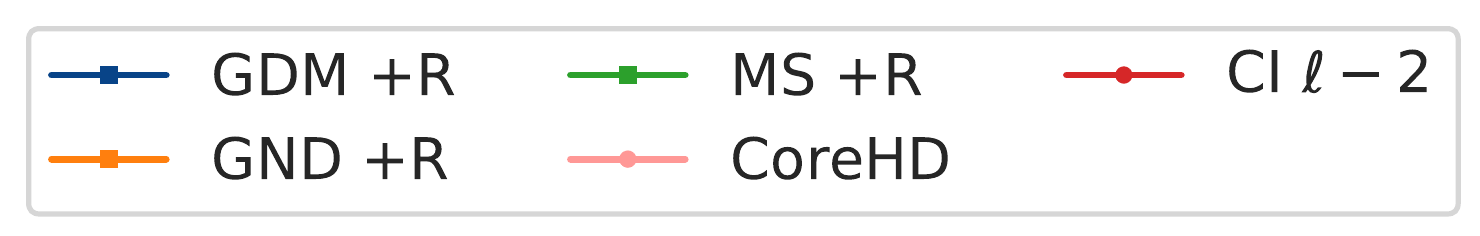}
		\hfill
		\includegraphics[width=\textwidth]{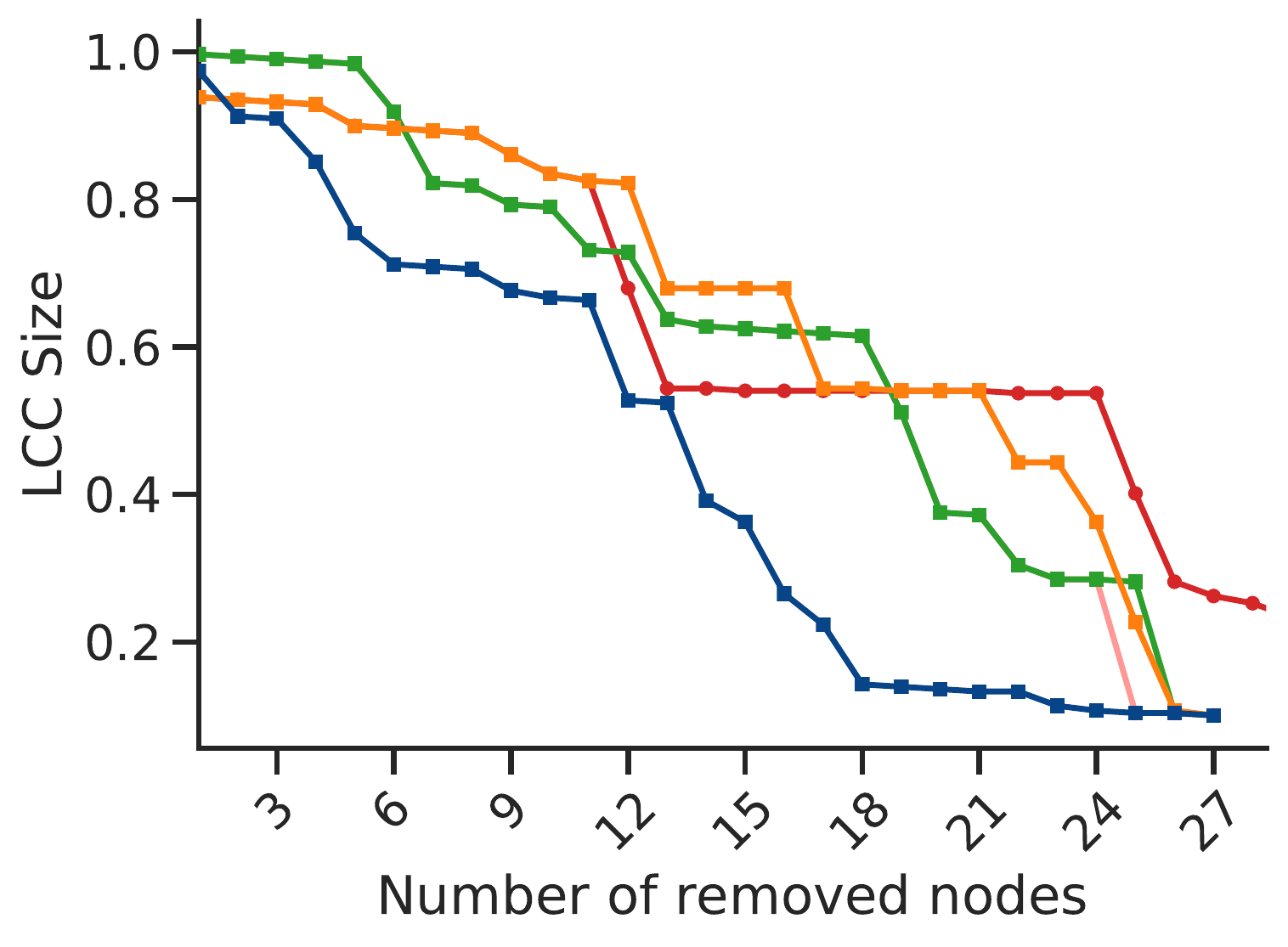}
		\caption{Dismantling process.}
		\label{f:corruption_dismantling}
	\end{subfigure}%
	\begin{subfigure}{0.33\textwidth}
		\centering
		\includegraphics[width=\textwidth]{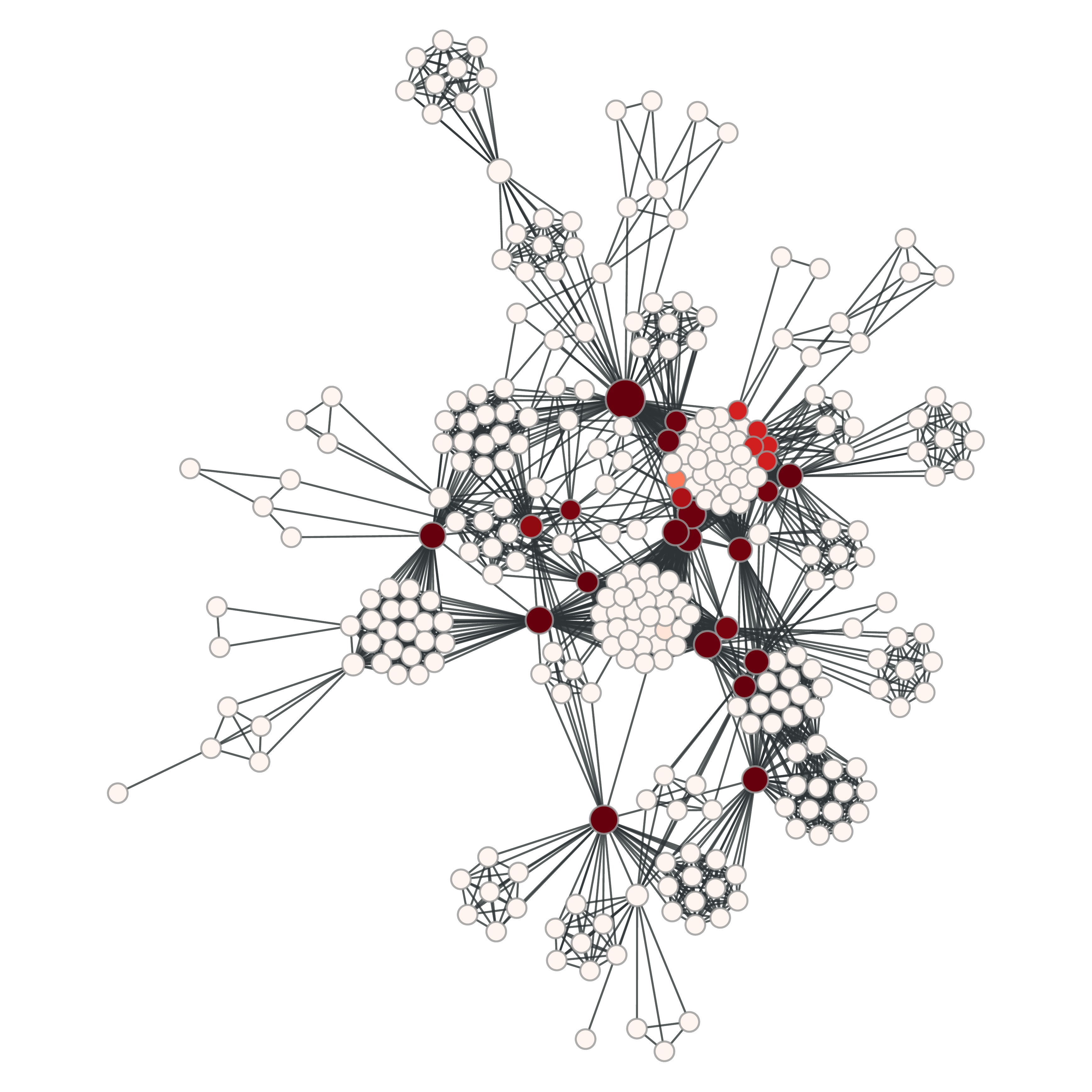}
    \caption{Original network.}
    \label{f:corruption_original}
	\end{subfigure}%
	\begin{subfigure}{0.33\textwidth}
		\centering
		\includegraphics[width=\textwidth]{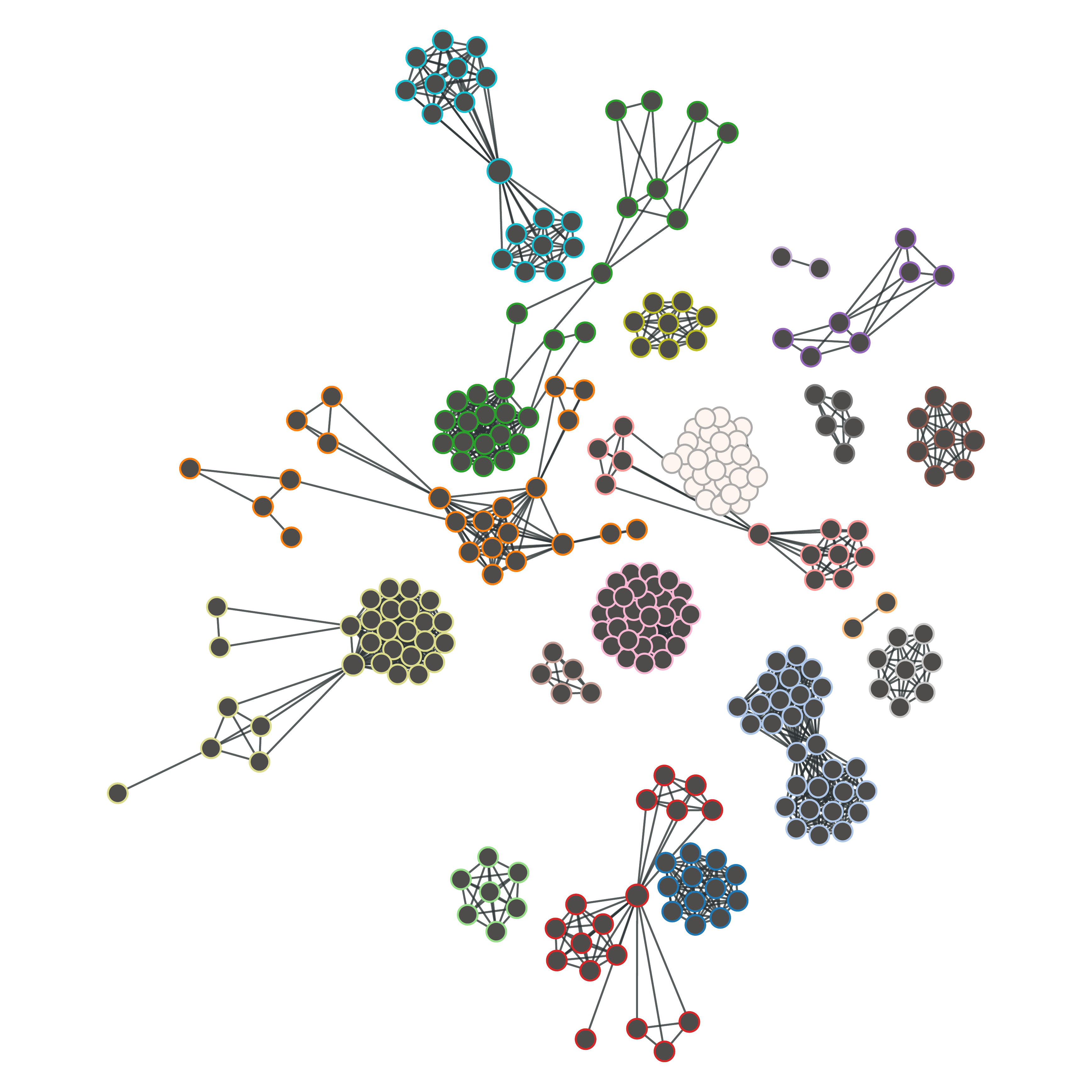}
		\caption{Attacked network.}
		\label{f:corruption_attacked}
	\end{subfigure}%
 \caption{\textbf{Dismantling the Brazilian corruption network.} (a) GDM and state-of-the-art algorithms with reinsertion of the nodes are compared. The network before (b) and after (c) a GDM attack is shown. The color of the nodes represents (from dark red to white) the attack order, while their size represents their betweenness value. In the attacked network, darker nodes do not belong to the LCC, and their contour color represents the component they belong to.}
 \label{f:corruption_dismantling_graph}
\end{figure}

As a representative example, we show in
Figure~\ref{f:corruption_dismantling}
  the result of the dismantling process for the \emph{corruption} network~\cite{10.1093/comnet/cny002}, built from $65$ corruption scandals in Brazil, as a function of the number of removed units. Results are shown for GDM and the two cutting-edge algorithms mentioned above. In
Figures~\ref{f:corruption_original} and~\ref{f:corruption_attacked}, instead, we show the structure before and after dismantling, respectively. Our framework disintegrates the network faster than other methods: to verify if this feature is general, we perform a thorough analysis of several empirical systems.

\begin{figure}[htb!]
	\centering
 	\includegraphics[width=1.0\textwidth]{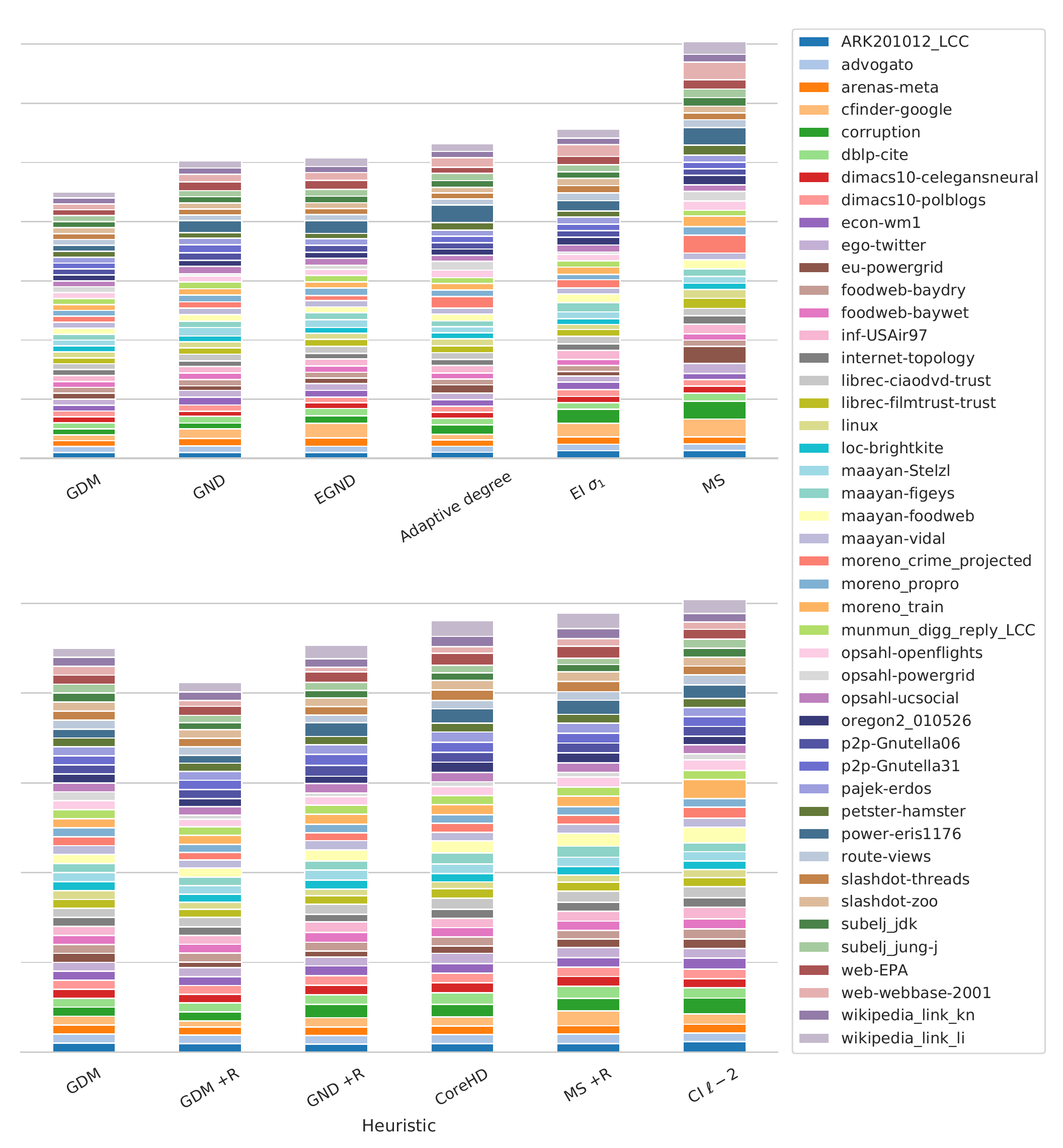}
 	\caption{\textbf{Dismantling empirical complex systems.} Per-method cumulative area under the curve (AUC) of real-world networks dismantling. The lower the better. The dismantling target for each method is $10\%$ of the network size. Each value is scaled to the one of our approach (GDM) for the same network. \emph{GND} stands for \emph{Generalized Network Dismantling} (with cost matrix $\mathbf{W} = \mathbf{I}$), \emph{MS} stands for \emph{Min-Sum}, \emph{EI} stands for \emph{Explosive Immunization} and \emph{CI} for \emph{Collective Influence}. +R means that the reinsertion phase is performed. \emph{CoreHD} and \emph{CI} are compared to other +R algorithms as they include the reinsertion phase. Also, note that some values are clipped (limited) to 3x for the \emph{MS} heuristic to improve visualization.}
 	\label{f:empirical_results}
\end{figure}

Figure~\ref{f:empirical_results} shows the performance of each dismantling method on each empirical system considered in this study, allowing for an overall comparison. On average, our approach outperforms the others. For instance, Generalized Network Dismantling's cumulative AUC is $\sim12\%$ higher and the Min-Sum algorithm is outscored by a significant margin, which is remarkable considering that our approach is static -- i.e., predictions are made at the beginning of the attack -- while the other ones are dynamic -- i.e., structural importance of the nodes is (re)computed during the attacks.
For a more extensive comparison with these approaches, we also introduce a node reinsertion phase using a greedy algorithm which reinserts, a posteriori, those nodes that belong to smaller components of the (virtually) dismantled system and which removal is not actually needed in order to reach the desired target~\cite{braunstein2016network}. Once again, our approach outperforms the other algorithms: even without accounting for the reinsertion phase, GDM performs comparably with GND + reinsertion and outscores the others, highlighting how it is able to identify the more critical nodes of a network.


An interesting feature of our framework is that it can enhance existing heuristics based on node descriptors, by employing the same measure as the only node feature. It is plausible to assess that our framework learns correlations among node features. To probe this hypothesis, in the Supplementary Materials we analyze the configuration models of the same networks analyzed so far: those models keep the observed connectivity distribution while destroying topological correlations.
We observe that dismantling performance drop on these models, confirming that the existing topological correlations are learned and, consequently, exploited by the machine.



\subsection*{Early-warning signals of systemic collapse}
Another relevant output of our method is the calculation of a damage score that can be used to predict the impact of future attacks to the system.
Accordingly, we introduce an estimator of early warning that can be used for inform policy and decision making in applications where complex interconnected systems -- such as water management systems, power grids, communication systems and public transportation networks -- are subject to potential failures or targeted attacks.
%
We define $\Omega$, namely \emph{Early Warning}, as a value between $0$ and $1$, calculated as follows.
We first simulate the dismantling of the target network using our approach and call $S_o$ the set of virtually removed nodes that cause the percolation of the network. Then, we sum the $p_n$ values predicted by our model for each node $n \in S_o$ and define

$$
	\Omega_{m} = \sum\limits_{n\in S_o}^{}{p_n}.
$$
The value of the Early Warning $\Omega$ for the network after the removal of a generic set $S$ of nodes is given by
$$
	\Omega = \begin{cases}
            \Omega_{s}/\Omega_{m}				& \text{if} \ \Omega_{s} \leq \Omega_{m} \\
            1 										&	\text{otherwise}
        	\end{cases}
$$
where $\Omega_{s} = \sum\limits_{n\in S}{p_n}$.

The rationale behind this definition is that the system will tolerate a certain amount of damage before it collapses: this value is captured by $\Omega_{m}$.
$\Omega$ will quickly reach values close to $1$ when nodes with key-role in the integrity of the system are removed.
Of course, the system could be heavily harmed by removing many less relevant nodes (e.g., the peripheral ones) with an attack that causes a small decrease in LCC size over time, and probably get a low value of $\Omega$. However, this kind of attacks does not need an early-warning signal since they do not cause an abrupt disruption of the system and can be easily detected. 

\begin{figure}[!ht]
	\centering
	\begin{subfigure}{\textwidth}
		\centering
		\includegraphics[width=0.50\textwidth]{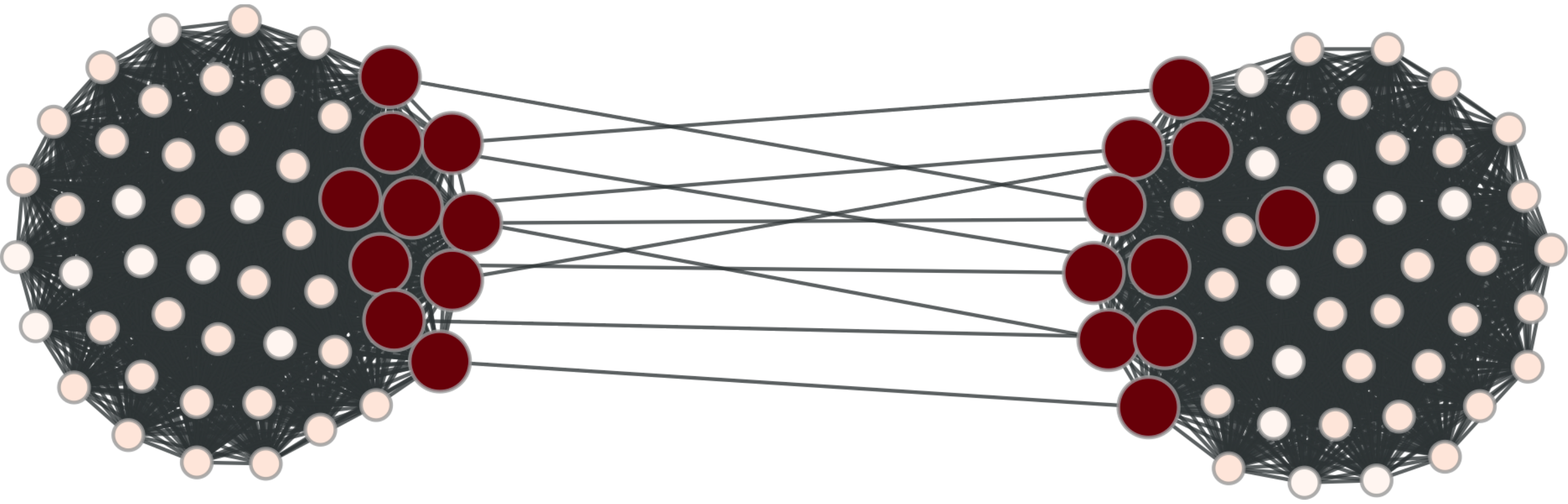}
		\caption{Toy-example network composed of two cliques connected by $10$ bridges. The size of the nodes represents their betweenness value and the color (from dark red to white) represents their importance to the system's health according to our method.}
		\label{f:early_warning_toy_example_net}
	\end{subfigure}%
	\vspace{0.5cm}
	\hfill
	\begin{subfigure}{\textwidth}
		\centering
		\includegraphics[width=0.5\textwidth]{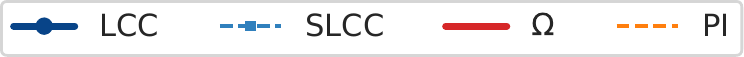}
	\end{subfigure}%
	\hfill
	\begin{subfigure}{0.50\textwidth}
		\centering
		\includegraphics[width=\textwidth]{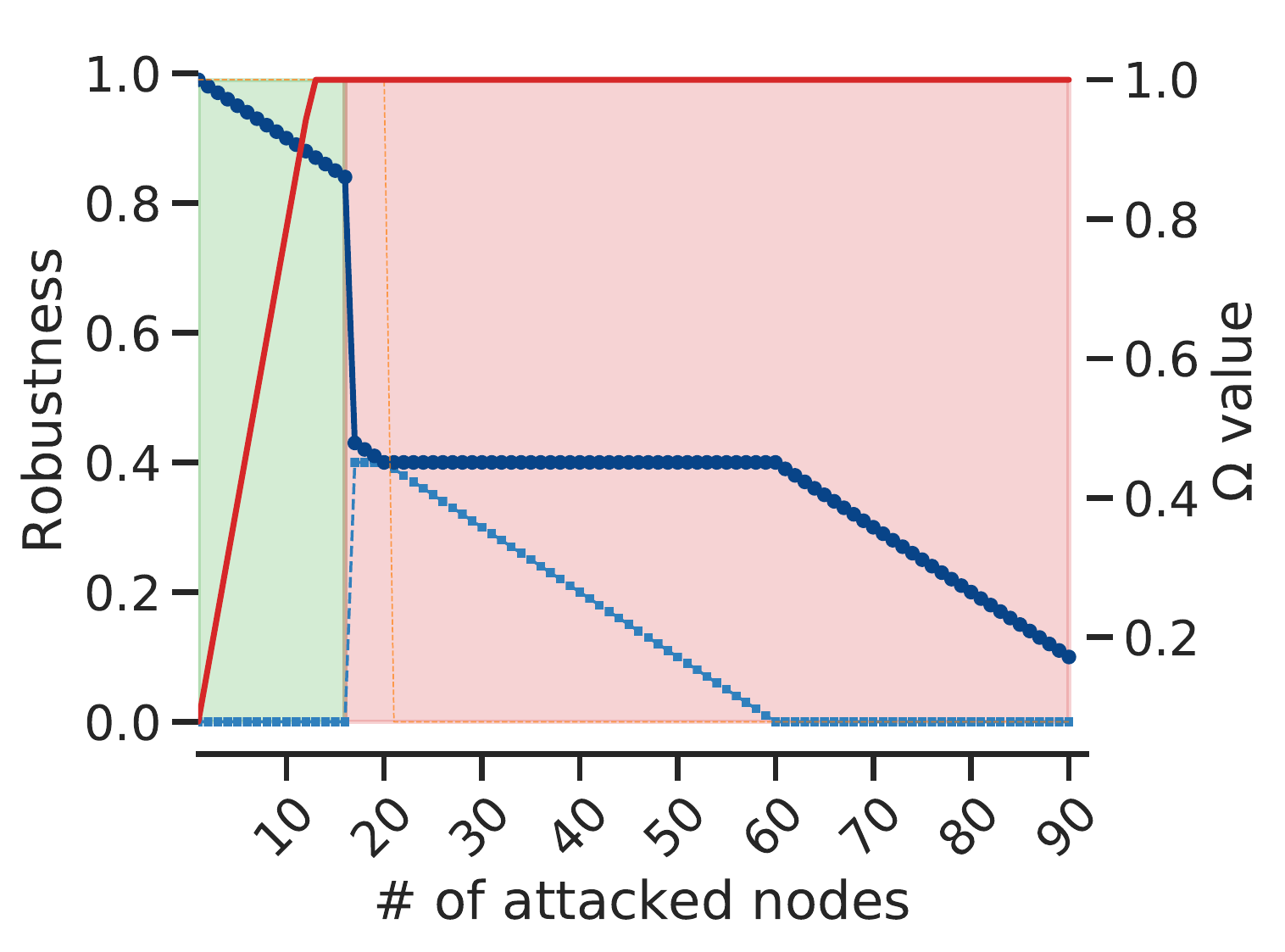}
		\caption{Degree or betweenness based attack.}
		\label{f:early_warning_toy_example_ew}
	\end{subfigure}%
	\begin{subfigure}{0.50\textwidth}
		\centering
		\includegraphics[width=\textwidth]{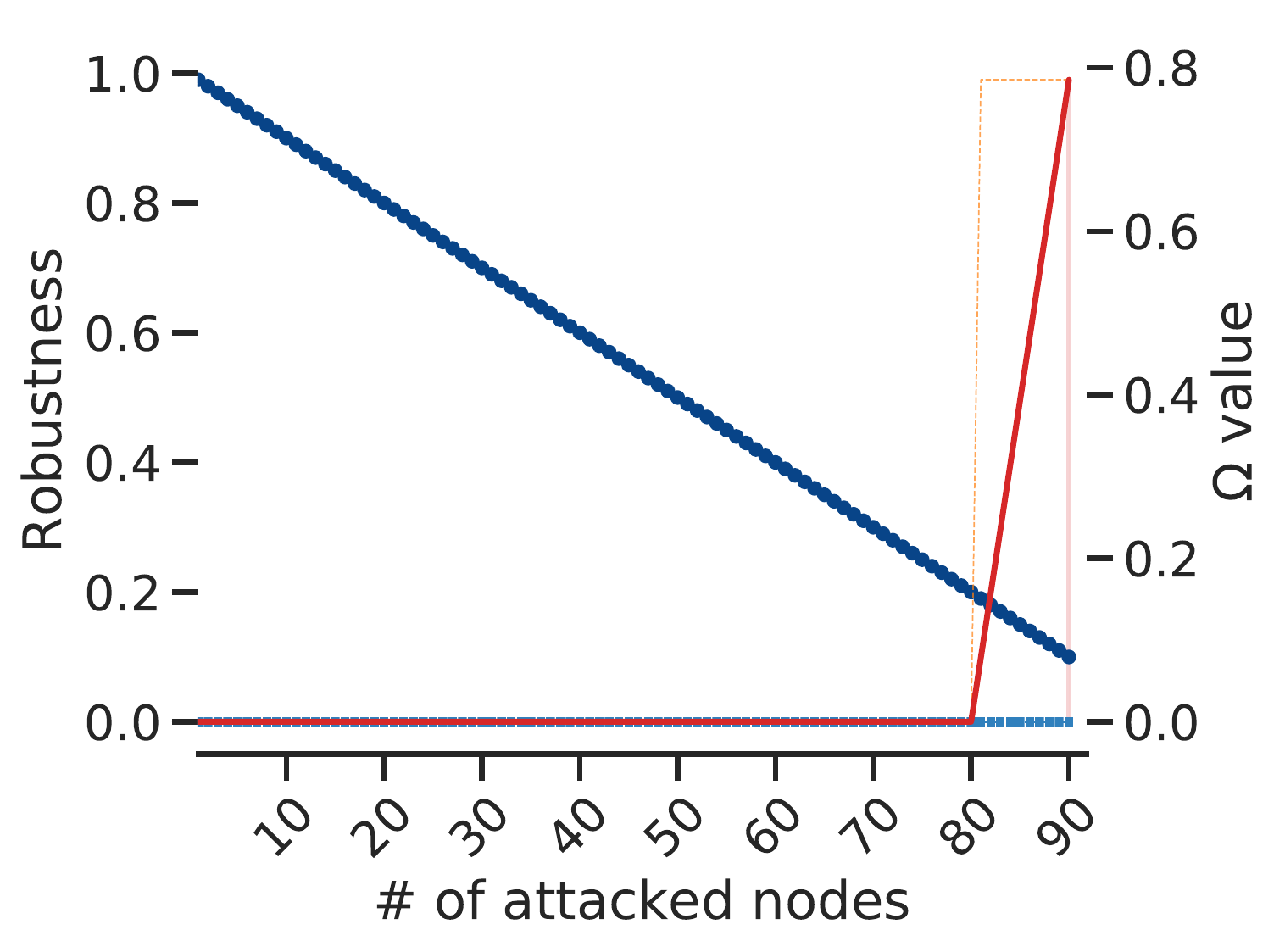}
		\caption{Inverse degree or betweenness based attack.}
		\label{f:early_warning_toy_example_ew_inv}
	\end{subfigure}%
	\caption{Toy-example meant to explain why the LCC is not sufficient to evaluate the state of the system. The LCC decreases at the same rate during the initial part of both the attacks shown. Instead, $\Omega$ values do not and reach warning levels before the system suddenly collapses.}
  \label{f:early_warning_toy_example}
\end{figure}

\paragraph*{Why do we need an Early Warning signal?}
In
Figure~\ref{f:early_warning_toy_example}
we show a toy-example meant to explain why the Largest Connected Component size may not be enough to determine the state of a system.
The toy-example network in
Figure~\ref{f:early_warning_toy_example_net}
is composed of two cliques (fully connected sub-networks) connected by a few border nodes (bridges) that also belong to the respective cliques.
Many dismantling approaches (like the degree and betweenness-based heuristics, or even ours) would remove those bridge nodes first, meaning that the network would eventually break in two, as shown in
Figure~\ref{f:early_warning_toy_example_ew}.
Now, when most of the bridge nodes are removed (e.g., after $~16$ removals), the LCC is still quite large as it includes more than $80\%$ of the nodes, but it takes just a few more removals of the bridges to break the network in two.
While $\Omega$ is able to capture the imminent system disruption (i.e., the $\Omega$ value gets closer to $1$ very fast), the LCC size is not, and one would notice when it is too late.
Moreover, the LCC curve during the initial part of the attack is exactly the same as the one in
Figure~\ref{f:early_warning_toy_example_ew_inv},
showing the removal of nodes in inverse degree (or betweenness) order, which does not cause the percolation of the system.
Again, $\Omega$ captures this difference and does not grow, meaning that a slow degradation should be expected.

\begin{figure}[htbp!]
	\centering
	\includegraphics[width=\textwidth]{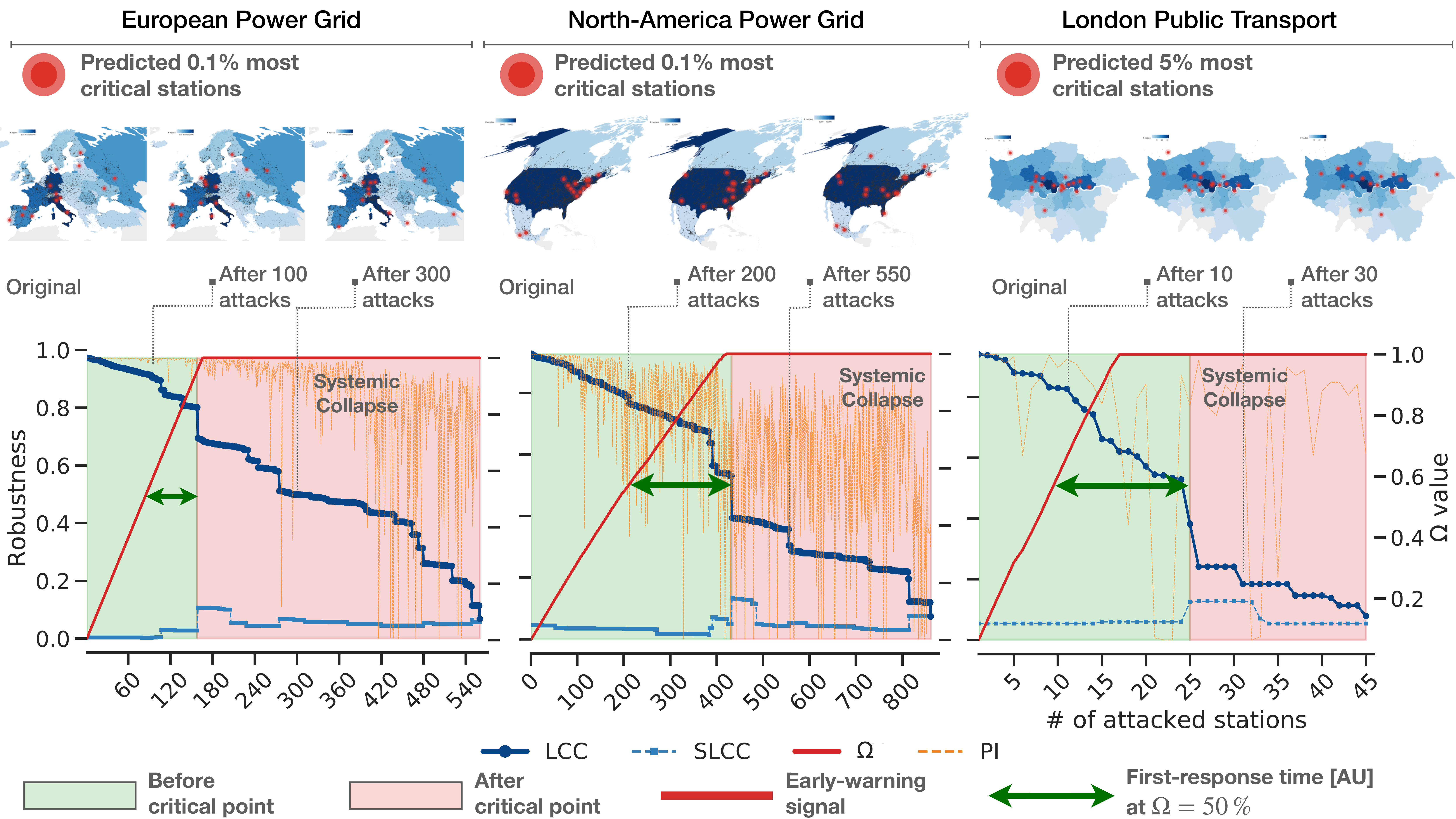}
	\caption{\textbf{Early warning due to network dismantling of real infrastructures.} Three empirical systems, namely the European power grid (left), the North-American power grid (middle) and the London public transport (right), are repeatedly attacked using a degree-based heuristics, i.e., hubs are damaged first. A fraction of the most vulnerable stations is shown for the original systems and some representative damaged states (i.e., before and after the critical point for system's collapse), in the top of the figure. The plots show the behavior of the largest (LCC) and second-largest (SLCC) connected components, as well as the behavior of $\Omega$, the Early Warning descriptor introduced in this study and the $p_n$ value of each removed node (PI). Transitions between green and red areas indicate the percolation point of the corresponding systems, found through the SLCC peak. We also show the first response time in arbitrary units (AU), to highlight how our framework allows to anticipate system's collapse, allowing for timely emergency response.}
	\label{f:early_warning}
\end{figure}

\paragraph*{Tests on real-world systems}
We test our method on key infrastructure networks and predict the collapse of the system under various attack strategies (see Fig.~\ref{f:early_warning} for details).
Remarkably, while the LCC size decreases slowly without providing a clear alarm signal until the system is heavily damaged and collapses, $\Omega$ grows faster when critical nodes are successfully attacked, reaching warning levels way before the system is disrupted, as highlighted by the First Response Time, defined as the time occurring between system's collapse and an early-warning signal of 50\% (i.e., $\Omega=0.5$). Moreover, the first order derivative ${\Omega}_{s}'$ tracks the importance of nodes that are being attacked, providing a measure of the attack intensity over time.

\section*{Discussion}
Our results show that using machine learning to learn network dismantling comes with a series of advantages. While the ultimate theoretical framework is still missing, our framework allows one to learn directly from the data, at variance with traditional approaches which rely on the definition of new heuristics, metrics or algorithms. An important advantage of our method, typical of data-driven modeling, is that it can be further improved by simply retuning the parameters of the underlying model and training again: conversely, existing approaches require the (re)definition of heuristics and algorithms which are more demanding in terms of human efforts. Remarkably, the computational complexity of dismantling networks with our framework is considerably low: just $O(N + E)$, where $N$ is system's size and $E$ the number of connections -- which drops to $O(N)$ for sparse networks. This feature allows for applications to systems consisting of millions of nodes while keeping excellent performance in terms of computing time and accuracy.
Last but not least, from a methodological perspective, it is worth remarking that our framework is general enough to be adapted and applied to other interesting NP-hard problems on networks, opening the door for new opportunities and promising research directions in complexity science, together with very recent results employing machine learning, for instance, to predict extreme events~\cite{qi2020using}.

The impact of our results is broad. On the one hand, we provide a framework which disintegrates real systems more efficiently and faster than state-of-the-art approaches: for instance, applications to covert networks might allow to hinder communications and information exchange between harmful individuals. On the other hand, we provide a quantitative descriptor of damage which is more predictive than existing ones, such as the size of the largest connected component: our measure allows to estimate the potential system's collapse due to subsequent damages, providing policy and decision makers with a quantitative early-warning signal for triggering a timely response to systemic emergencies in water management systems, power grids, communication and public transportation networks.




\bibliographystyle{plain}
\bibliography{scibib}

%
%



\end{document}



\baselineskip24pt

\maketitle

\section*{Supplementary Materials}
%
Here, we provide additional examples and details about the results we discussed in the main text.
We begin detailing the architecture of the employed deep learning model and the way it is trained.
Secondly, we discuss the computational complexity of our framework and then we show more toy and real-world examples on the dismantling process and on the Early Warning $\Omega$.
Lastly, we list the test networks used to evaluate our approach.


\subsection*{Deep Learning Model}

\paragraph*{How the model works}
A simplistic but more practical understanding of how a model with $L$ GAT network layers assigns the $p_n$ value to each node $n$ can be achieved by considering the $L$-hop neighborhood of the node.
Consider a tree with height $L$ with node $n$ as root and where each node's neighbors are its children.
That is, each level $l+1$ of the tree is populated with the neighbors of nodes at level $l$.
For instance, at level $1$ we have $n$'s neighbors.

Now, $n$'s high-level node features ($h^{L}_n$) are computed by aggregating the information from the $L$-hop neighborhood in a bottom up fashion. Each GAT network layer processes a level of the three, so the deeper the model, the farther the information comes from.
That is, the model starts from the bottom of the tree (i.e., the nodes at $L$ hops from $n$) to compute the high-level node features of each node at layer $L$ and goes up until the root (node $n$) is reached.

This means that the model is able to aggregate the information in the whole $n$'s $L$-hop neighborhood in $h^{L}_n$, which also accounts for the different importance each node has in that neighborhood thanks to the \emph{GAT}'s self-attention mechanism.
The basic idea is somehow similar to the \emph{Collective Influence} approach, with the main differences being that the geometric deep learning model learns a weighted sum function from the training data to aggregate many node features, whereas the \emph{Collective Influence} just sums the degrees, and also that the model aggregates the whole $L$-hop neighborhood ball, not just its frontier.

These high-level features ($h^{L}_n$) are  then fed to a regressor that returns $p_n$, the node's structural importance indicator used in our work.

The actual implementation of our model relies on \emph{PyTorch Geometric} library~\cite{Fey/Lenssen/2019} on-top of \emph{PyTorch}~\cite{paszke2017automatic}, while the handling of the graphs (i.e., implementation of the data structures, removal of the nodes and the computation of the connected components) is performed using \emph{graph-tool}~\cite{peixoto_graph-tool_2014}.

\paragraph*{Training}
We train our models in a supervised manner.
Our training data is composed of small synthetic networks ($25$ nodes each) generated using the Barab\'{a}si-Albert (BA), the Erd\H{o}s-R\'{e}nyi (ER) and the Static Power law generational models that are implemented in \emph{iGraph}~\cite{igraph} and \emph{NetworkX}~\cite{hagberg-2008-exploring}.
Each synthetic network is dismantled optimally using brute-force and nodes are assigned a numeric label (the learning target) that depends on their presence in the optimal dismantling set(s).
That is, all combinations of increasing length of nodes are removed until we find at least one that shrinks the Largest Connected Component (LCC) to a given target size, $\sim18\%$ in our tests; then, the label of each node is computed as the number of optimal sets it belongs to, divided by the total number of optimal sets.
For example, if there is only a set of optimal size, we assign a label value of $1$ to the nodes in that set and $0$ to all other nodes.
This is meant to teach the model that some nodes are more critical than others since they belong to many optimal dismantling sets.

We stress that the training label is arbitrary and others may work better for other training sets or targets.
Moreover, while we train on a generic purpose dataset that includes both power law and ER networks, the training networks can also be chosen to fit the target networks, e.g., by using networks from similar domains or with similar characteristics.

\paragraph*{Node features}
Considering that the model can process any features combination, one could just choose to stuff every suitable node metrics that comes to his mind and, since it is proven that Deep Neural Networks learn the feature importance, let them do the rest.
On the other hand, it could also be tempting to use no features at all (e.g, a constant value for every node) since Kipf et al.~\cite{kipf2016semi} showed that their \emph{Graph Convolutional Network (GCN)}, a particular type of convolutional-style graph neural networks, can learn to linearly separate the communities based on the network structure alone and on minimal supervision (one labelled node per community), meaning that convolutional-style neural networks can leverage the network topology to assign a higher-level node feature that describes its role in the network.

We argue that, while the first idea could make sense for scenarios where training data is abundant and the features are cheap to compute, and while the second shows worse (with respect to models with simple features) but still interesting performance, it makes sense to perform some feature selection a priori to keep the computational complexity of the attack low and also to speed-up the learning process.
With that in mind, we pick node degree (plus its \emph{Pearson’s chi-square statistic}, $\chi^2$, over the neighborhood), $k$--coreness and local clustering coefficient as node features.


\paragraph*{Parameters}
We run a grid search to test various combination of model parameters, which are reported here, and select the models that better fit the dismantling target (i.e., lower area under the curve or lower number of removals).
\begin{itemize}
  \item Convolutional-style layers: \emph{Graph Attention Network} layers.
  \begin{itemize}
    \item Number of layers: from $1$ to $4$;
    \item Output channels for each layer: $5, 10, 20, 30, 40$ or $50$, sometimes with a decreasing value between consecutive layers;
    \item Multi-head attentions: $1, 5, 10, 15, 20$ or $30$ concatenated heads;
    \item Dropout probability: fixed to $0.3$;
    \item Leaky ReLU angle of the negative slope: fixed to $0.2$;
    \item Each layer learns an additive bias;
    \item Each layer is coupled with a linear layer with the same number of input and output channels;
    \item Activation function: Exponential Linear Unit (ELU). The input at each convolutional layer is the sum between the output of the GAT and the linear layers;
  \end{itemize}
  \item Regressor:
  Multi Layer Perceptron
  \begin{itemize}
    \item Number of layers: from $1$ to $4$;
    \item Number of neurons per layer: $20, 30, 40, 50$ or $100$, sometimes with a decreasing value between consecutive layers.
  \end{itemize}
   \item Learning rate: fixed to $10^{-5}$;
   \item Epochs: we train each model for $50$ epochs;
\end{itemize}

\subsection*{Computational complexity}
The computational complexity of our approach mainly depends on two elements: 1) the computational complexity of the node features used and 2) the computational complexity of the convolutional-style layers in the model.
In particular, the convolutional-style layers that we employ, i.e., the \emph{Graph Attention Network}, scale as $O(N+E)$ where $N$ is the number of nodes and $E$ is the number of edges in the network.
Considering that real-world networks are usually sparse, we assume that $O(E)\approx O(N)$, so $O(N+E) \approx O(N)$, and the computational complexity of our approach is the maximum between this and the computational complexity of the features.
Given that, the most expensive feature we compute in our experiments is the $k$--coreness, that is $O(N+E)$, so the computational complexity of the approach detailed above is $O(N)$.
For what concerns the computational complexity of the brute-force performed during the training set generation, it is irrelevant as it is a highly parallelizable one-time task that is performed on very small networks.
Moreover, since the neural models can generalize, there is no need to train them for each dismantling, and the actual time spent training is negligible.

\subsection*{Other Results}

\paragraph*{Understanding GDM's behavior}
Before testing on real-world networks, we investigate the behavior of our approach by dismantling some toy-example networks.
To this aim, we employ the same low computational complexity node features from the main paper (that are also detailed above).

\begin{figure}[htbp!]
	\centering
	\begin{subfigure}{0.33\textwidth}
		\centering
		\includegraphics[width=\textwidth]{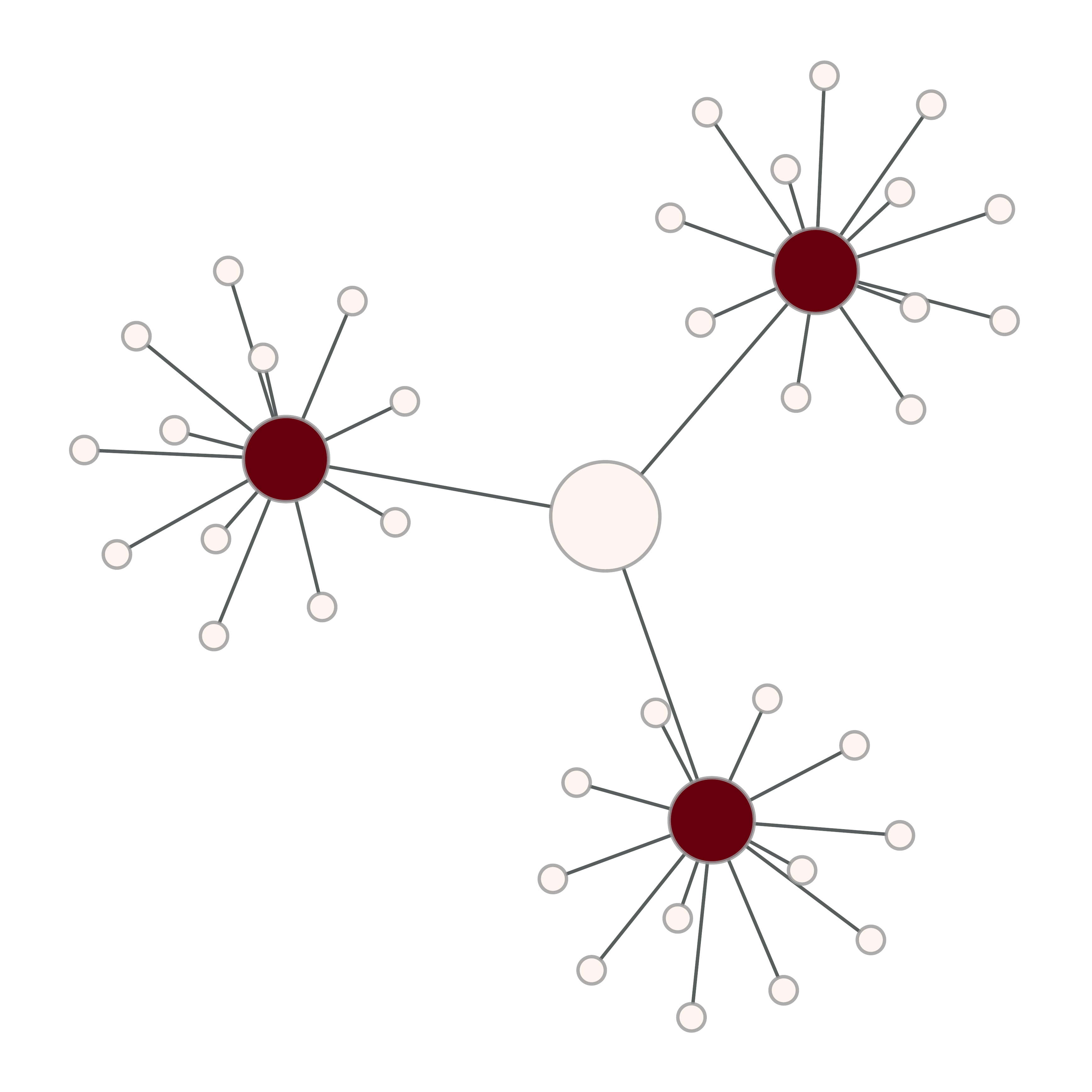}
		\caption{Three bridged ego networks.}
		\label{f:toy_example_1}
	\end{subfigure}%
	\begin{subfigure}{0.33\textwidth}
		\centering
		\includegraphics[width=\textwidth]{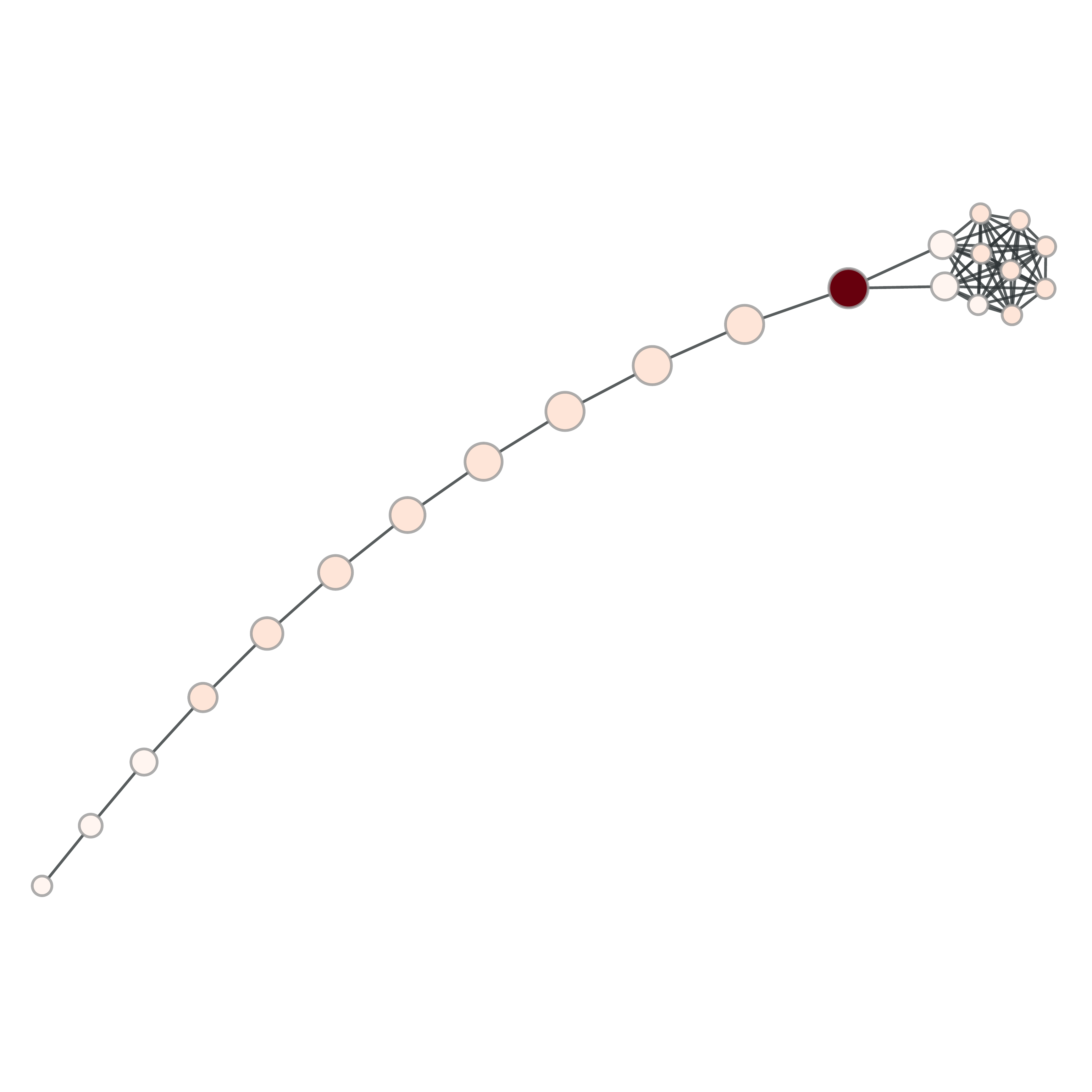}
    \caption{Tailed clique}
    \label{f:toy_example_2}
	\end{subfigure}%
	\caption{Toy examples. The color of the nodes represents (from dark red to white) the removal order of predicted strategy, while their size represents their betweenness value.}
\end{figure}

\begin{figure}[!ht]
	\centering
	\begin{subfigure}{0.4\textwidth}
		\centering
		\includegraphics[width=\textwidth]{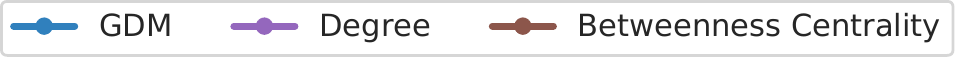}
	\end{subfigure}%
	\hfill
	\begin{subfigure}{0.5\textwidth}
		\centering
		\includegraphics[width=\textwidth]{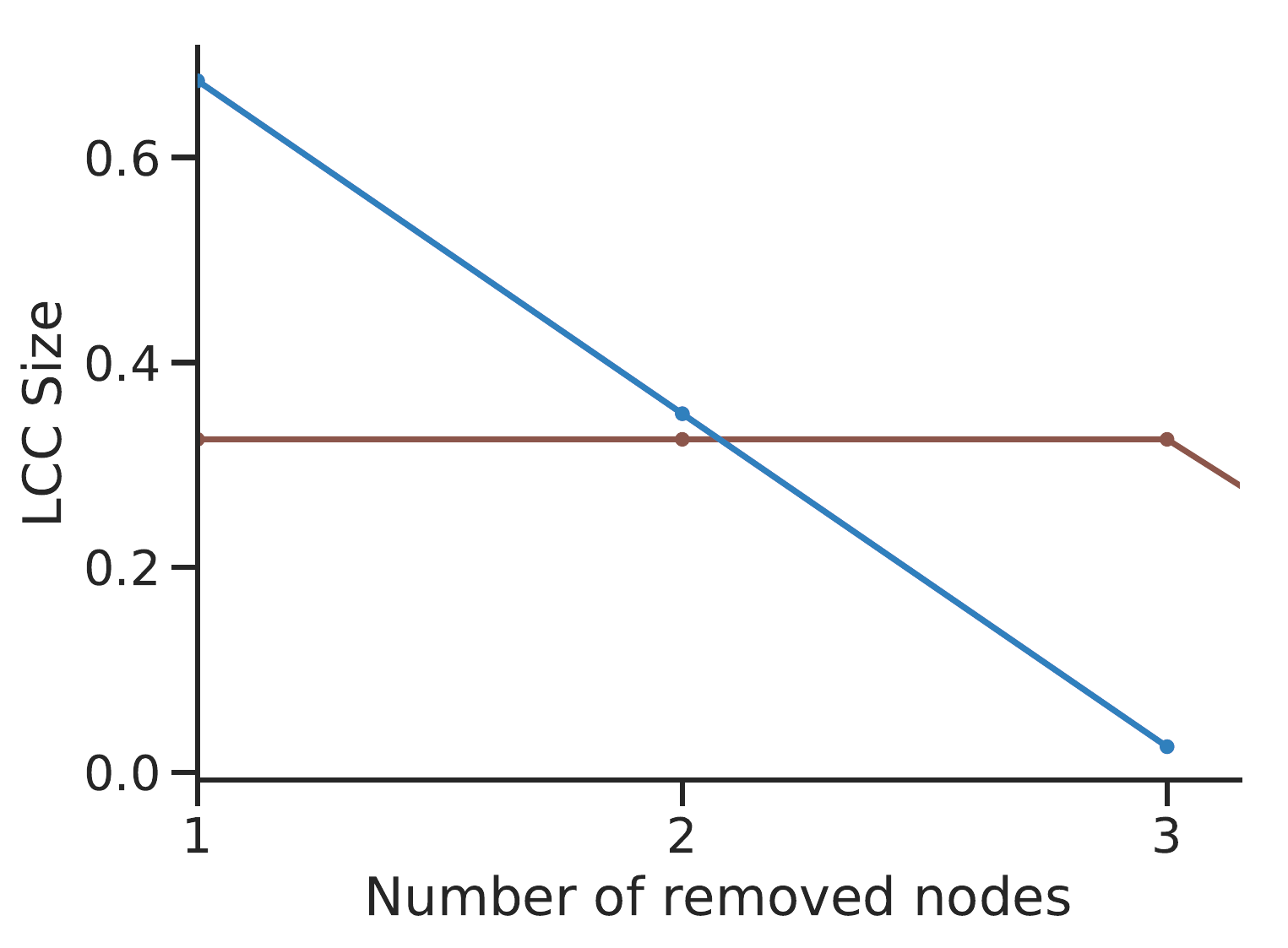}
		\caption{Three bridged ego networks.}
		\label{f:dismantling_toy_example_degree_bridged_ego}
	\end{subfigure}%
	\begin{subfigure}{0.5\textwidth}
		\centering
		\includegraphics[width=\textwidth]{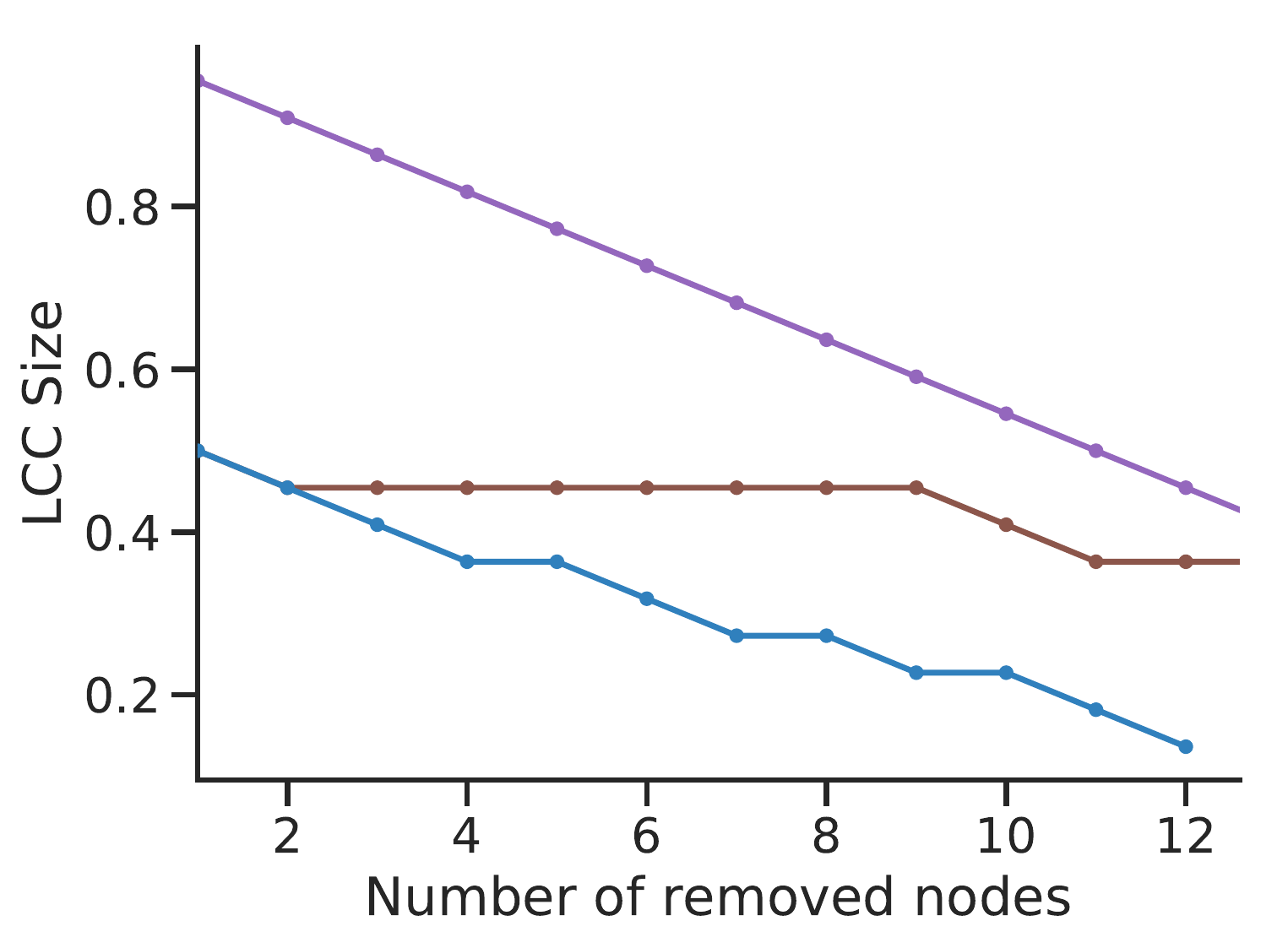}
		\caption{Tailed clique}
		\label{f:dismantling_toy_example_degree_chained_clique_5}
	\end{subfigure}%
	\caption{Dismantling the toy example networks using our approach, GDM, and the degree and betweenness based heuristics as comparison.}
  \label{f:dismantling_toy_example}
\end{figure}

The first toy example, shown in figure
\ref{f:toy_example_1},
is a network built from three ego-networks joined by a bridge.
The betweenness based heuristics\footnote{The removal of nodes by descending betweenness centrality order. The node betweenness is a node centrality measure that captures the importance of the node to the shortest paths through the network.}, and also our common sense, would suggest to remove the bridge first, reducing the LCC size to one third of the initial value, and then remove the nodes at the center of the unconnected ego networks left, for a total of four removals.
Instead, our model predicts a different strategy and removes only the cores of the ego sub-networks, reaching the same LCC size with just three removals, as shown in
Figure~\ref{f:dismantling_toy_example_degree_bridged_ego}.

At this point, we want to probe if the model is just learning to remove the nodes in descending degree order as the previous example would suggest.
If that is the case, in our second toy example network, composed of a clique with an appended tail as illustrated in
Figure~\ref{f:toy_example_2},
the model would remove the nodes in the clique first, given their high degree.
Instead, the tail is detached first, meaning that the predicted strategy differs from the degree based one, and both the degree and betweenness-based heuristics are outperformed, as shown in
Figure~\ref{f:dismantling_toy_example_degree_chained_clique_5}.


\paragraph*{Dismantling real-world complex systems}

\begin{figure}[!ht]
	\centering
	\begin{subfigure}{0.5\textwidth}
		\centering
		\includegraphics[width=\textwidth]{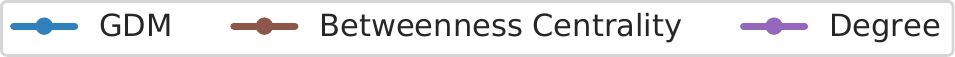}
	\end{subfigure}%
	\hfill
 	\begin{subfigure}{0.50\textwidth}
		\centering
		\includegraphics[width=\textwidth]{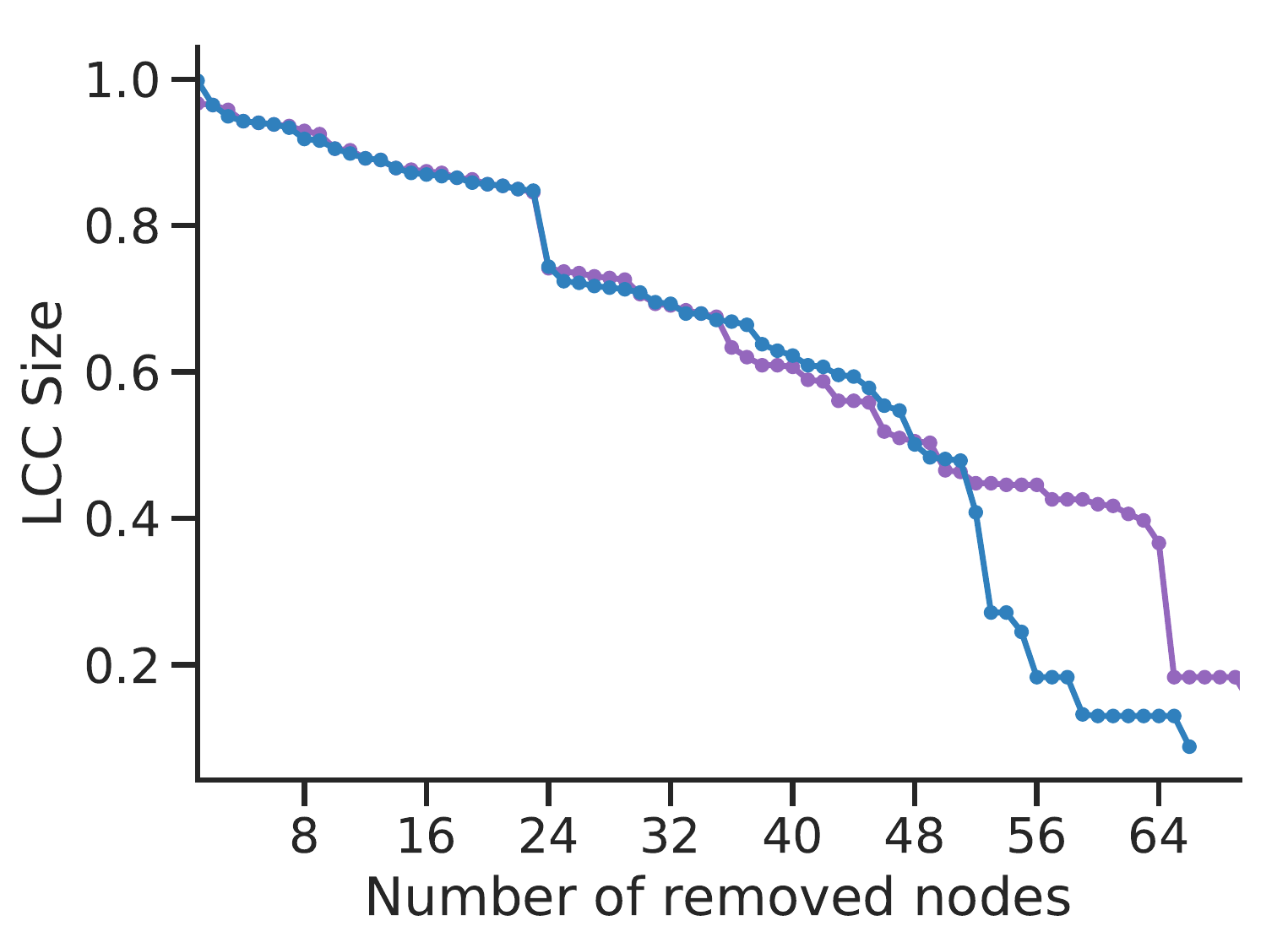}
		\caption{arenas-meta degree}
		\label{f:arenas-meta_heuristics_degree_improvement}
	\end{subfigure}%
	\begin{subfigure}{0.50\textwidth}
		\centering
		\includegraphics[width=\textwidth]{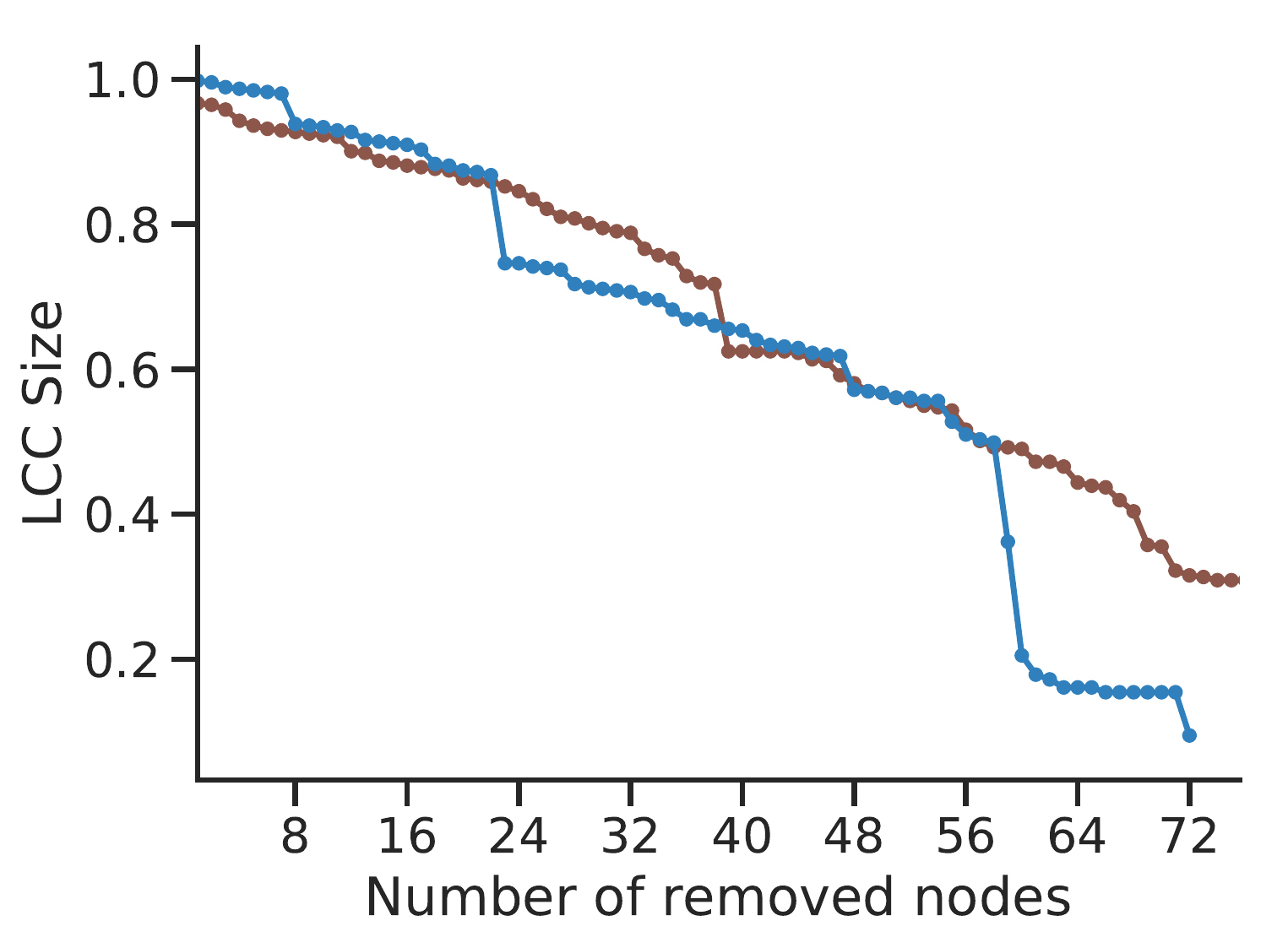}
		\caption{arenas-meta betweenness}
		\label{f:arenas-meta_heuristics_betweenness_improvement}
	\end{subfigure}%
	\hfill
	\begin{subfigure}{0.50\textwidth}
		\centering
		\includegraphics[width=\textwidth]{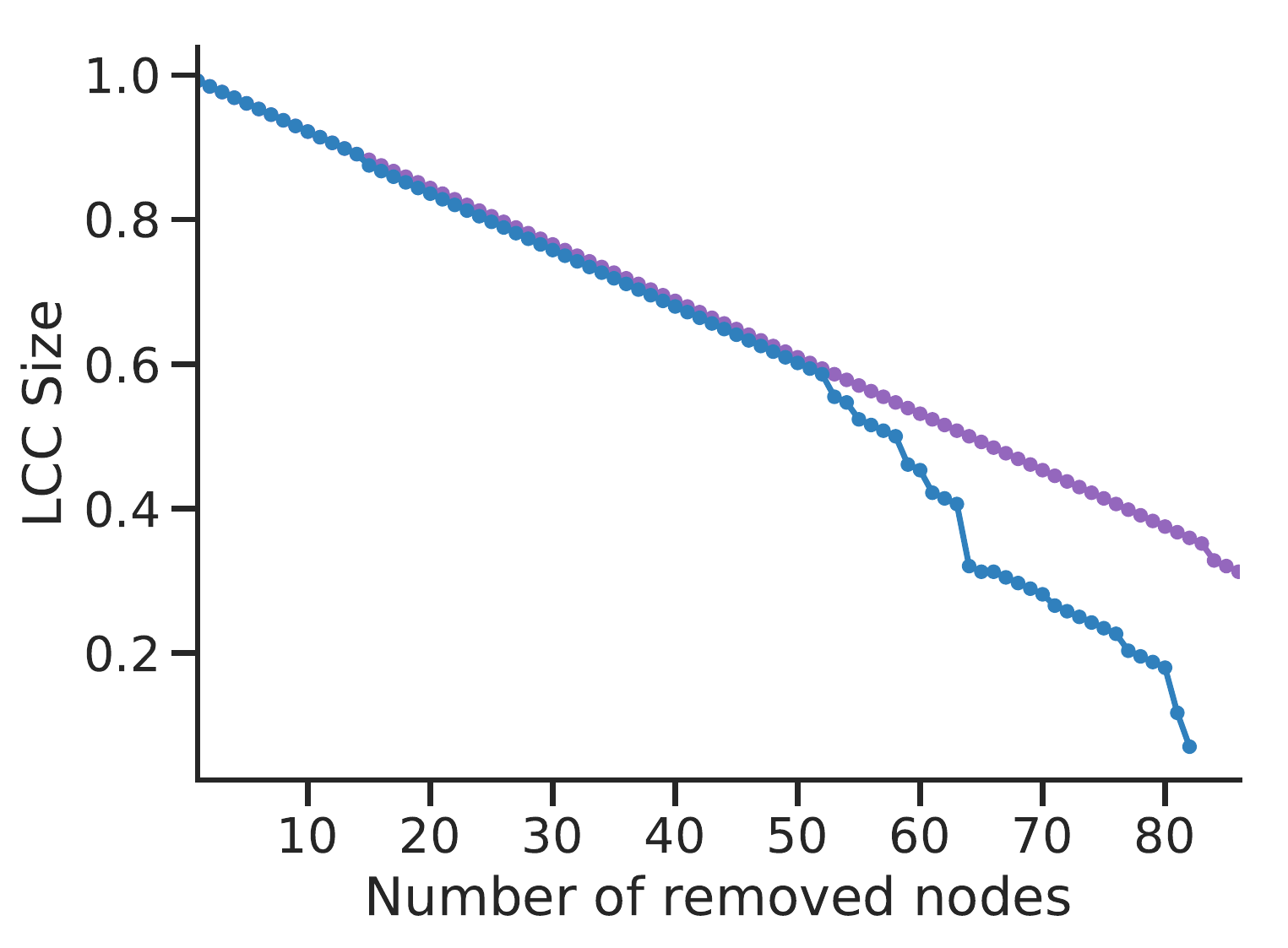}
		\caption{foodweb-baywet degree}
		\label{f:foodweb-baywet_heuristics_degree_improvement}
	\end{subfigure}%
	\begin{subfigure}{0.50\textwidth}
		\centering
		\includegraphics[width=\textwidth]{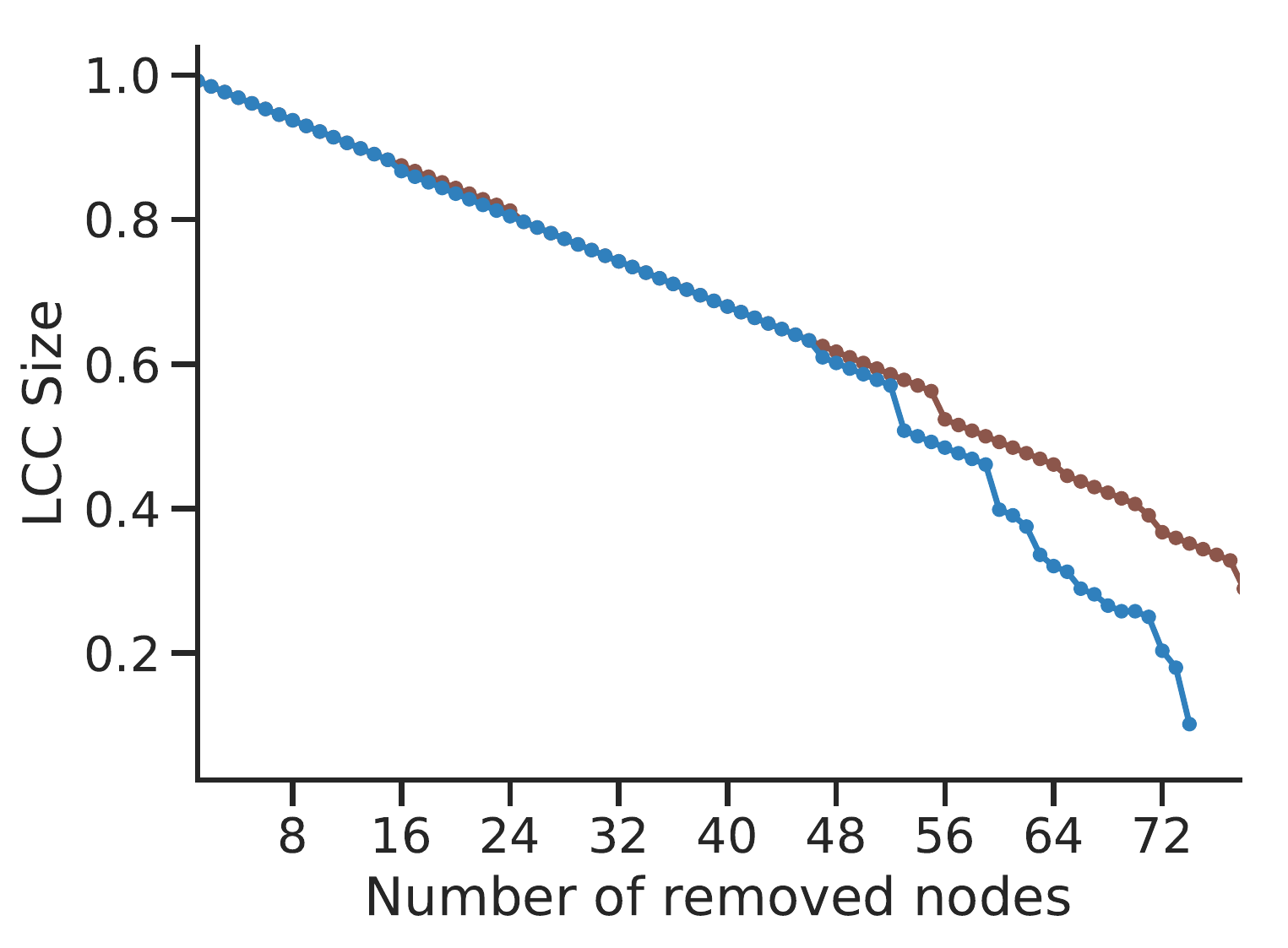}
		\caption{foodweb-baywet betweenness}
		\label{f:foodweb-baywet_heuristics_betweenness_improvement}
	\end{subfigure}%
	\hfill
	\begin{subfigure}{0.50\textwidth}
		\centering
		\includegraphics[width=\textwidth]{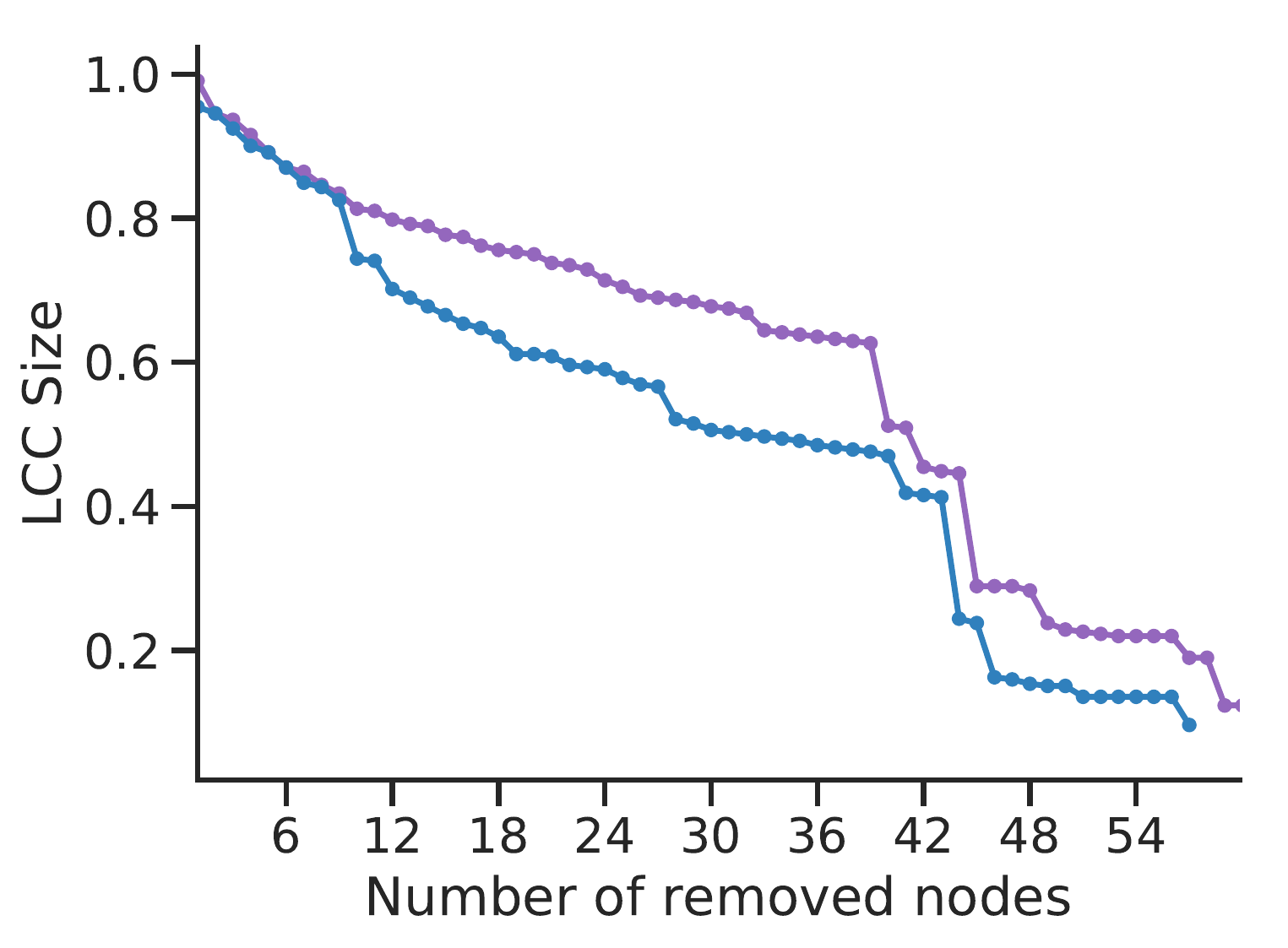}
		\caption{inf-USAir97 degree}
		\label{f:inf-USAir97_heuristics_degree_improvement}
	\end{subfigure}%
	\begin{subfigure}{0.50\textwidth}
		\centering
		\includegraphics[width=\textwidth]{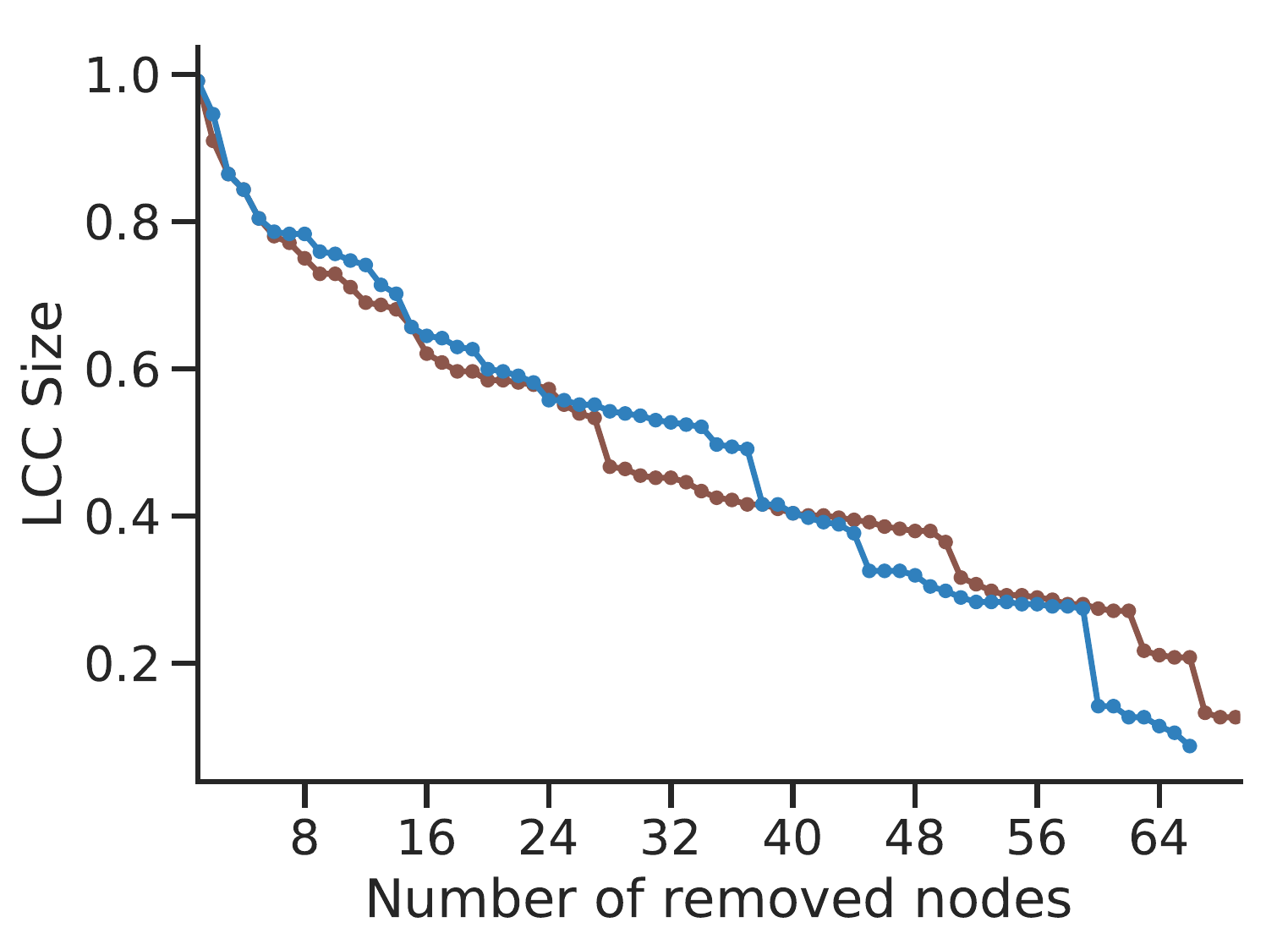}
		\caption{inf-USAir97 betweenness}
		\label{f:inf-USAir97_heuristics_betweenness_improvement}
	\end{subfigure}%
	\caption{Comparison of degree and betweenness vanilla heuristics with their GDM-enhanced versions on the arenas-meta, foodweb-baywet and inf-USAir97 networks.}
   \label{f:heuristics_improvement}
\end{figure}

\subparagraph*{Enhancement of metric-based heuristics}
In order to better understand how our framework is able to outperform cutting-edge algorithms, we compare existing node metric-based heuristics (e.g., removal of nodes in degree order) against GDM models that employ the corresponding node metric as the only node feature.
As an example, in
Figure~\ref{f:heuristics_improvement}
we display the enhancement of the degree and the betweenness based heuristics in the left and right columns respectively.
These GDM-enhanced heuristics effectively outperform the vanilla ones, highlighting the fact that the model is able to capture the importance of the nodes thanks to the feature propagation discussed before.
This also gives an important insight as the model seems to learn correlations between node features.

\begin{figure}[htbp!]
	\centering
	\begin{subfigure}{0.50\textwidth}
		\centering
   		\includegraphics[width=\textwidth]{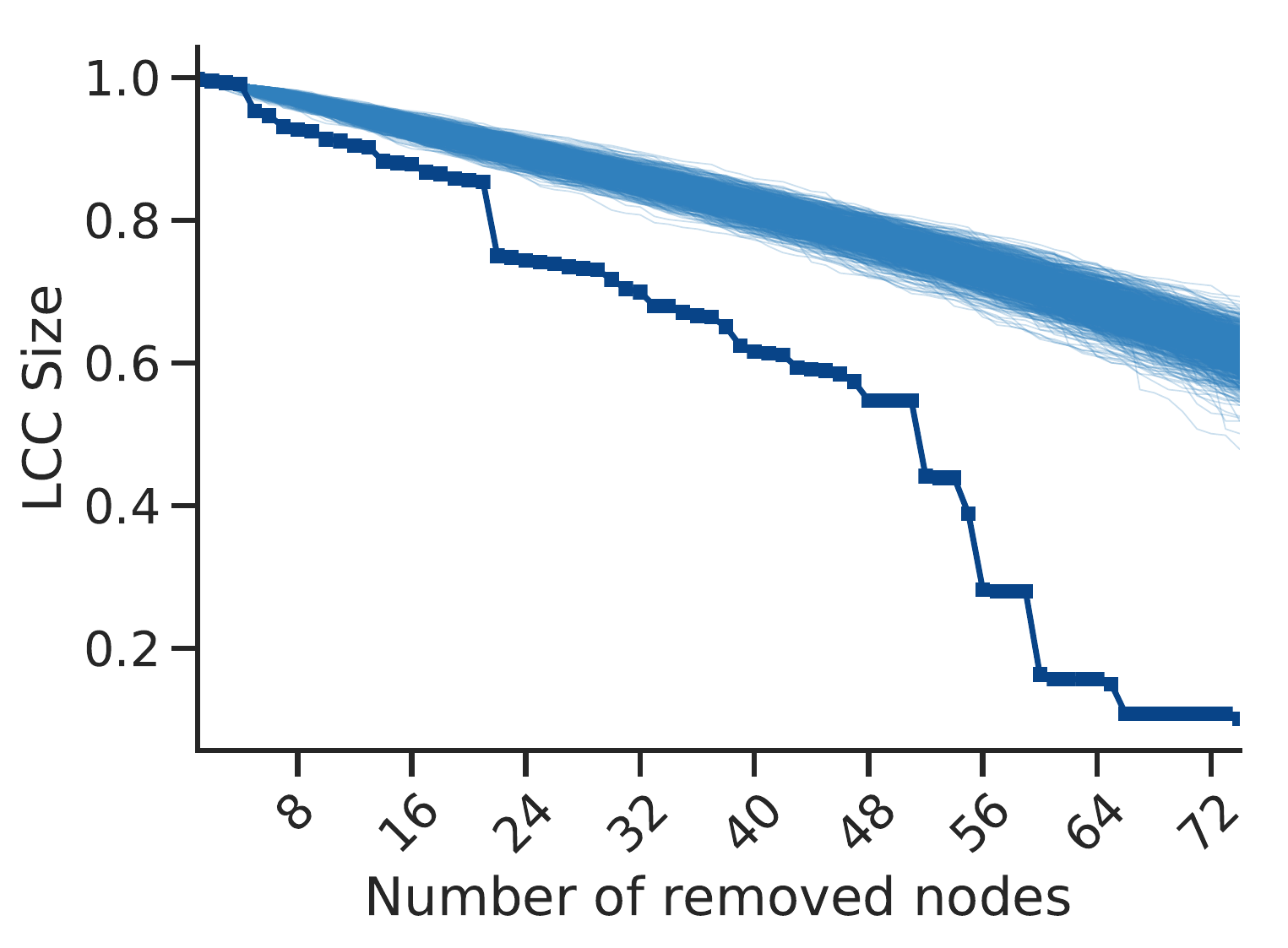}
   		\caption{arenas-meta network}
   		\label{f:arenas_rewirings}
	\end{subfigure}%
	\begin{subfigure}{0.50\textwidth}
		\centering
   		\includegraphics[width=\textwidth]{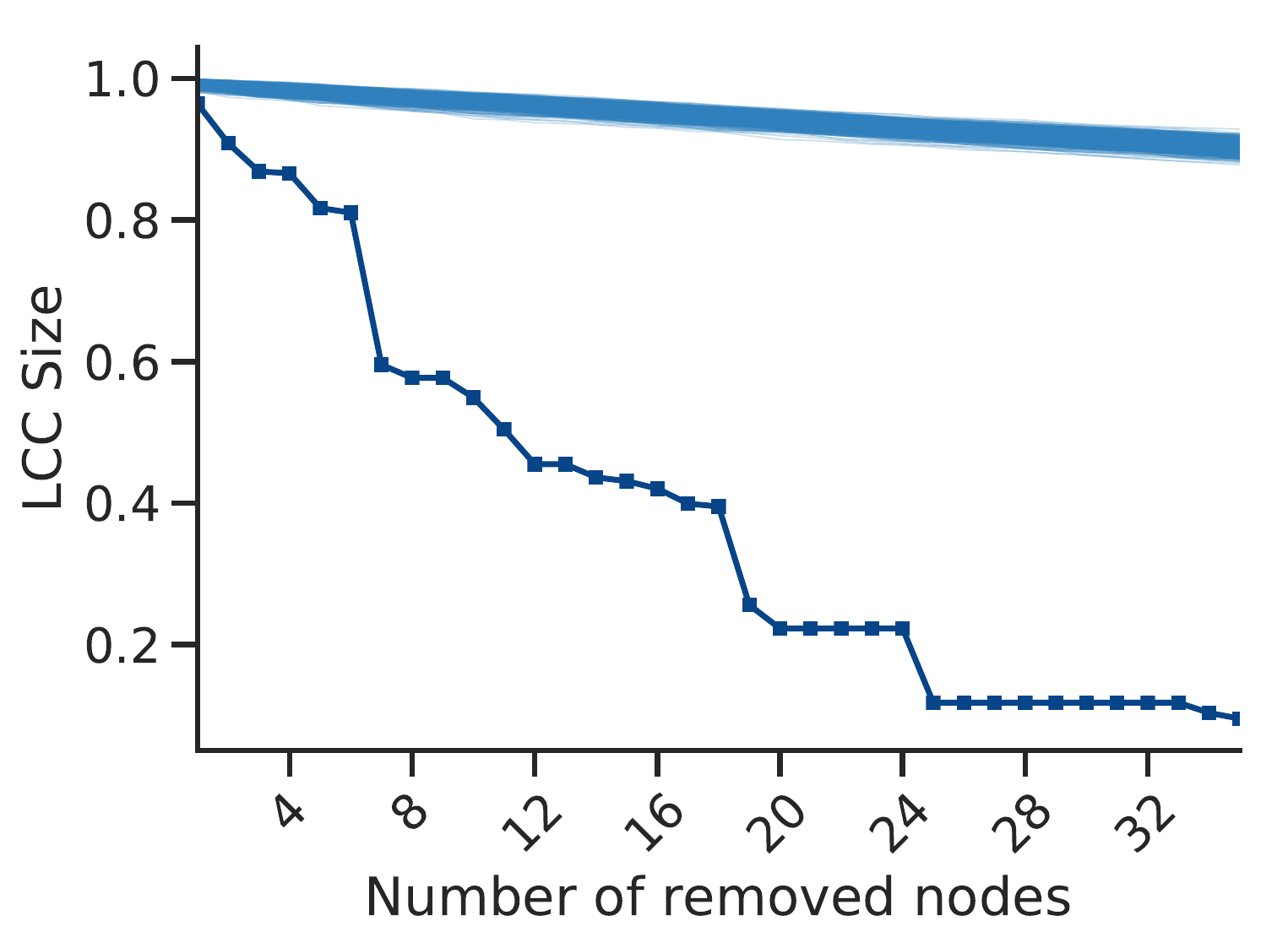}
   		\caption{Moreno crime network}
   		\label{f:moreno_rewirings}
	\end{subfigure}%
	\hfill
	\begin{subfigure}{0.50\textwidth}
		\centering
			\includegraphics[width=\textwidth]{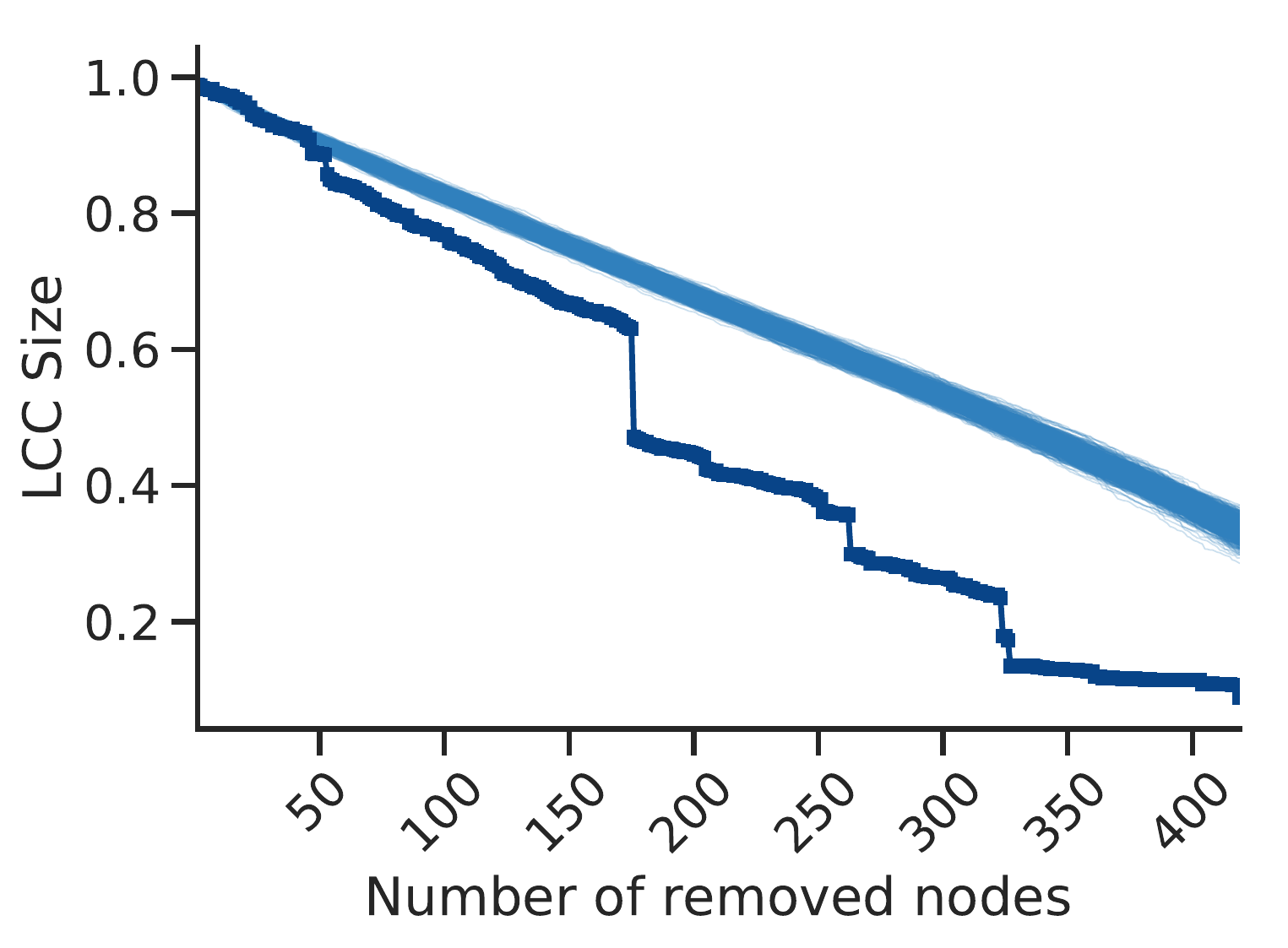}
			\caption{opsahl-openflights network}
			\label{f:opsahl-openflights_rewirings}
	\end{subfigure}%
	\begin{subfigure}{0.50\textwidth}
		\centering
			\includegraphics[width=\textwidth]{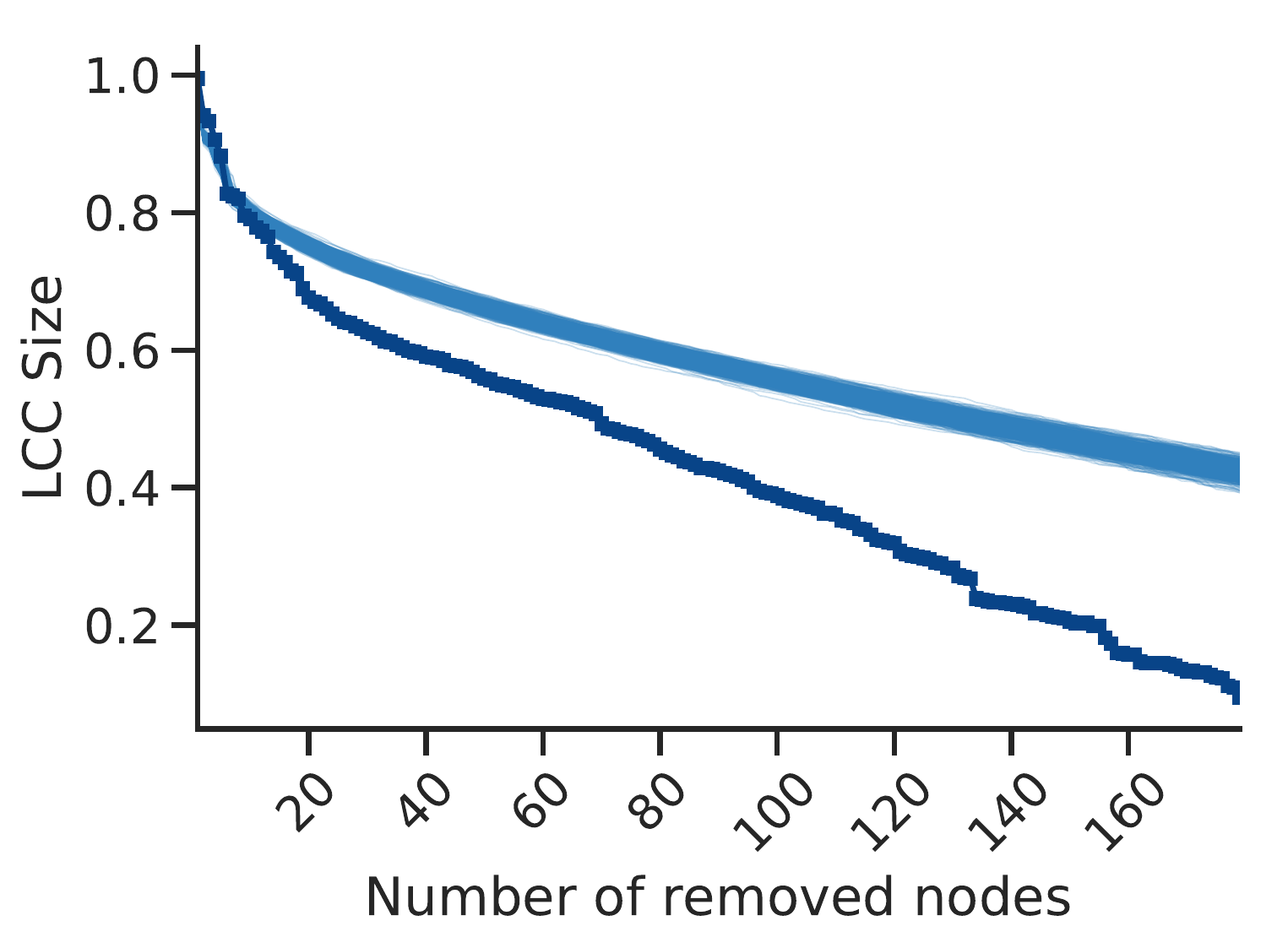}
			\caption{route-views network}
			\label{f:route-views_rewirings}
	\end{subfigure}
	\caption{Dismantling of original networks (dark blue) and $1000$ configuration model rewirings for each (light blue).}
	\label{f:configuration_model_rewirings}
\end{figure}

\subparagraph*{Dismantling of configuration model rewired networks}
We investigate further if the model is learning correlations among node features by dismantling the configuration model rewirings\footnote{The configuration model of a network keeps the observed connectivity distribution while destroying
topological correlations, meaning that feature correlations are lost.} of the networks in our test set.
If that is the case, the dismantling power of our approach on the rewirings should be heavily affected.
In
Figure~\ref{f:configuration_model_rewirings}
we show, for each network, the dismantling of $1000$ configuration models and also the original instance as comparison.
In all the tested networks, there is a severe performance drop.
For instance, in
Figures~\ref{f:moreno_rewirings}
it takes just $\sim35$ removals to dismantle the original instance of the Moreno crime network, while the LCC size of the rewired networks after the same number of removals is still very large (i.e., $\sim95\%$).
This result confirms our insight.
That is, existing topological correlations are learned and, consequently, exploited by the machine.

\subparagraph*{Dismantling results}
In the main paper we compare our approach with the state-of-the-art algorithms.
In Table~\ref{t:full_test_network_table} we report the same results in numerical form.

The table also includes other commonly used static attack approaches that remove the nodes in descending importance order according to some node centrality metric.
While many heuristics fall in this category, we compare with the removal of nodes in descending degree~\cite{PhysRevE.65.056109}, betweenness~\cite{PhysRevE.65.056109} and PageRank~\cite{page1999pagerank}.
Our approach outperforms all these static approaches with a significant margin, even the ones with higher computational complexity (e.g., the betweenness-based one).



\begin{table}[ht]
	\begin{adjustbox}{width=\textwidth}
	  \begin{tabular}{| l | c | c | c | c | c | c | c | c | c | c | c | c | c | c | c | c |}
		  \hline
Heuristic &    GDM &    GND &   EGND &  Adaptive degree &  EI $\sigma_1$ &  Pagerank &  Degree &  Betweenness &      MS &  EI $\sigma_2$ &  GDM +R &  GND +R &  CoreHD &  MS +R &  CI $\ell-2$ \\
Network                 &        &        &        &                   &                &           &         &                         &         &                &         &         &         &        &                                \\
\hline
ARK201012\_LCC           &  100.0 &   99.7 &  100.1 &             103.3 &          128.4 &     103.1 &   104.9 &                   123.3 &   130.9 &         3883.7 &    94.5 &    87.6 &    92.6 &   95.8 &                          114.6 \\
advogato                &  100.0 &  108.0 &  105.5 &             101.8 &          111.6 &     150.1 &   113.4 &                   114.8 &   112.6 &          494.5 &    94.8 &    97.5 &   102.1 &  102.7 &                           98.8 \\
arenas-meta             &  100.0 &  129.0 &  141.9 &             103.6 &          120.4 &     114.8 &   116.4 &                   142.4 &   120.5 &          579.3 &    90.8 &    92.5 &    95.5 &   94.9 &                           96.9 \\
cfinder-google          &  100.0 &  160.4 &  246.5 &              99.5 &          233.7 &     113.5 &   141.3 &                   377.9 &   682.8 &         1609.3 &    67.5 &   105.6 &   101.0 &  166.9 &                          114.0 \\
corruption              &  100.0 &   99.3 &  126.9 &             157.3 &          236.3 &     147.5 &   400.1 &                   166.7 &   864.8 &         1141.9 &    97.6 &   147.4 &   138.6 &  139.6 &                          176.6 \\
dblp-cite               &  100.0 &  113.3 &  121.7 &             113.5 &          111.7 &     114.7 &   131.6 &                   119.0 &   139.8 &          533.5 &   103.9 &   108.5 &   132.3 &  132.5 &                          117.1 \\
dimacs10-celegansneural &  100.0 &   85.0 &   95.9 &             103.1 &          105.7 &     116.4 &   120.8 &                   125.1 &   117.5 &          182.2 &    94.2 &   103.8 &   111.6 &  110.3 &                           99.7 \\
dimacs10-polblogs       &  100.0 &  107.5 &   97.1 &             102.1 &          115.5 &     112.5 &   117.9 &                   114.8 &   107.5 &          262.3 &    98.4 &   108.4 &   106.0 &  104.9 &                          104.6 \\
econ-wm1                &  100.0 &  130.3 &  114.4 &             109.8 &          128.0 &     131.0 &   129.4 &                   132.7 &   107.7 &          309.3 &    99.6 &   109.4 &   106.0 &  105.9 &                          126.3 \\
ego-twitter             &  100.0 &  116.8 &  115.8 &             108.9 &          103.0 &     107.8 &   108.8 &                   133.3 &   167.3 &         6017.4 &    98.8 &    98.2 &   114.4 &  111.7 &                          103.9 \\
eu-powergrid            &  100.0 &   75.9 &   89.1 &             138.8 &           73.8 &     180.1 &   163.5 &                   174.5 &   290.9 &         3313.0 &    64.4 &    66.5 &    83.4 &   92.8 &                          109.4 \\
foodweb-baydry          &  100.0 &  104.5 &   99.5 &              98.1 &          103.0 &     120.5 &   122.3 &                   109.4 &   104.4 &          125.2 &    97.8 &    98.0 &   101.2 &   99.3 &                          110.6 \\
foodweb-baywet          &  100.0 &  110.2 &  108.4 &              99.6 &          103.9 &     123.6 &   125.4 &                   112.9 &   106.8 &          128.3 &    98.5 &   108.5 &   102.1 &  101.8 &                          113.0 \\
inf-USAir97             &  100.0 &  112.4 &  117.8 &             130.4 &          147.0 &     117.1 &   139.1 &                   128.6 &   164.0 &          633.6 &   100.1 &   117.2 &   103.7 &  107.6 &                          129.8 \\
internet-topology       &  100.0 &   95.6 &   95.8 &              99.1 &          113.9 &     109.2 &   131.4 &                   122.9 &   138.6 &         3879.9 &    94.8 &    84.7 &   100.2 &  101.7 &                          103.0 \\
librec-ciaodvd-trust    &  100.0 &  113.1 &  115.5 &             117.6 &          129.4 &     120.5 &   139.8 &                   114.9 &   126.6 &          634.5 &   104.3 &   114.4 &   124.4 &  126.3 &                          126.1 \\
librec-filmtrust-trust  &  100.0 &  108.9 &  118.3 &             117.7 &          112.8 &     131.8 &   148.4 &                   158.9 &   168.7 &         1308.2 &    89.7 &    95.5 &   106.8 &   98.6 &                           98.0 \\
linux                   &  100.0 &   97.9 &  101.1 &             116.2 &           84.5 &     176.0 &   190.8 &                   365.1 &   150.0 &         1035.2 &    78.3 &    71.4 &    74.1 &   80.1 &                           92.1 \\
loc-brightkite          &  100.0 &  100.2 &  100.3 &              98.6 &           97.7 &     104.3 &   110.9 &                   122.1 &   106.7 &          593.9 &    89.5 &    99.7 &    92.1 &   92.4 &                           93.0 \\
maayan-Stelzl           &  100.0 &  144.1 &  133.0 &             102.5 &          114.3 &     113.4 &   127.7 &                   137.0 &   111.7 &         1269.6 &    96.3 &   113.4 &   107.1 &  105.2 &                          105.4 \\
maayan-figeys           &  100.0 &  104.3 &  120.2 &             100.7 &          155.9 &     127.3 &   146.9 &                   153.4 &   129.5 &         1656.6 &    98.0 &   100.1 &   123.7 &  123.4 &                           99.5 \\
maayan-foodweb          &  100.0 &  111.5 &   94.6 &             114.7 &          147.8 &     118.9 &   123.8 &                   126.2 &   154.6 &          268.7 &   100.0 &   125.5 &   136.1 &  144.4 &                          173.9 \\
maayan-vidal            &  100.0 &  111.0 &  106.7 &             103.3 &          101.6 &     109.1 &   110.6 &                   123.9 &   114.1 &          843.9 &    90.1 &   102.5 &    95.6 &   97.9 &                           97.3 \\
moreno\_crime\_projected  &  100.0 &  105.8 &   86.0 &             191.2 &          139.2 &     157.6 &   218.8 &                   180.6 &   976.7 &         2103.3 &    82.7 &    88.8 &   100.3 &  104.1 &                          126.2 \\
moreno\_propro           &  100.0 &  115.9 &  123.6 &             115.6 &           87.9 &     126.1 &   123.5 &                   146.7 &   145.2 &         1985.3 &    90.7 &    94.6 &    92.2 &   93.1 &                           96.3 \\
moreno\_train            &  100.0 &  104.9 &  104.9 &             107.1 &          124.0 &     149.5 &   156.0 &                   134.7 &   176.9 &          408.8 &   100.0 &   109.7 &   115.6 &  120.3 &                          211.6 \\
munmun\_digg\_reply\_LCC   &  100.0 &  116.3 &  108.6 &              98.5 &          109.4 &     106.5 &   108.3 &                   117.5 &    98.9 &          556.8 &    95.6 &   104.0 &    99.0 &   98.4 &                           98.5 \\
opsahl-openflights      &  100.0 &  101.2 &  106.2 &             127.2 &          109.9 &     123.2 &   135.4 &                   123.6 &   157.3 &          807.7 &    84.4 &    92.0 &   102.6 &  111.3 &                          120.9 \\
opsahl-powergrid        &  100.0 &   36.9 &   69.4 &             148.6 &           37.0 &     173.4 &   180.9 &                   183.9 &   164.3 &         1508.1 &    43.1 &    42.1 &    51.4 &   52.5 &                           65.6 \\
opsahl-ucsocial         &  100.0 &  122.1 &  116.1 &              99.9 &          118.5 &     105.9 &   109.9 &                   109.8 &   108.8 &          342.0 &    97.0 &   106.1 &   105.8 &  106.0 &                          101.7 \\
oregon2\_010526          &  100.0 &  106.8 &  101.5 &             108.8 &          131.1 &     101.6 &   130.5 &                   114.6 &   162.0 &         3247.5 &    90.0 &    80.5 &   113.0 &  112.8 &                           95.1 \\
p2p-Gnutella06          &  100.0 &  128.5 &  120.4 &             108.5 &          108.6 &     111.6 &   125.1 &                   118.4 &   108.7 &          274.0 &   101.4 &   120.4 &   110.1 &  108.4 &                          109.1 \\
p2p-Gnutella31          &  100.0 &  133.6 &    NaN &             109.1 &          112.7 &     110.3 &   123.1 &                   129.5 &   109.2 &          474.4 &   102.3 &   121.6 &   110.4 &  108.8 &                          109.8 \\
pajek-erdos             &  100.0 &  112.2 &  107.5 &             103.3 &          119.9 &     103.3 &   104.6 &                   106.7 &   122.8 &         2790.7 &    98.2 &   106.9 &   116.7 &  113.9 &                          101.0 \\
petster-hamster         &  100.0 &   92.5 &   90.9 &             122.7 &          103.8 &     135.1 &   127.2 &                   123.8 &   166.7 &          402.6 &    91.5 &    93.3 &    96.2 &   96.5 &                           98.6 \\
power-eris1176          &  100.0 &  199.1 &  218.0 &             340.2 &          171.7 &     253.5 &   622.5 &                   430.2 &   632.6 &         1957.4 &    86.6 &   154.8 &   161.3 &  157.8 &                          153.7 \\
route-views             &  100.0 &   99.3 &   99.1 &             101.8 &          133.2 &     103.5 &   103.5 &                   112.3 &   131.5 &         4340.9 &    94.0 &    82.0 &    93.0 &   95.2 &                          112.5 \\
slashdot-threads        &  100.0 &  100.1 &  102.2 &              99.5 &          122.5 &     104.6 &   105.4 &                   114.2 &   117.6 &         1495.8 &    96.1 &    95.8 &   115.7 &  115.1 &                           97.9 \\
slashdot-zoo            &  100.0 &   99.2 &  100.1 &              95.6 &          120.9 &     103.3 &   106.8 &                   124.0 &   112.4 &          683.8 &    95.2 &    97.7 &   106.9 &  105.9 &                           96.5 \\
subelj\_jdk              &  100.0 &  107.6 &  110.2 &             115.1 &          113.0 &     144.2 &   181.5 &                   346.9 &   144.6 &         1275.7 &    80.9 &    84.7 &    84.8 &   81.0 &                          103.4 \\
subelj\_jung-j           &  100.0 &  102.1 &  111.6 &             122.0 &          118.4 &     150.7 &   185.7 &                   334.6 &   143.1 &         1295.0 &    80.1 &    88.5 &    82.9 &   72.2 &                          101.5 \\
web-EPA                 &  100.0 &  148.4 &  157.6 &             102.2 &          141.1 &     104.9 &   109.7 &                   137.7 &   158.1 &         1471.9 &   101.1 &   115.6 &   133.8 &  132.8 &                          107.3 \\
web-webbase-2001        &  100.0 &  127.6 &  130.0 &             165.0 &          196.6 &     216.3 &   165.4 &                   207.7 &  3603.1 &        55066.4 &    64.7 &    50.1 &    76.6 &   82.6 &                           80.9 \\
wikipedia\_link\_kn       &  100.0 &  107.3 &  102.6 &             103.7 &          113.2 &     124.8 &   143.6 &                   140.3 &   128.8 &            NaN &    92.9 &    98.0 &   113.9 &  113.5 &                           96.8 \\
wikipedia\_link\_li       &  100.0 &  120.3 &  145.4 &             132.8 &          151.8 &     120.5 &   165.2 &                   110.6 &   211.9 &         1049.4 &   107.2 &   151.0 &   177.5 &  174.5 &                          157.8 \\
\hline
Average                 &  100.0 &  111.7 &  115.4 &             119.1 &          123.6 &     128.7 &   151.1 &                   158.9 &   273.3 &         2596.4 &    91.5 &   100.8 &   106.9 &  108.7 &                          112.1 \\
\hline
\end{tabular}%
\end{adjustbox}
\caption{Per-method area under the curve (AUC) of real-world networks dismantling. The lower the better. The dismantling target for each method is $10\%$ of the network size. We compute the AUC value by integrating the $LCC(x)/|N|$ values using Simpson’s rule, and each value is scaled to the one of our approach (GDM) for the same network. +R means that the reinsertion phase is performed. CoreHD and CI are compared to other +R algorithms as they include the reinsertion phase. EGND for p2p-Gnutella31 is missing as the computation was killed after 10d.}
\label{t:full_test_network_table}
\end{table}%

\subparagraph{Dismantling of large networks}
While in the main paper we compare our approach on small and medium size networks, in this section we extend the comparison against the more promising state-of-the-art algorithms (\emph{GND} and \emph{MS} with and without reinsertion, and \emph{CoreHD}) to $12$ large networks with up to $1.8$M nodes and up to $2.8$M edges.

\begin{figure}[htb!]
	\centering
 	\includegraphics[width=1.0\textwidth]{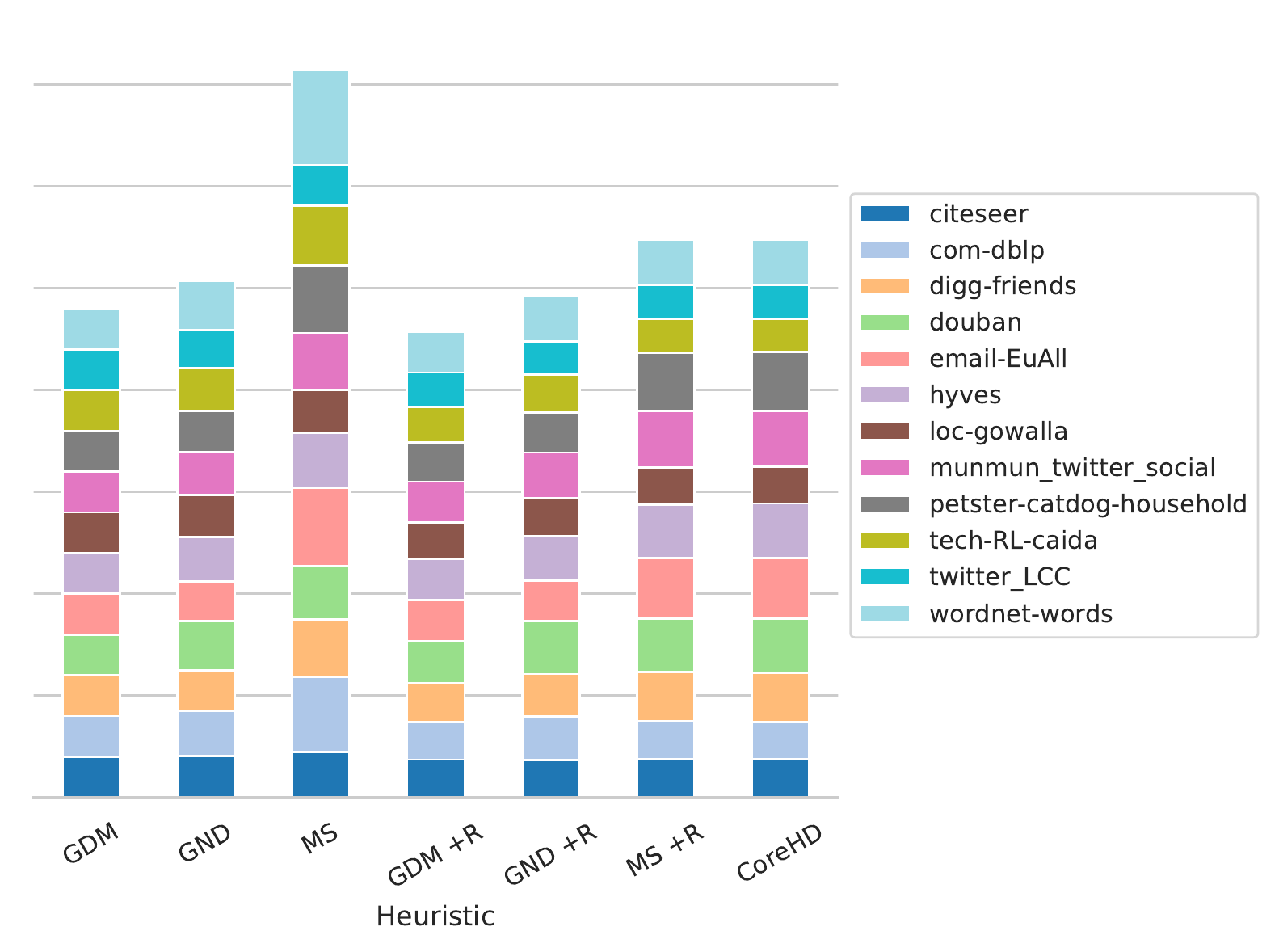}
 	\caption{\textbf{Dismantling empirical complex large systems.} Per-method cumulative area under the curve (AUC) of real-world networks dismantling. The lower the better. The dismantling target for each method is $10\%$ of the network size. We compute the AUC value by integrating the $LCC(x) / |N|$ values using Simpson's rule, and each value is scaled to the one of our approach (GDM) for the same network. \emph{GND} stands for \emph{Generalized Network Dismantling} (with cost matrix $\mathbf{W} = \mathbf{I}$) and \emph{MS} stands for \emph{Min-Sum}. +R means that the reinsertion phase is performed. Also, note that some values are clipped (limited) to 3x for the \emph{MS} heuristic to improve visualization.}
 	\label{f:empirical_large_results}
\end{figure}

As shown in Figure~\ref{f:empirical_large_results} and in Table~\ref{t:large_test_networks}, the results from the main paper are confirmed even for these networks, although with smaller margins.
This is still impressive as the proposed approach is static while the others recompute the nodes' structural importance during the dismantling process, which involves many removals for these networks (e.g., $70$K on \emph{hyves} network) and changes the network topology drastically, confirming the validity of our approach.


\begin{table}[ht]
	\begin{adjustbox}{width=\textwidth}
\begin{tabular}{| l | c | c | c | c | c | c | c |}
\hline
Heuristic &    GDM &    GND &     MS &  GDM +R &  GND +R &  MS +R &  CoreHD \\
Network                  &        &        &        &         &         &        &         \\
\hline
citeseer                 &  100.0 &  102.2 &  111.2 &    92.8 &    91.3 &   95.0 &    94.3 \\
com-dblp                 &  100.0 &  109.6 &  184.5 &    91.5 &   108.3 &   92.4 &    91.2 \\
digg-friends             &  100.0 &  100.9 &  140.5 &    97.0 &   103.6 &  120.7 &   121.2 \\
douban                   &  100.0 &  120.8 &  132.7 &   102.6 &   129.3 &  131.6 &   132.9 \\
email-EuAll              &  100.0 &   97.0 &  192.1 &   100.0 &   100.0 &  147.4 &   148.5 \\
hyves                    &  100.0 &  109.3 &  133.6 &   101.6 &   109.6 &  131.9 &   133.6 \\
loc-gowalla              &  100.0 &  103.2 &  105.4 &    89.7 &    91.9 &   91.0 &    90.5 \\
munmun\_twitter\_social    &  100.0 &  105.2 &  140.5 &   100.2 &   112.4 &  138.5 &   137.3 \\
petster-catdog-household &  100.0 &  100.7 &  164.7 &    95.4 &    98.0 &  143.4 &   144.7 \\
tech-RL-caida            &  100.0 &  104.8 &  147.2 &    86.9 &    94.3 &   82.8 &    80.2 \\
twitter\_LCC              &  100.0 &   93.6 &   98.8 &    85.3 &    81.4 &   83.0 &    84.7 \\
wordnet-words            &  100.0 &  120.4 &  234.5 &   100.0 &   110.8 &  111.0 &   109.7 \\
\hline
Average                  &  100.0 &  105.6 &  148.8 &    95.3 &   102.6 &  114.1 &   114.1 \\
\hline
\end{tabular}%
\end{adjustbox}
\caption{Per-method area under the curve (AUC) of real-world large networks dismantling. The lower the better. The dismantling target for each method is $10\%$ of the network size. We compute the AUC value by integrating the $LCC(x)/|N|$ values using Simpson’s rule, and each value is scaled to the one of our approach (GDM) for the same network. +R means that the reinsertion phase is performed. CoreHD and CI are compared to other +R algorithms as they include the reinsertion phase.}
\label{t:large_test_networks}
\end{table}%

In Table~\ref{t:large_test_networks_timings}, we also report the prediction (if any) and dismantling time of each of the above mentioned methods to give a better idea on what their different computational complexities mean and translate into.
\begin{table}[ht]
\begin{adjustbox}{width=\textwidth}
\begin{tabular}{| l | c | c | c | c | c |}
    \hline
    {} & \multicolumn{2}{c}{Prediction time} & \multicolumn{3}{c}{Dismantle time} \\
    Heuristic &             GDM &          CoreHD &             GDM &             GND &              MS \\
    Network                  &                 &                 &                 &                 &                 \\
    \hline
    citeseer                 & 00:00:03.4 & 00:00:22.9 & 01:30:17.1 & 03:43:51.6 & 01:26:21.5 \\
    com-dblp                 & 00:00:02.9 & 00:00:14.9 & 00:22:30.7 & 04:57:25.6 & 00:59:38.4 \\
    digg-friends             & 00:00:02.8 & 00:00:19.9 & 00:08:01.9 & 00:30:55.5 & 01:11:37.4 \\
    douban                   & 00:00:01.3 & 00:00:06.1 & 00:01:10.1 & 00:03:34.8 & 00:11:40.4 \\
    email-EuAll              & 00:00:02.4 & 00:00:07.8 & 00:00:10.7 & 00:01:14.9 & 00:09:49.0 \\
    hyves                    & 00:00:13.5 & 00:00:36.6 & 03:08:02.7 & 08:21:22.8 & 02:03:26.9 \\
    loc-gowalla              & 00:00:02.0 & 00:00:15.9 & 00:17:22.3 & 01:27:28.0 & 00:46:15.0 \\
    munmun\_twitter\_social  & 00:00:04.3 & 00:00:14.3 & 00:00:53.5 & 00:07:53.4 & 00:29:13.9 \\
    petster-catdog-household & 00:00:03.9 & 00:00:40.6 & 00:44:20.5 & 03:58:17.1 & 02:16:02.8 \\
    tech-RL-caida            & 00:00:01.8 & 00:00:12.1 & 00:07:23.7 & 04:14:34.1 & 00:29:30.8 \\
    twitter\_LCC             & 00:00:04.4 & 00:00:13.0 & 00:32:01.0 & 05:33:36.3 & 00:19:18.8 \\
    wordnet-words            & 00:00:01.4 & 00:00:12.1 & 00:03:34.0 & 01:23:52.1 & 00:22:28.5 \\
    \hline
\end{tabular}
\end{adjustbox}
\caption{Real-world large networks dismantling timings. The lower the better. Time format is HH:MM:SS.s. \emph{MS} and \emph{GND} do not have prediction time as they refresh the predictions during the dismantling, while there is no \emph{CoreHD} dismantling column as we use our dismantler.}
\label{t:large_test_networks_timings}
\end{table}%

\subparagraph{Dismantling curves}
In
Figure~\ref{f:dismantling_curves_2},
we display the dismantling of most of our test networks and compare with the state-of-the-art algorithms and with the heuristics introduced in the previous paragraph.
As previously mentioned, one of the advantages of our approach is that we can choose the best model to reach a given objective.
As an example, we show the models that lower the area under the curve (GDM AUC) and the removals number (GDM \#Removals), which may overlap for some networks.
We also show the dismantling performing the reinsertion phase and compare with state-of-the-art algorithms in
Figure~\ref{f:dismantling_curves_reinserted}.

\begin{figure}[!ht]
	\centering
	\begin{subfigure}{\textwidth}
		\centering
		\includegraphics[width=\textwidth]{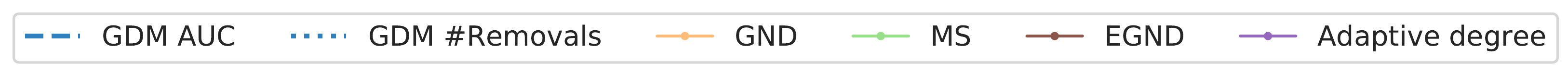}
	\end{subfigure}%
	\hfill
	\begin{subfigure}{0.5\textwidth}
		\centering
		\includegraphics[width=\textwidth]{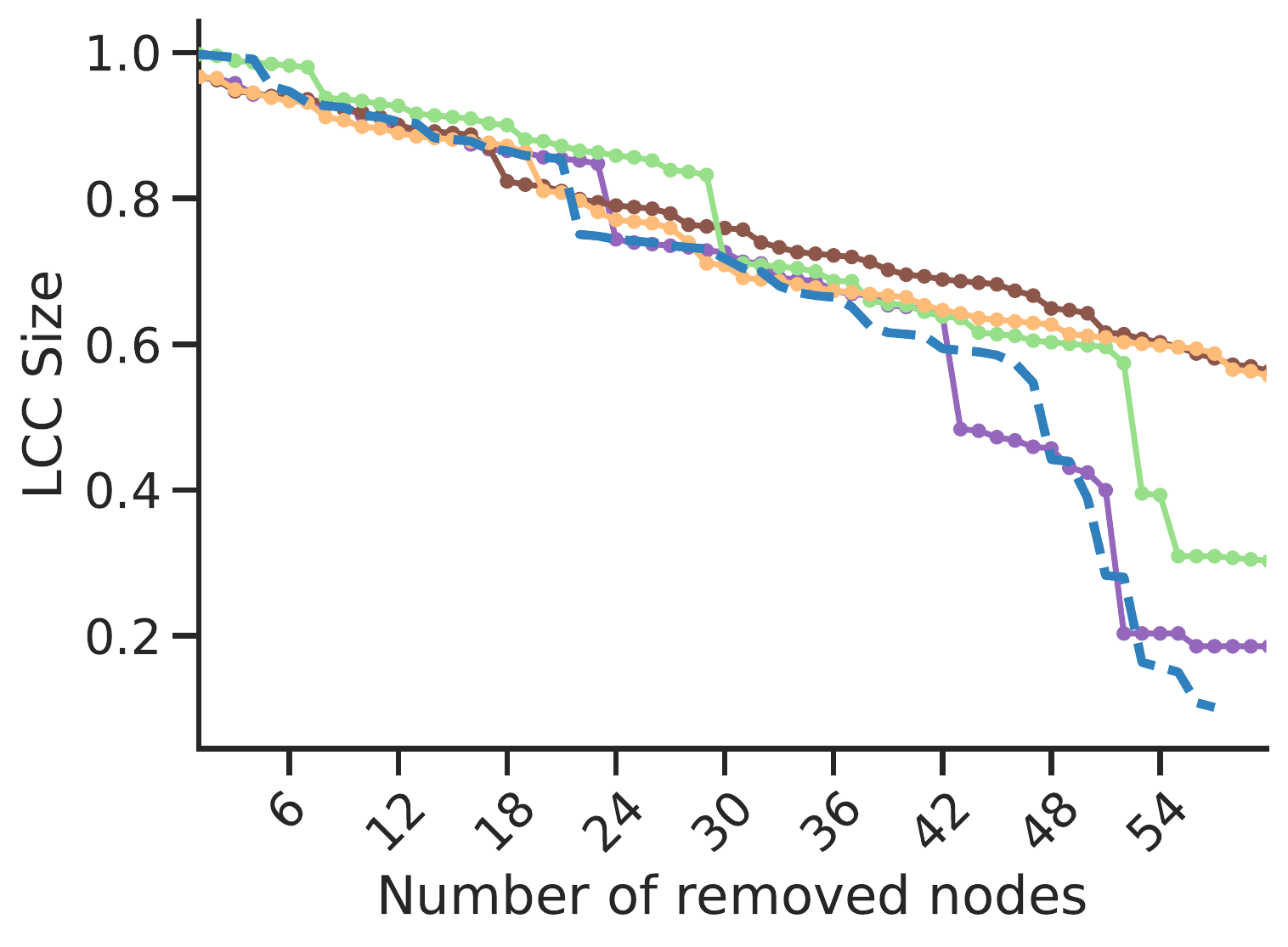}
		\caption{arenas-meta}
		\label{f:arenas_meta_dismantling}
	\end{subfigure}%
 	\begin{subfigure}{0.5\textwidth}
		\centering
		\includegraphics[width=\textwidth]{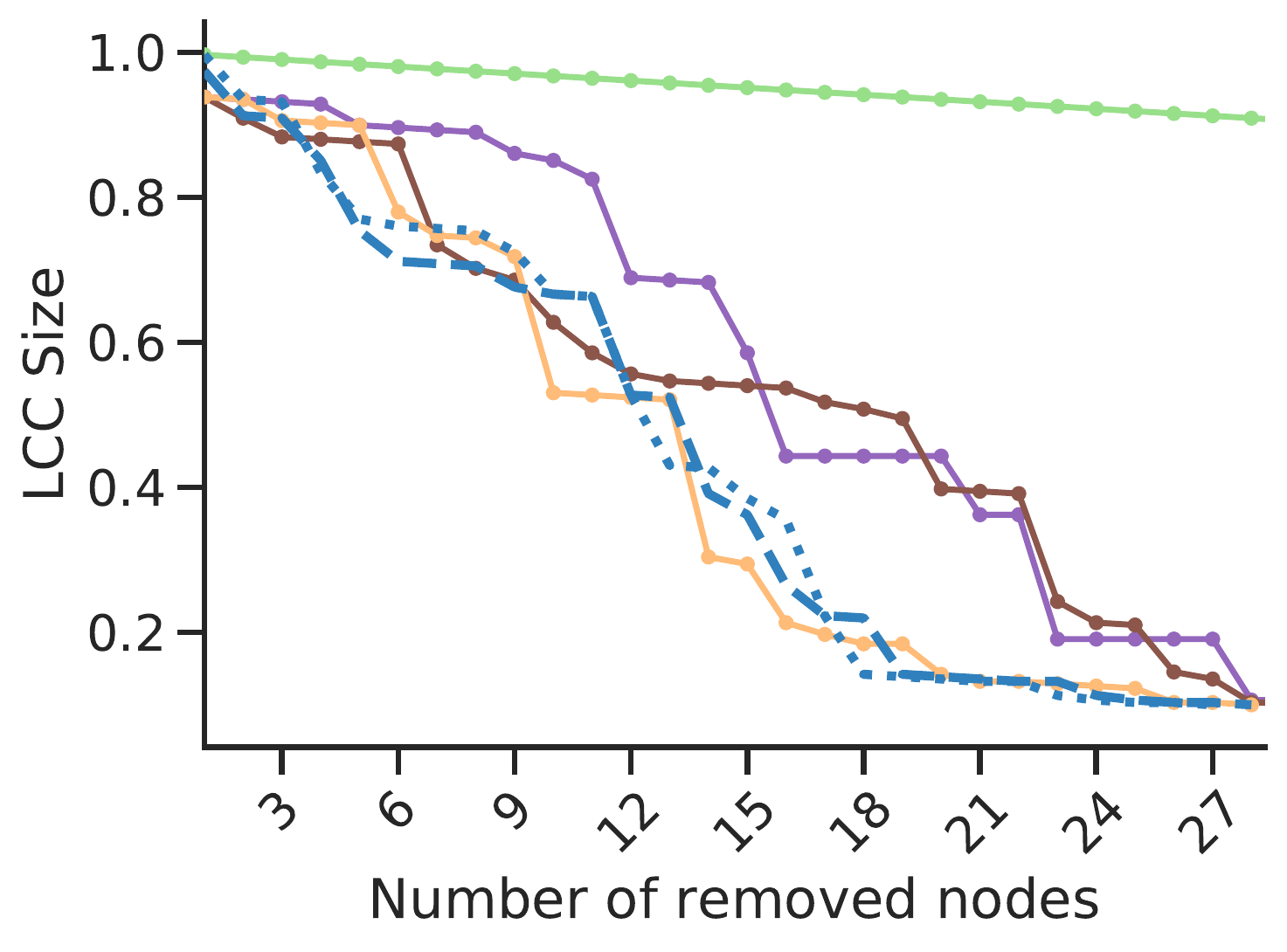}
		\caption{corruption}
		\label{f:corruption_dismantling}
	\end{subfigure}%
	\hfill
	\begin{subfigure}{0.5\textwidth}
		\centering
		\includegraphics[width=\textwidth]{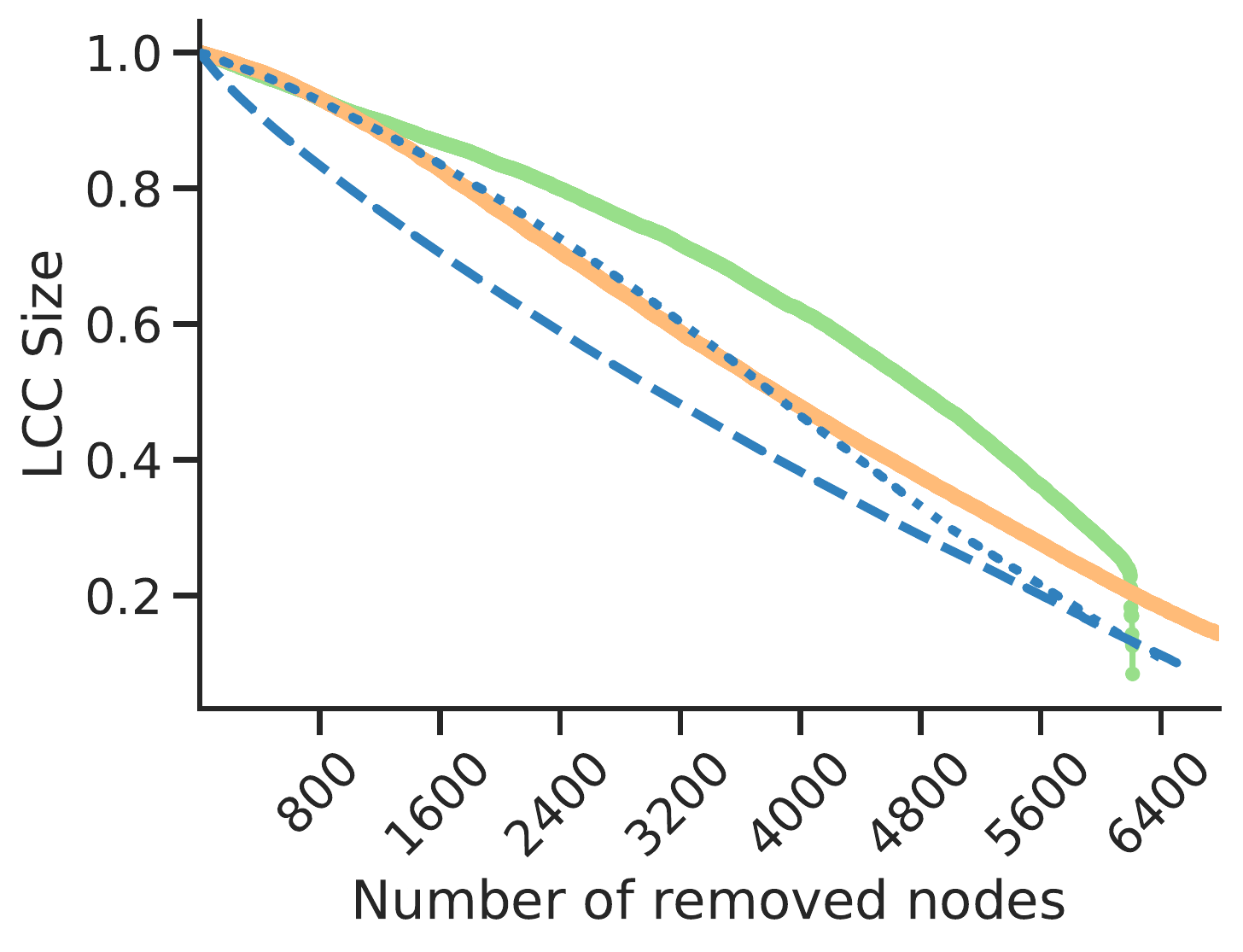}
		\caption{douban}
		\label{f:douban_dismantling}
	\end{subfigure}%
	\begin{subfigure}{0.5\textwidth}
		\centering
		\includegraphics[width=\textwidth]{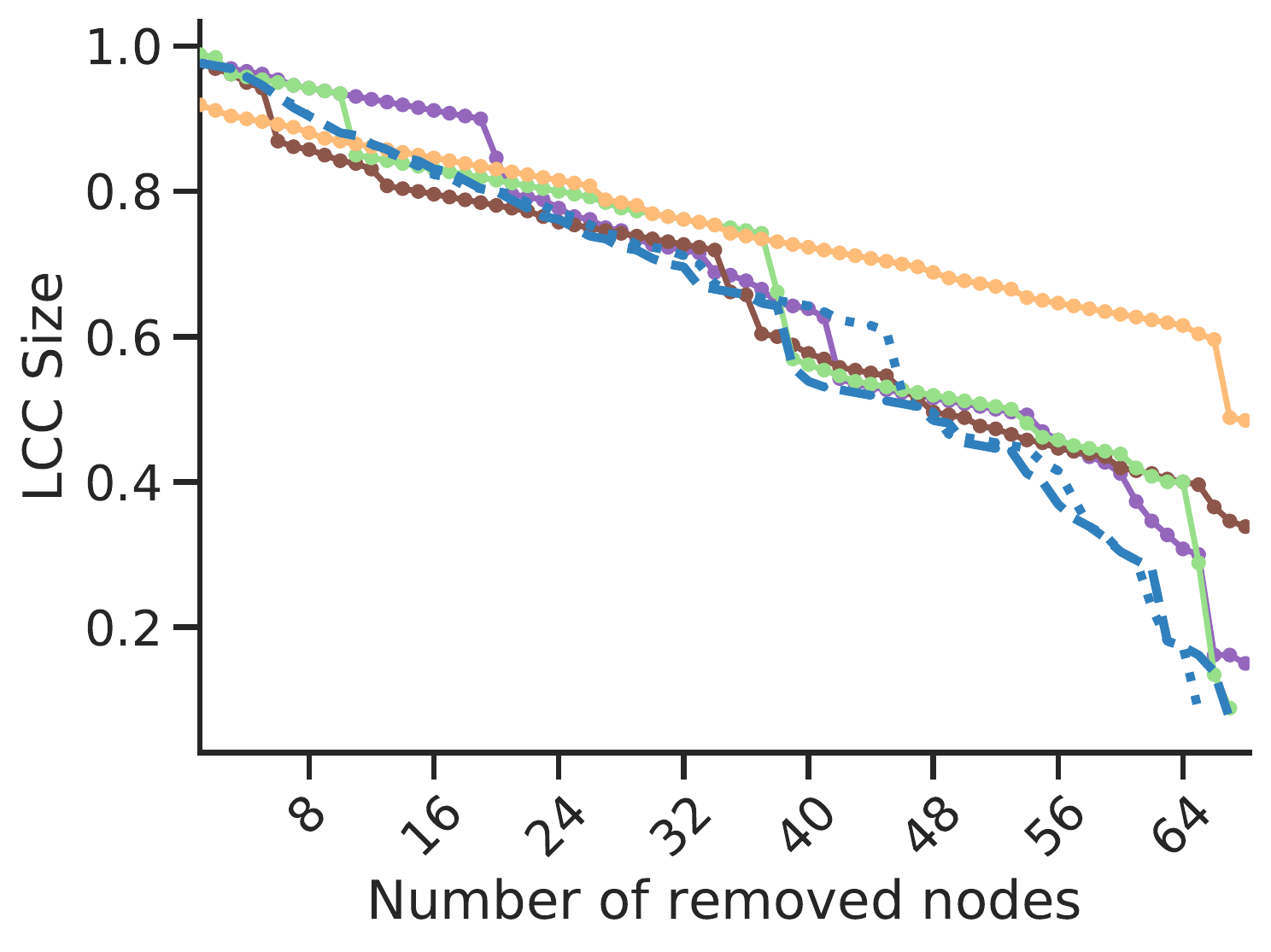}
		\caption{econ-wm1}
		\label{f:econ-wm1_dismantling}
	\end{subfigure}%
	\hfill
	\begin{subfigure}{0.5\textwidth}
		\centering
		\includegraphics[width=\textwidth]{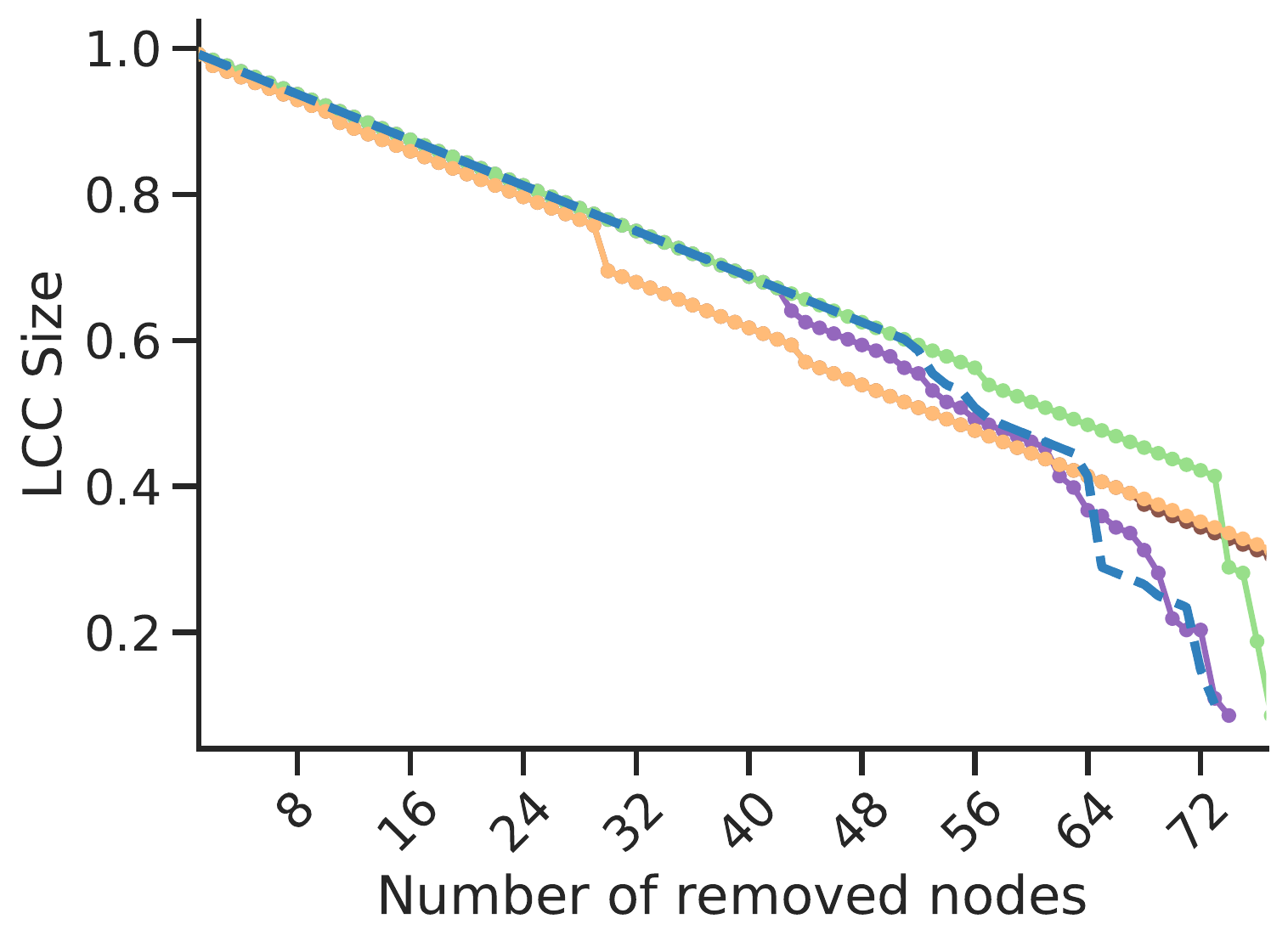}
		\caption{foodweb-baywet}
		\label{f:foodweb-baywet_dismantling}
	\end{subfigure}%
	\begin{subfigure}{0.5\textwidth}
		\centering
		\includegraphics[width=\textwidth]{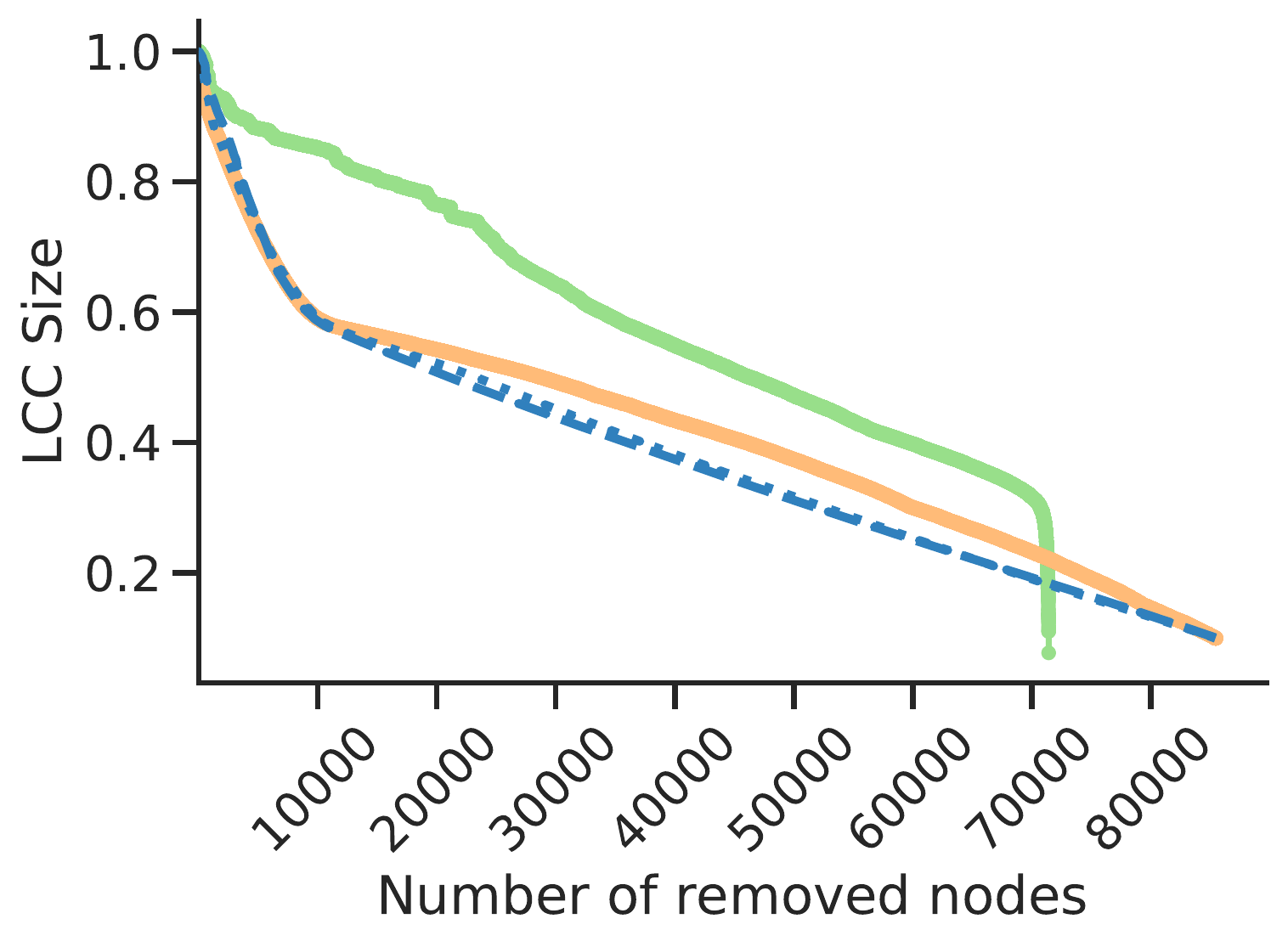}
		\caption{hyves}
		\label{f:hyves_dismantling}
	\end{subfigure}%
\end{figure}
\begin{figure}[!ht]\ContinuedFloat
	\centering
	\begin{subfigure}{0.5\textwidth}
		\centering
		\includegraphics[width=\textwidth]{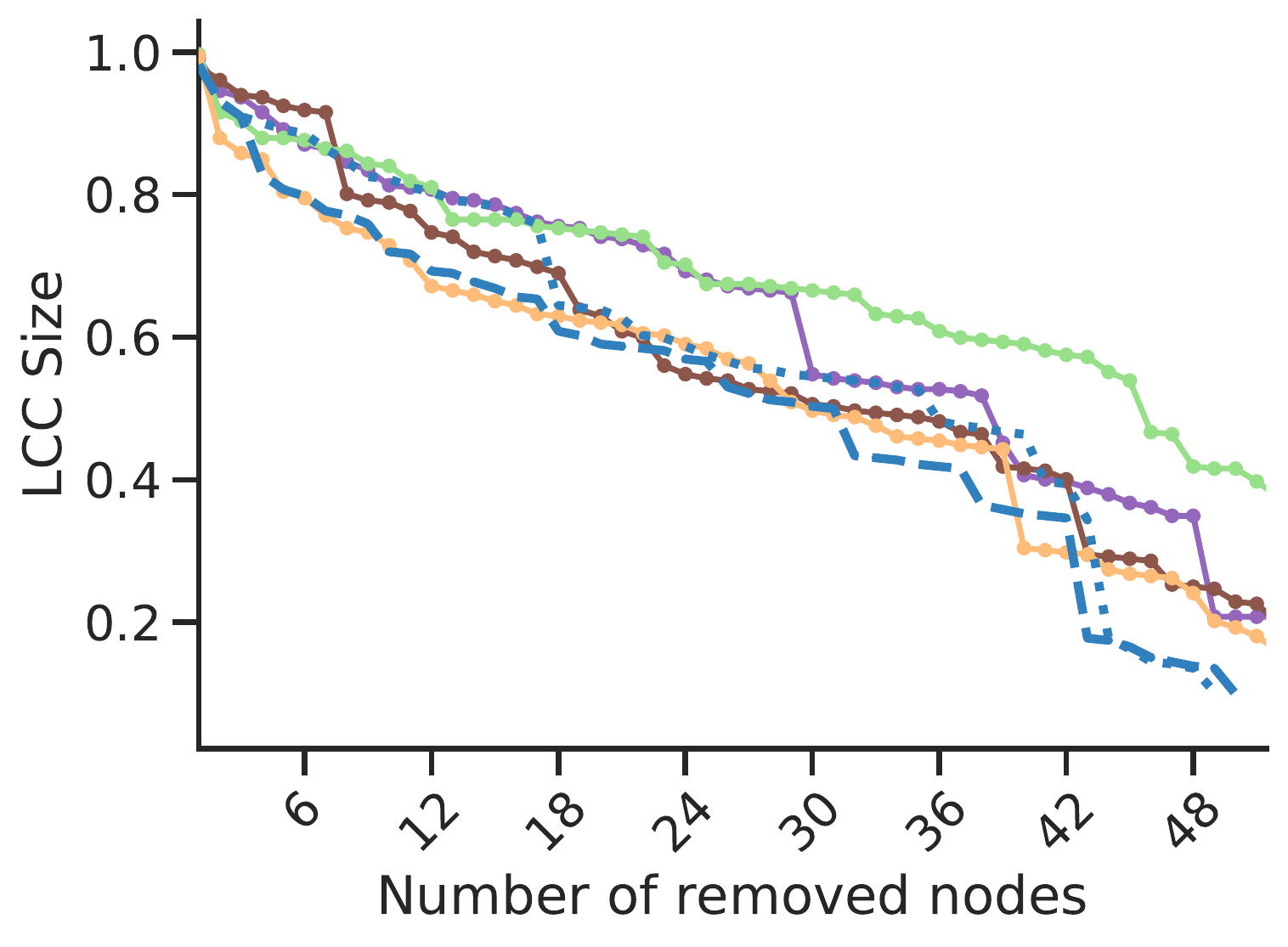}
		\caption{inf-USAir97}
		\label{f:inf-USAir97_dismantling}
	\end{subfigure}%
	\begin{subfigure}{0.5\textwidth}
	 \centering
	 \includegraphics[width=\textwidth]{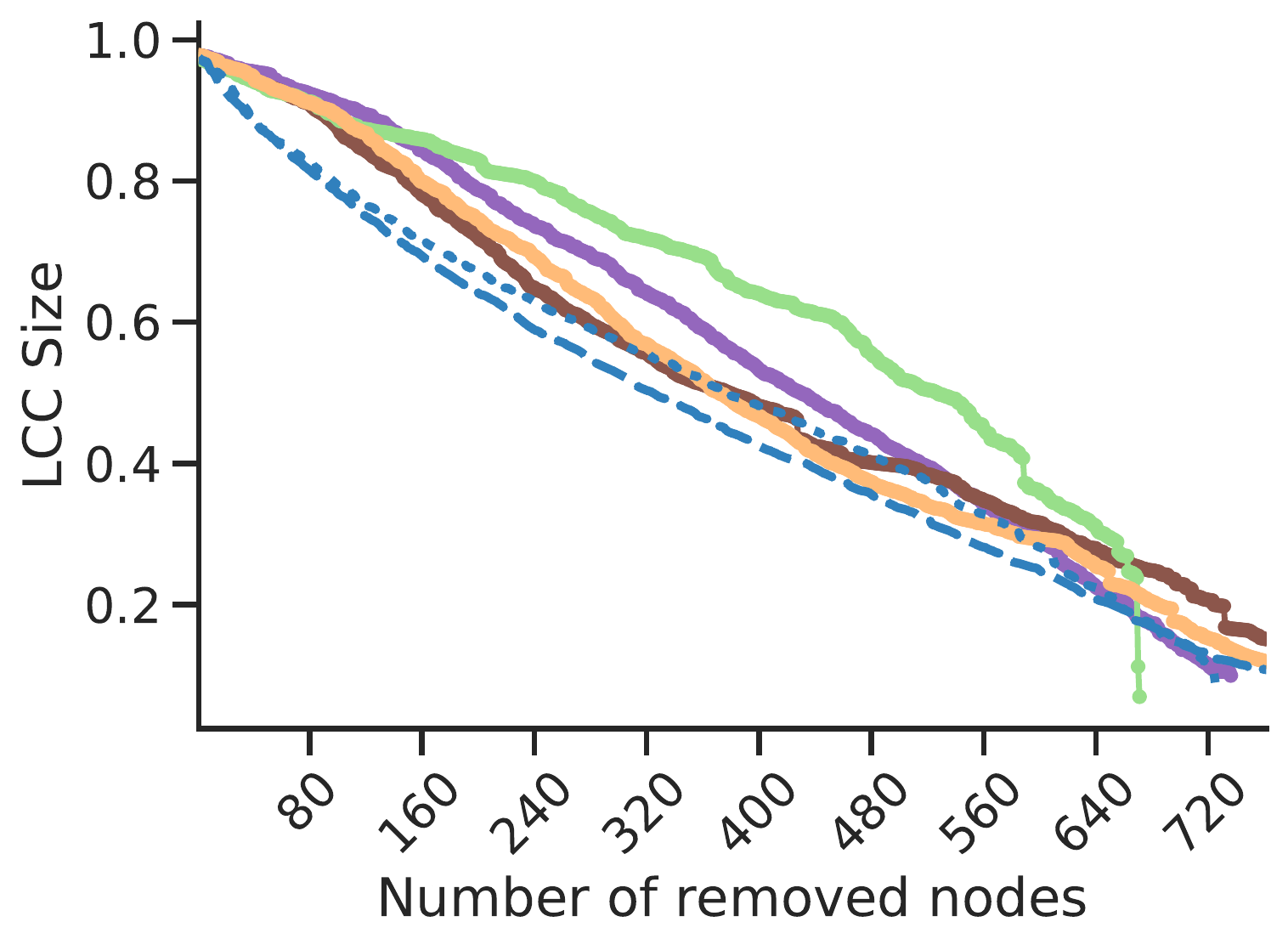}
	 \caption{librec-ciaodvd-trust}
	 \label{f:librec-ciaodvd-trust_dismantling}
	 \end{subfigure}%
\hfill
	\begin{subfigure}{0.5\textwidth}
		\centering
		\includegraphics[width=\textwidth]{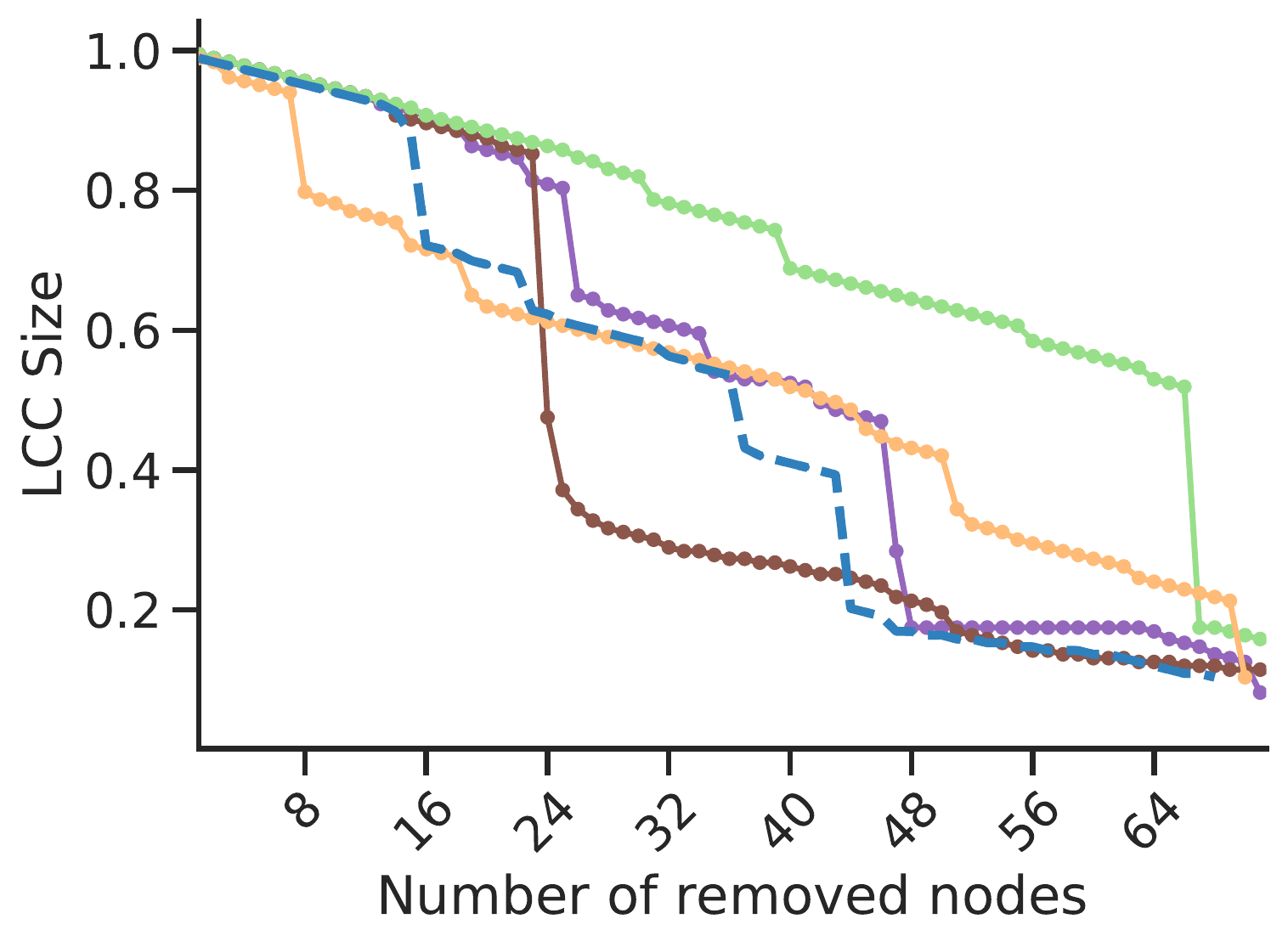}
		\caption{maayan-foodweb}
		\label{f:maayan-foodweb_dismantling}
	\end{subfigure}%
	\begin{subfigure}{0.5\textwidth}
		\centering
		\includegraphics[width=\textwidth]{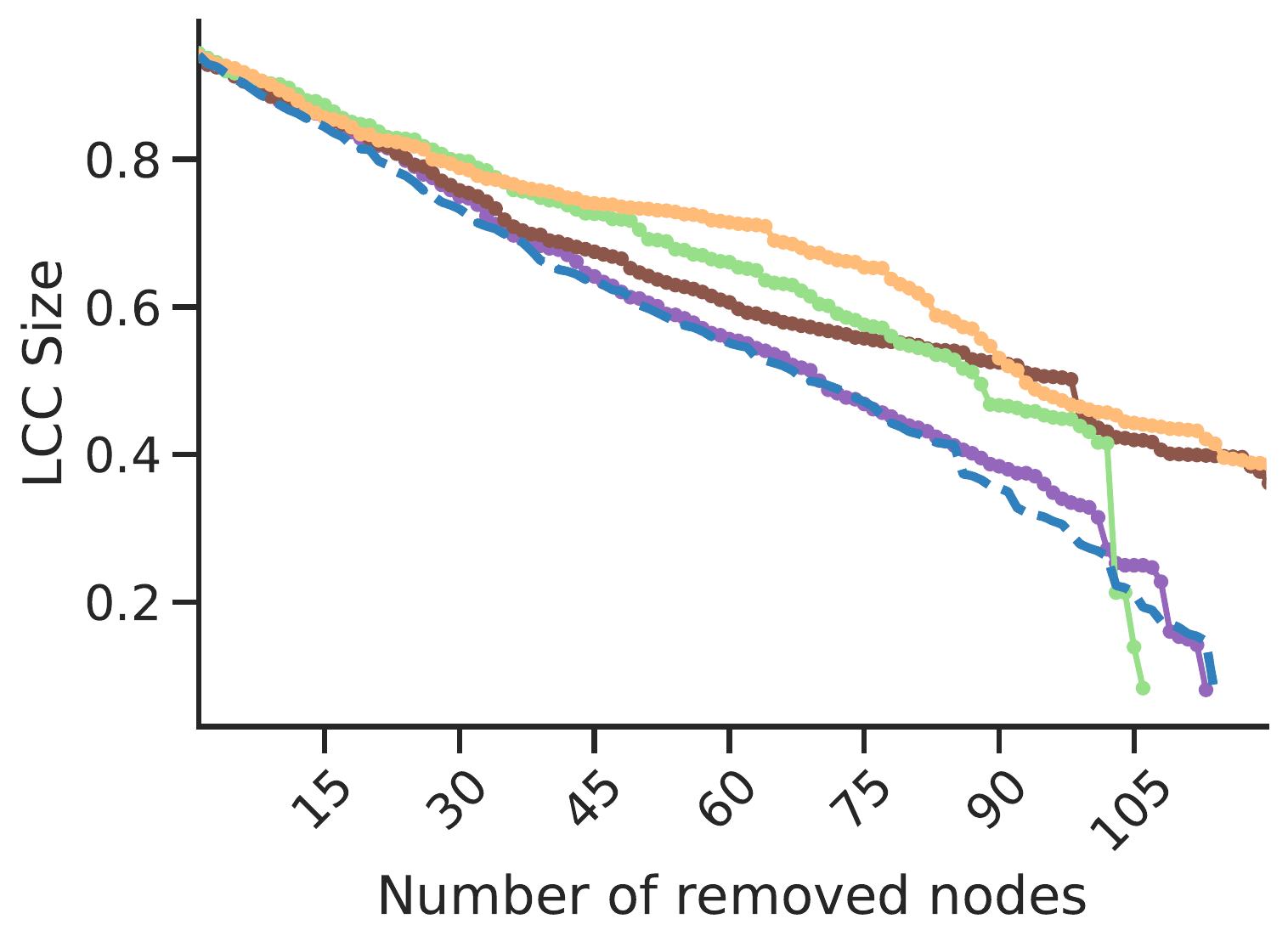}
		\caption{maayan-Stelzl}
		\label{f:maayan-Stelzl_dismantling}
	\end{subfigure}%

	\begin{subfigure}{0.5\textwidth}
		\centering
		\includegraphics[width=\textwidth]{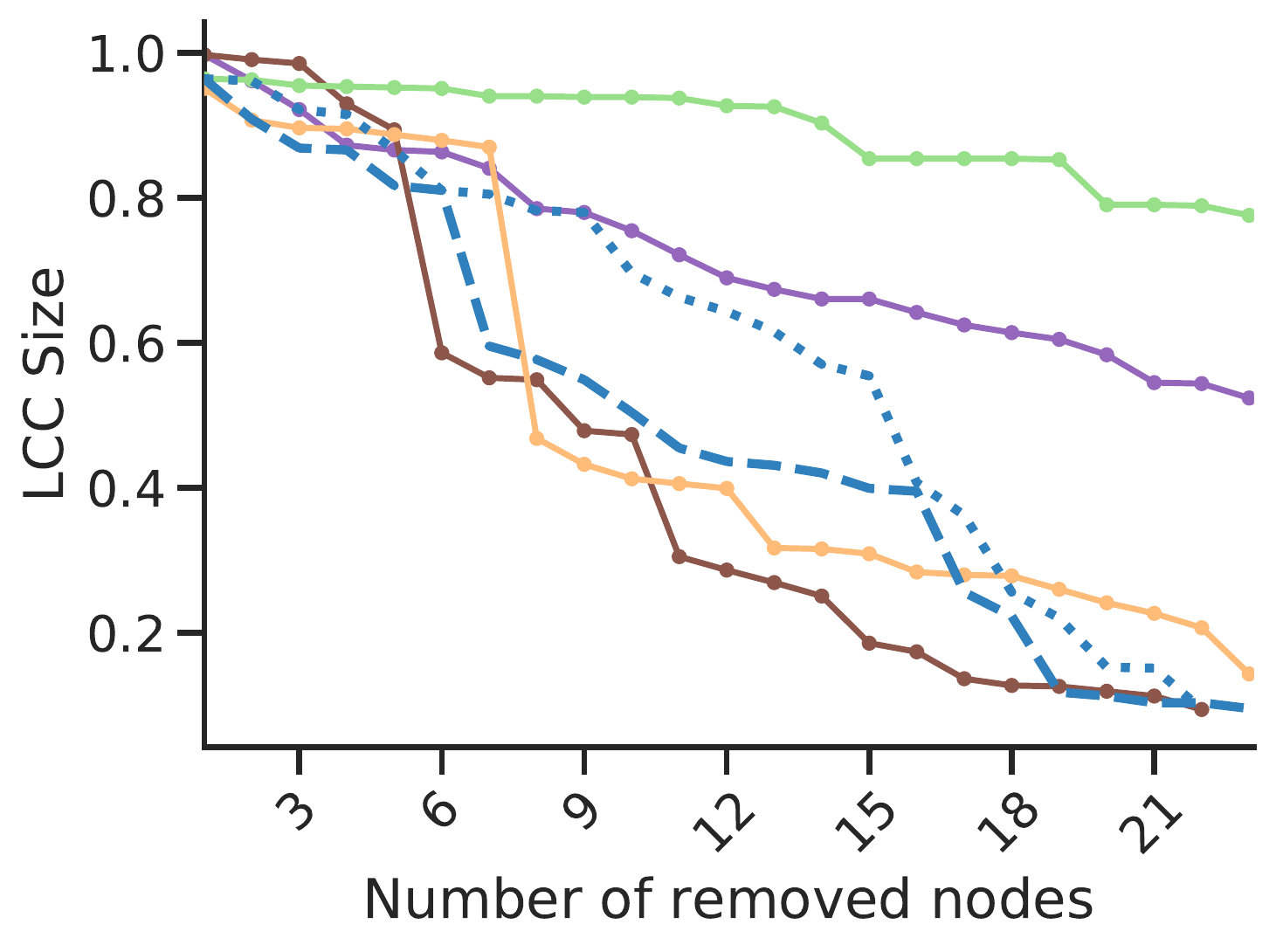}
		\caption{moreno-crime-projected}
		\label{f:moreno_crime_projected_dismantling}
	\end{subfigure}%
	\begin{subfigure}{0.5\textwidth}
		\centering
		\includegraphics[width=\textwidth]{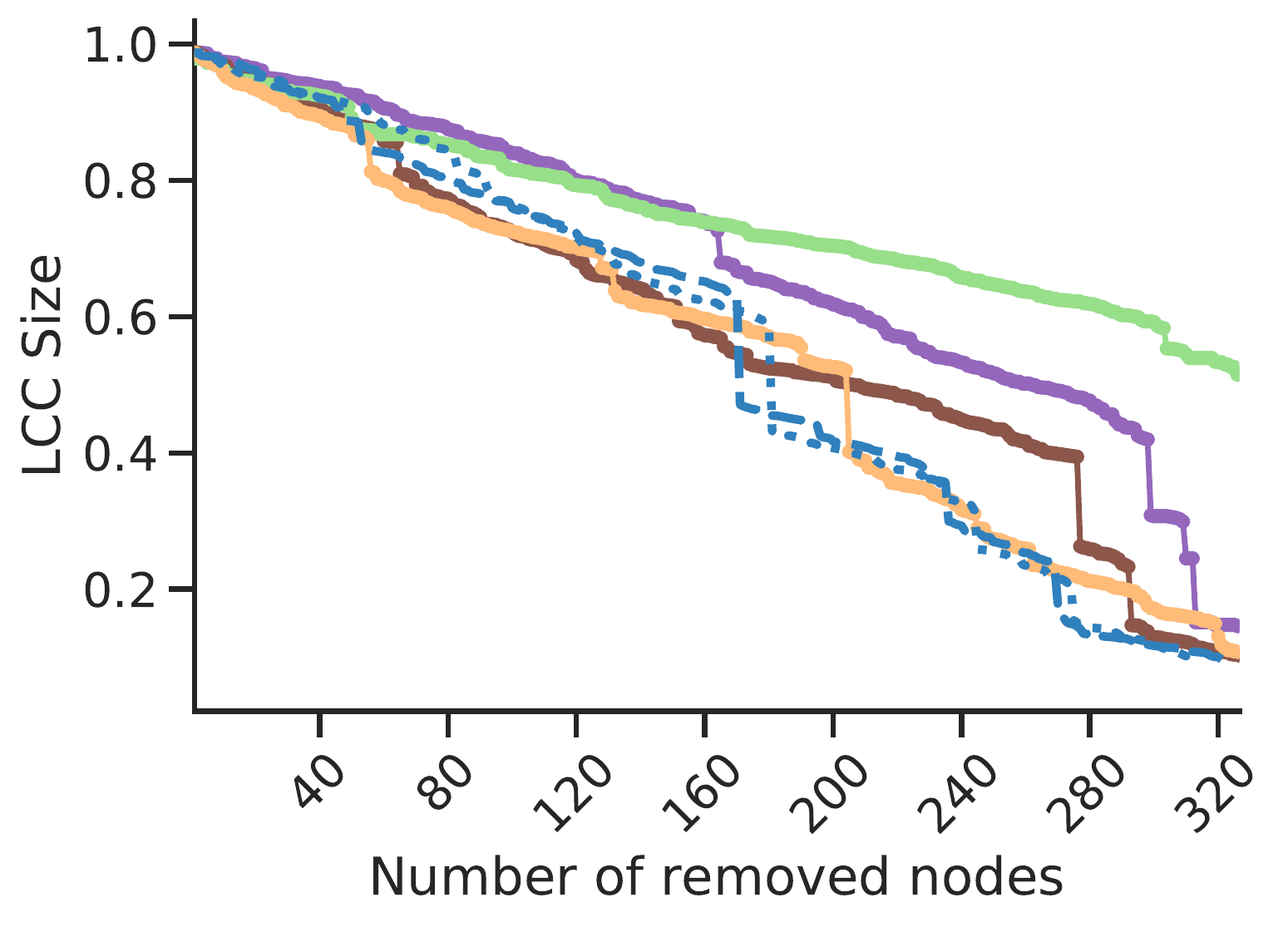}
		\caption{opsahl-openflights}
		\label{f:opsahl-openflights_dismantling}
	\end{subfigure}%
\end{figure}

\begin{figure}[!ht]\ContinuedFloat
	\centering
	\begin{subfigure}{0.5\textwidth}
		\centering
		\includegraphics[width=\textwidth]{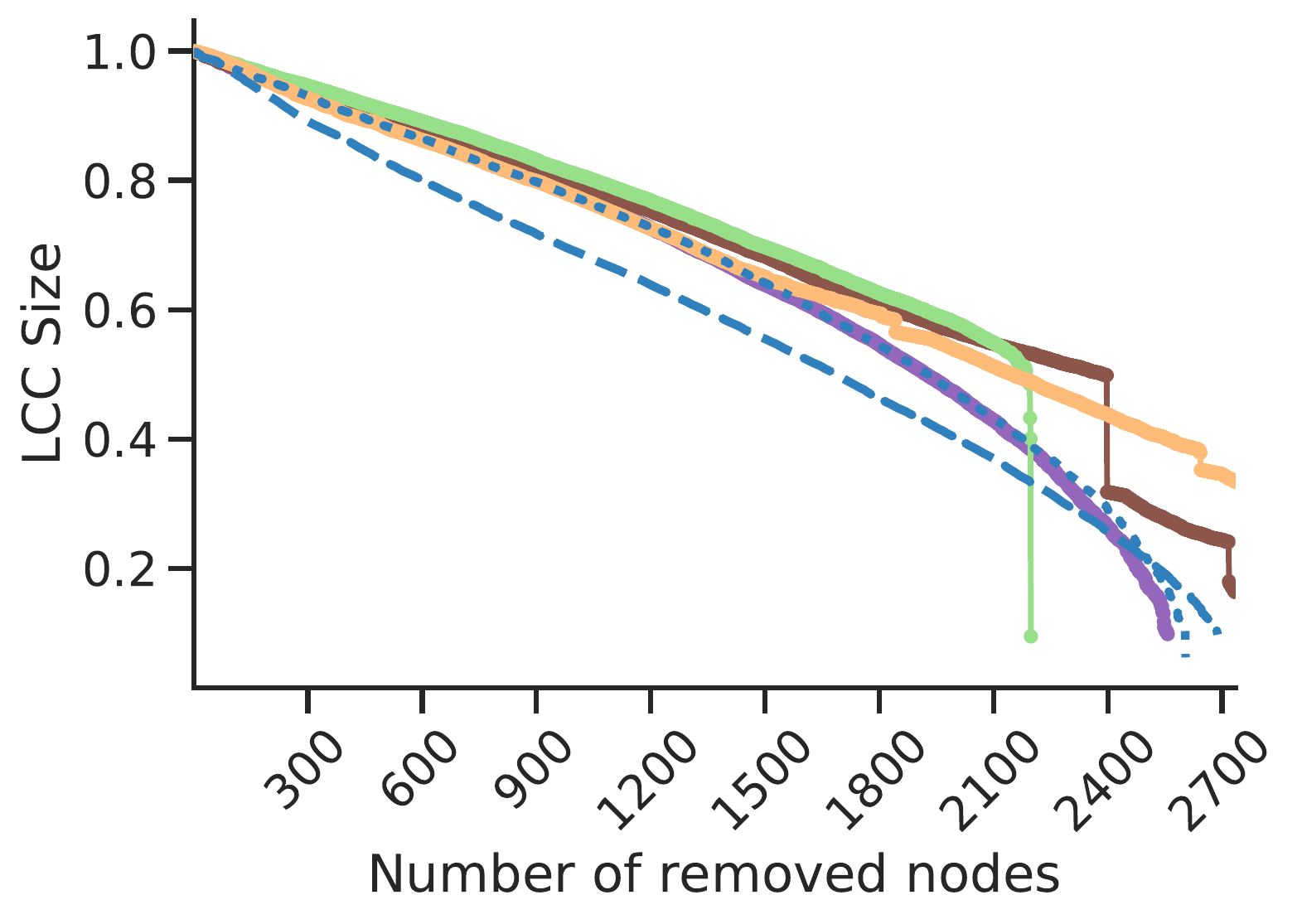}
		\caption{p2p-Gnutella06}
		\label{f:p2p-Gnutella06_dismantling}
	\end{subfigure}%
	\begin{subfigure}{0.5\textwidth}
		\centering
		\includegraphics[width=\textwidth]{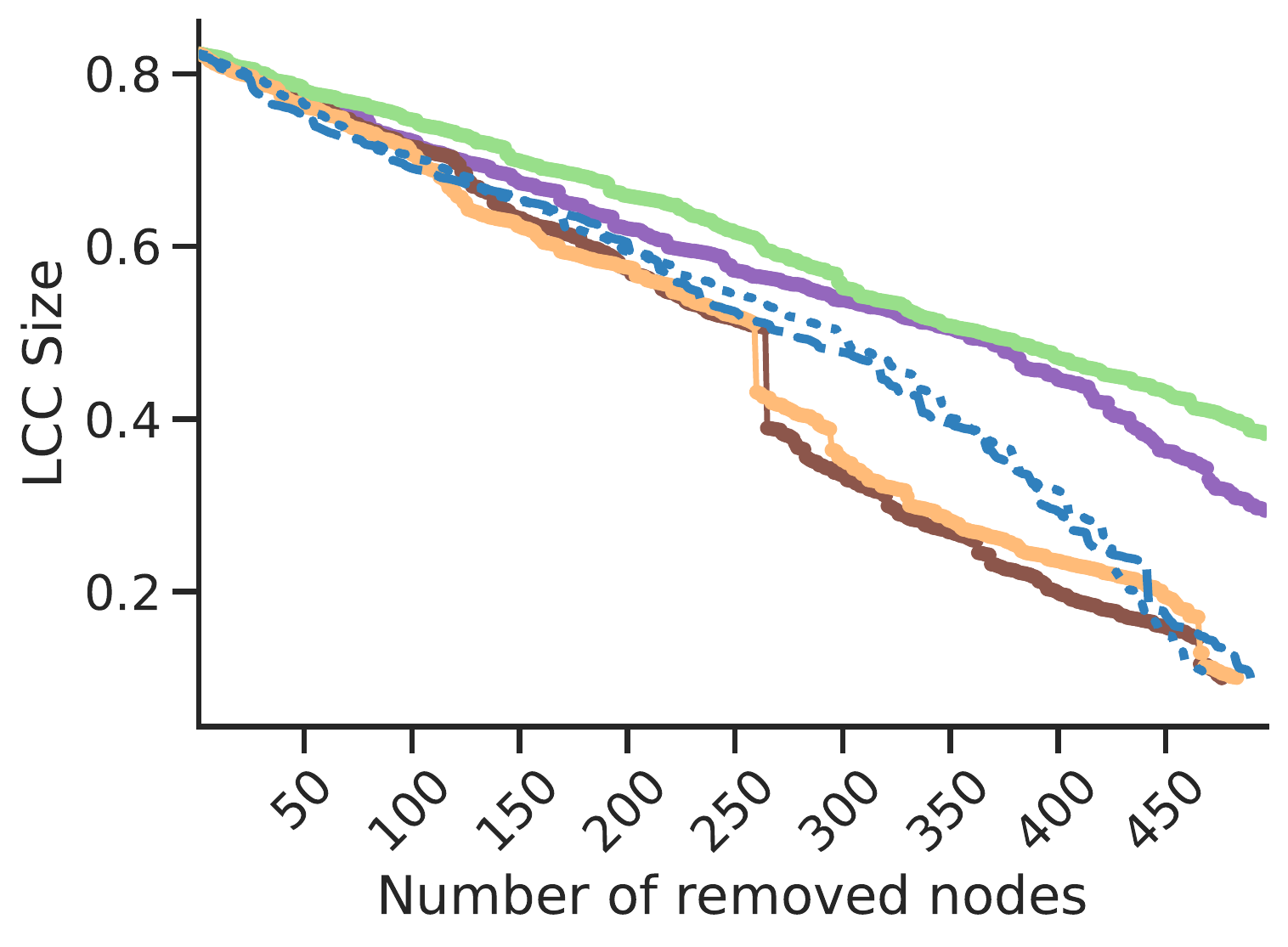}
		\caption{petster-hamster}
		\label{f:petster-hamster_dismantling}
	\end{subfigure}%
\hfill
	\begin{subfigure}{0.5\textwidth}
		\centering
		\includegraphics[width=\textwidth]{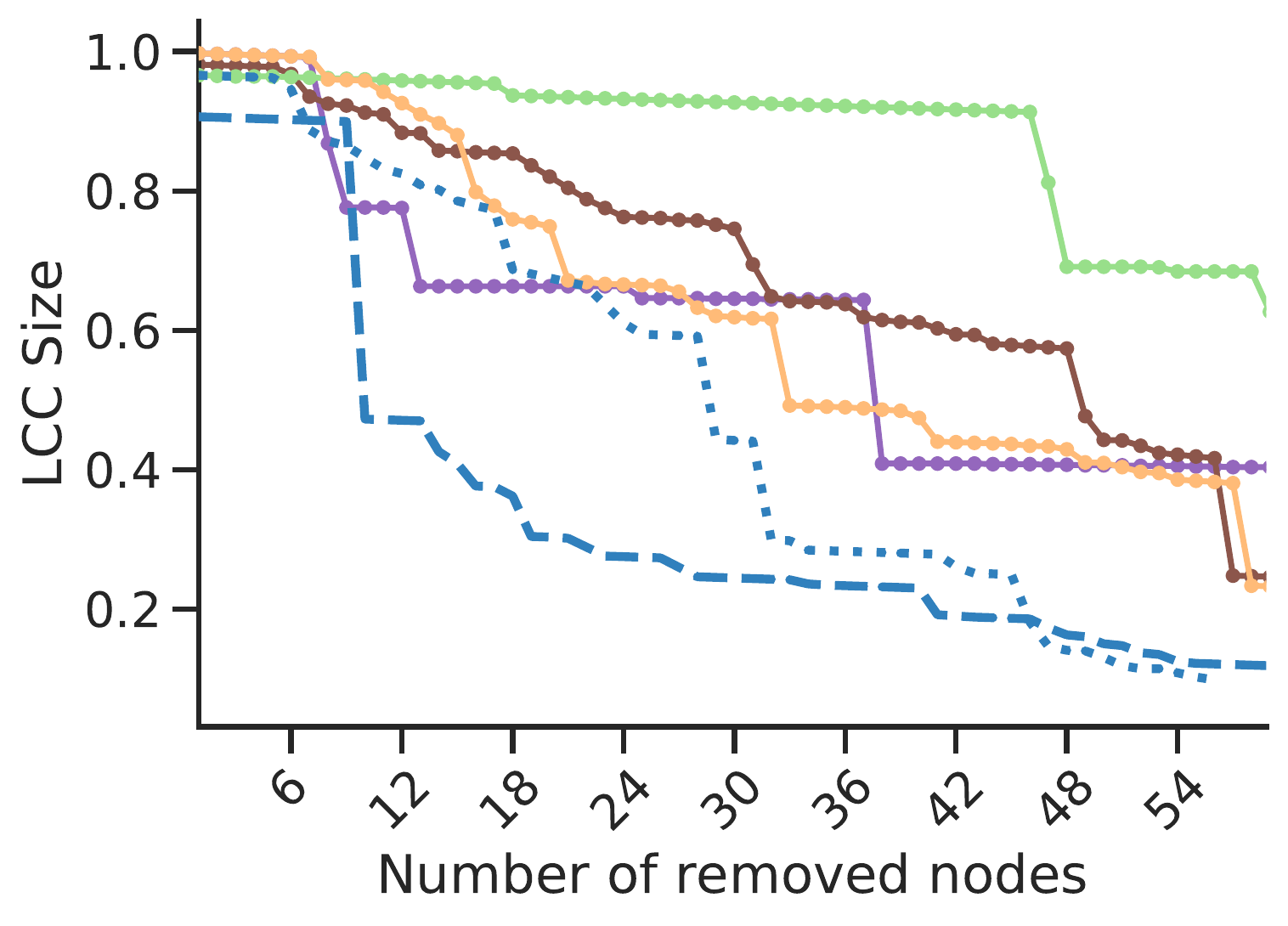}
		\caption{power-eris1176}
		\label{f:power-eris1176_dismantling}
	\end{subfigure}%
	\begin{subfigure}{0.5\textwidth}
		\centering
		\includegraphics[width=\textwidth]{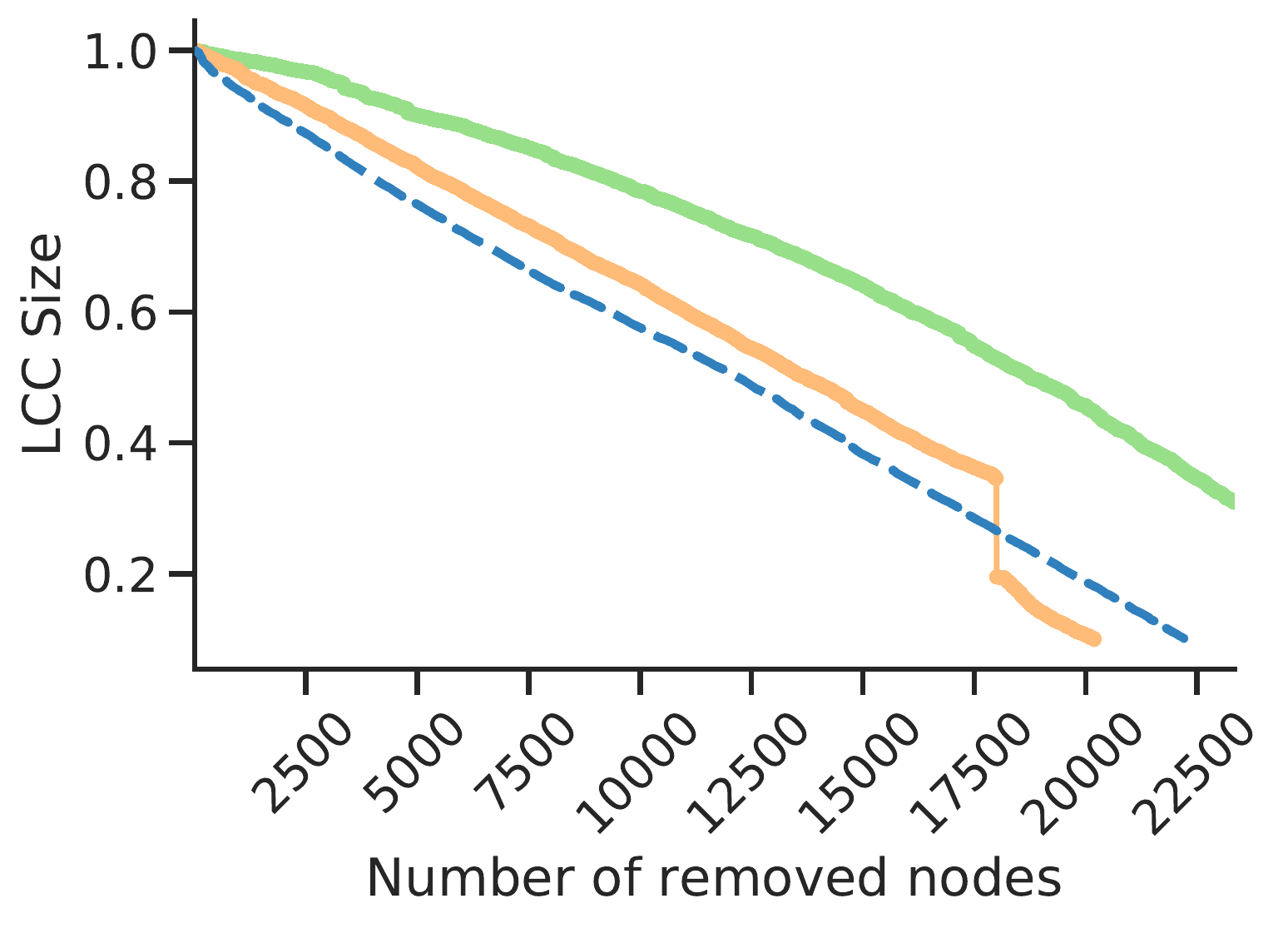}
		\caption{tech-RL-caida}
		\label{f:tech-RL-caida_dismantling}
	\end{subfigure}%
\hfill
	\begin{subfigure}{0.5\textwidth}
		\centering
		\includegraphics[width=\textwidth]{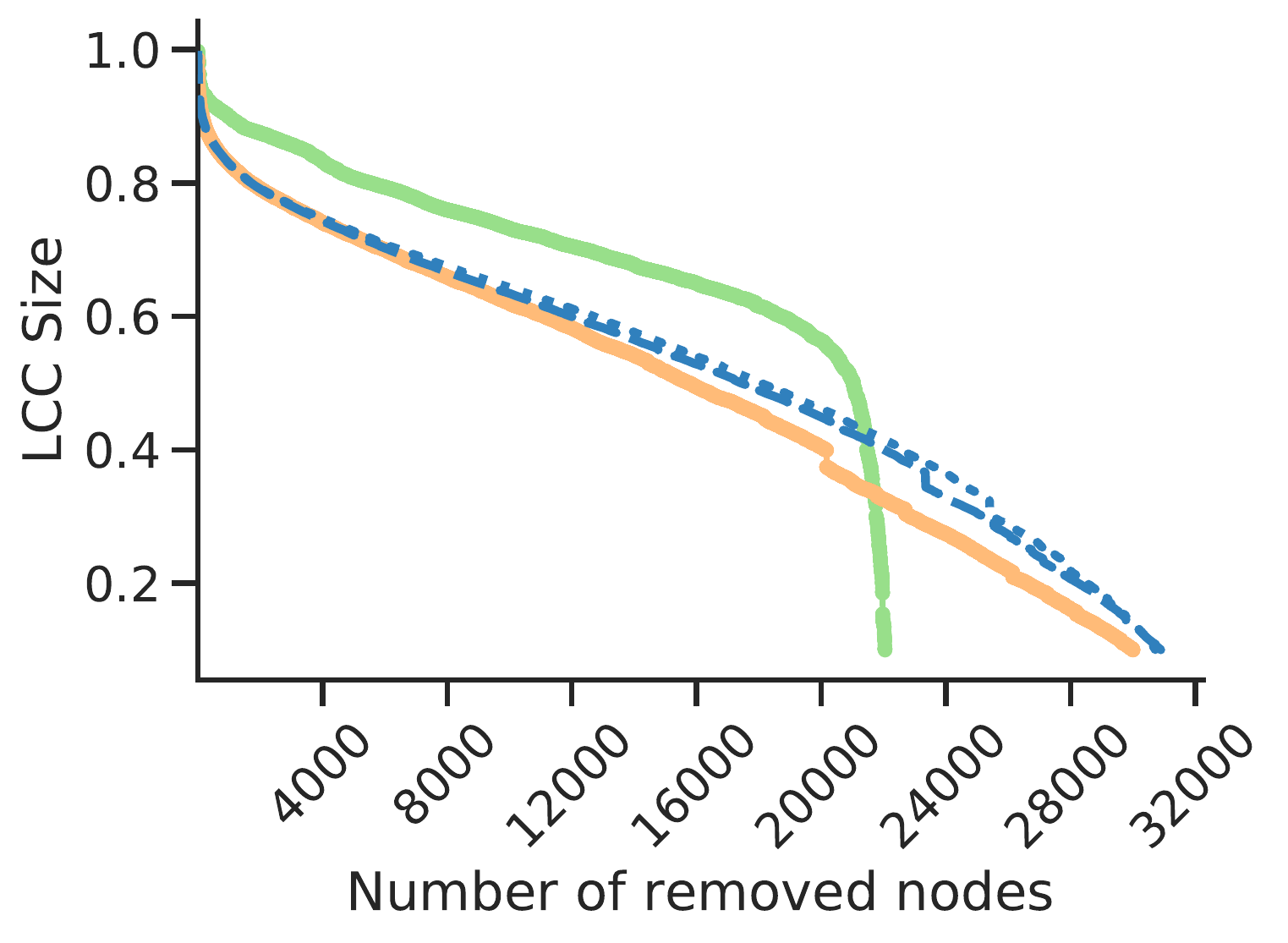}
		\caption{twitter\_LCC}
		\label{f:twitter_LCC_dismantling}
	\end{subfigure}%
	\begin{subfigure}{0.5\textwidth}
		\centering
		\includegraphics[width=\textwidth]{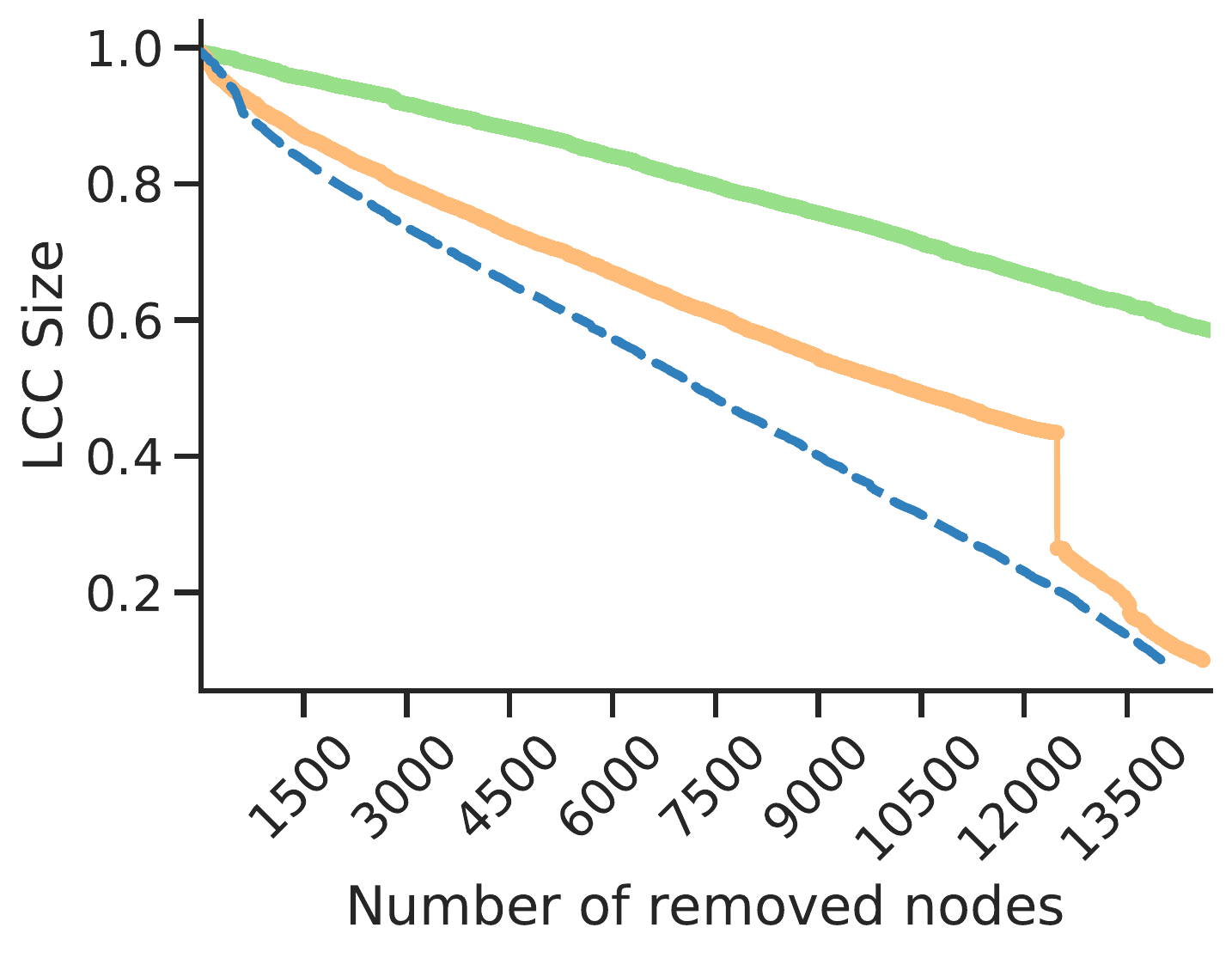}
		\caption{wordnet-words}
		\label{f:wordnet-words_dismantling}
	\end{subfigure}%
	\caption{Dismantling of some networks in our test set. We compare against the algorithms without reinsertion in Tables~\ref{t:full_test_network_table} and~\ref{t:large_test_networks} and show both the models with lower area under the curve (GDM AUC) and with lower number of removals (GDM \#Removals), which may overlap for some networks.}
   	\label{f:dismantling_curves_2}
\end{figure}

\begin{figure}[!ht]
	\centering
	\begin{subfigure}{\textwidth}
		\centering
		\includegraphics[width=\textwidth]{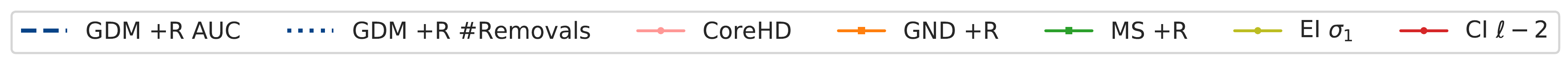}
	\end{subfigure}%
\hfill
	\begin{subfigure}{0.5\textwidth}
		\centering
		\includegraphics[width=\textwidth]{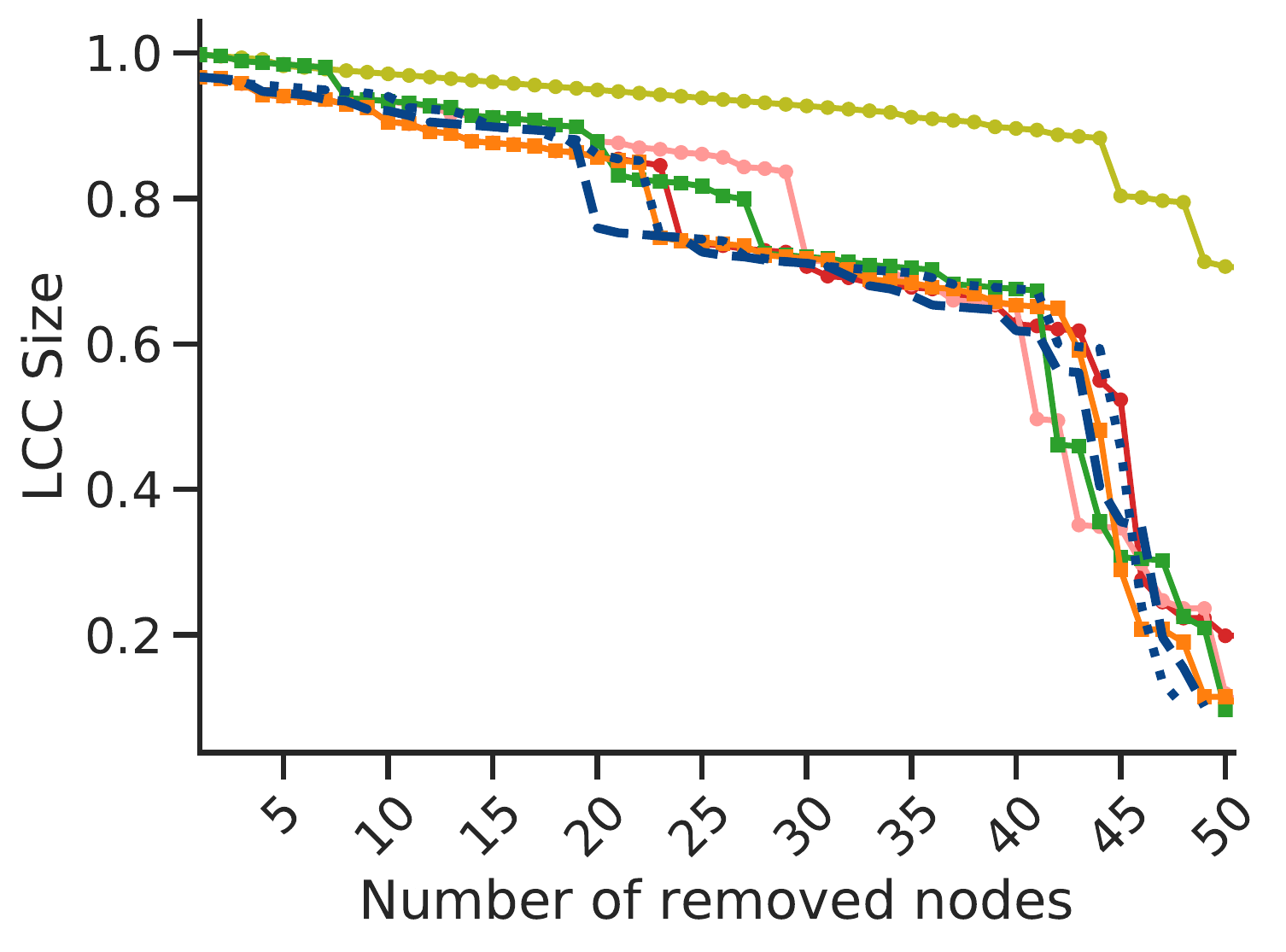}
		\caption{arenas-meta}
		\label{f:arenas_meta_reinserted_dismantling}
	\end{subfigure}%
 	\begin{subfigure}{0.5\textwidth}
		\centering
		\includegraphics[width=\textwidth]{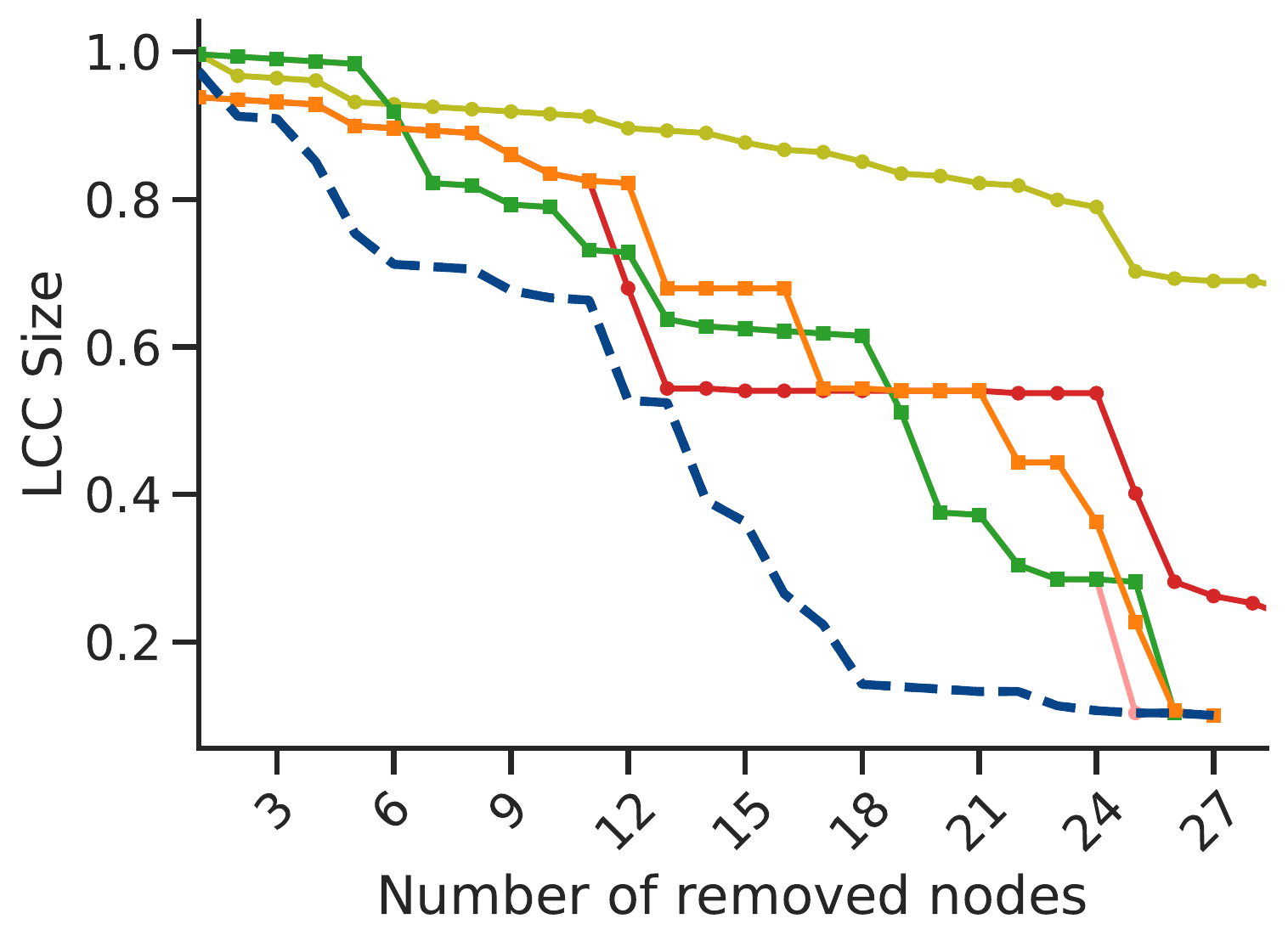}
		\caption{corruption}
		\label{f:corruption_reinserted_dismantling}
	\end{subfigure}%
	\hfill
	\begin{subfigure}{0.5\textwidth}
		\centering
		\includegraphics[width=\textwidth]{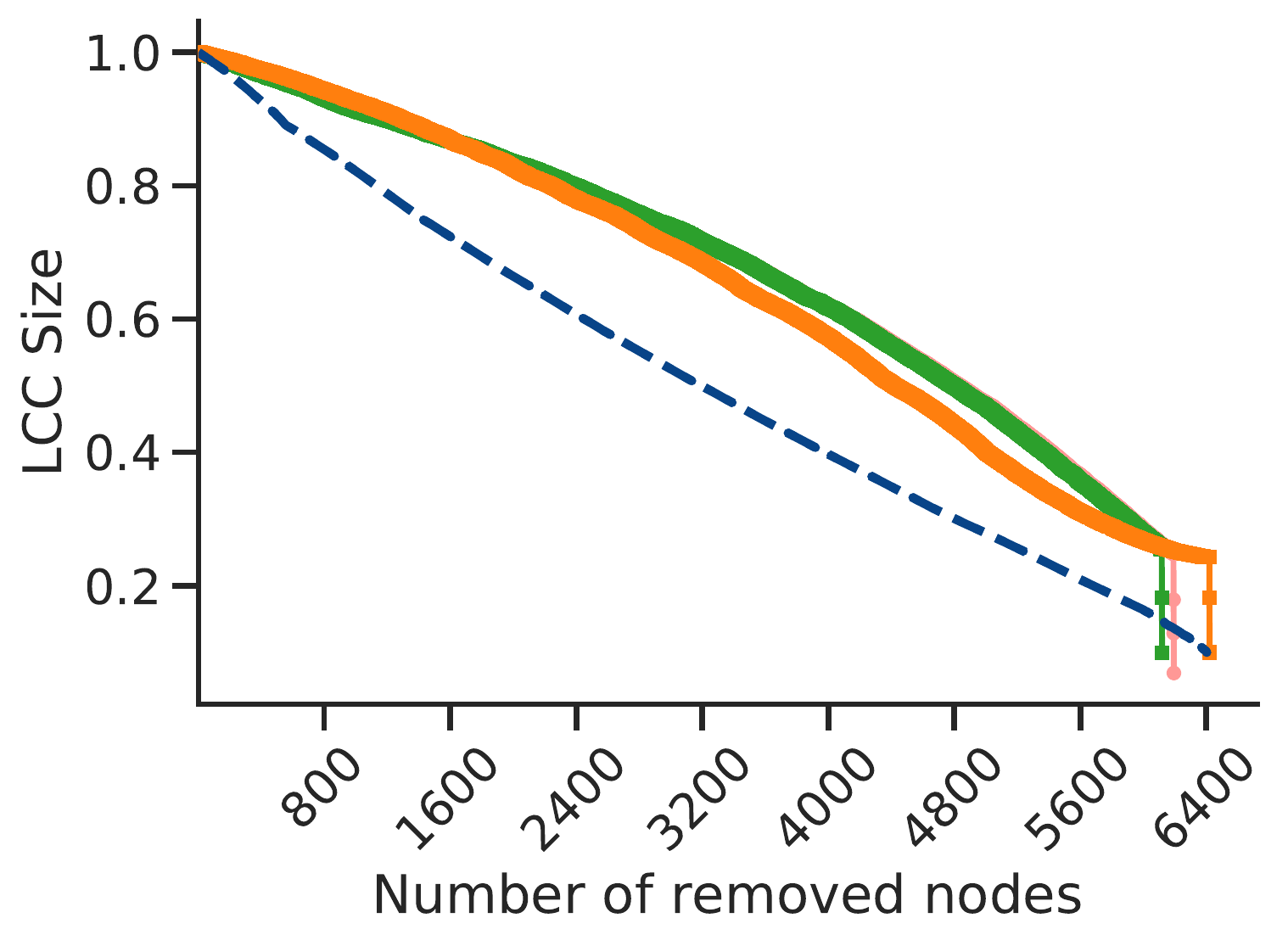}
		\caption{douban}
		\label{f:douban_reinserted_dismantling}
	\end{subfigure}%
	\begin{subfigure}{0.5\textwidth}
		\centering
		\includegraphics[width=\textwidth]{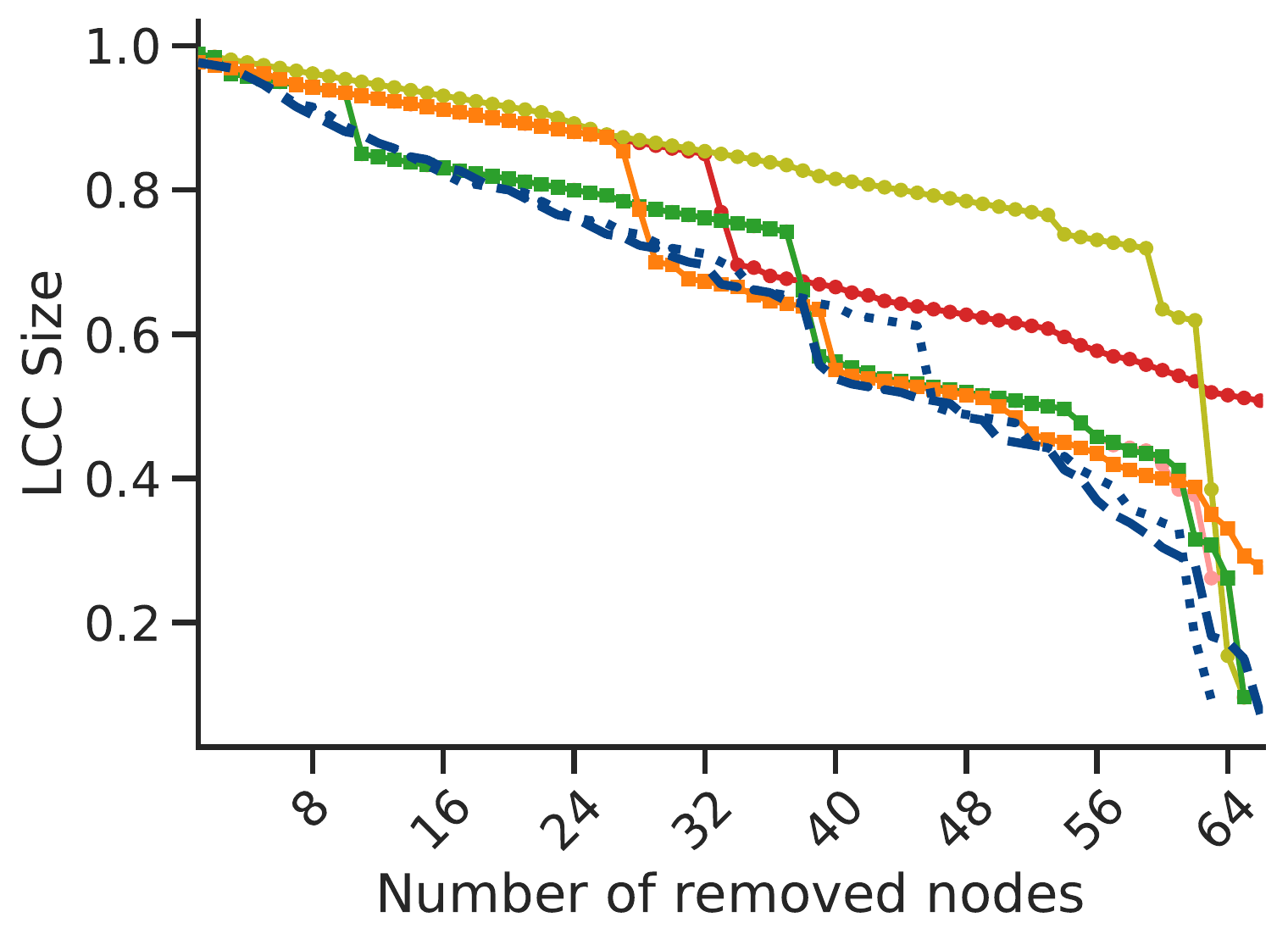}
		\caption{econ-wm1}
		\label{f:econ-wm1_reinserted_dismantling}
	\end{subfigure}%
	\hfill
	\begin{subfigure}{0.5\textwidth}
		\centering
		\includegraphics[width=\textwidth]{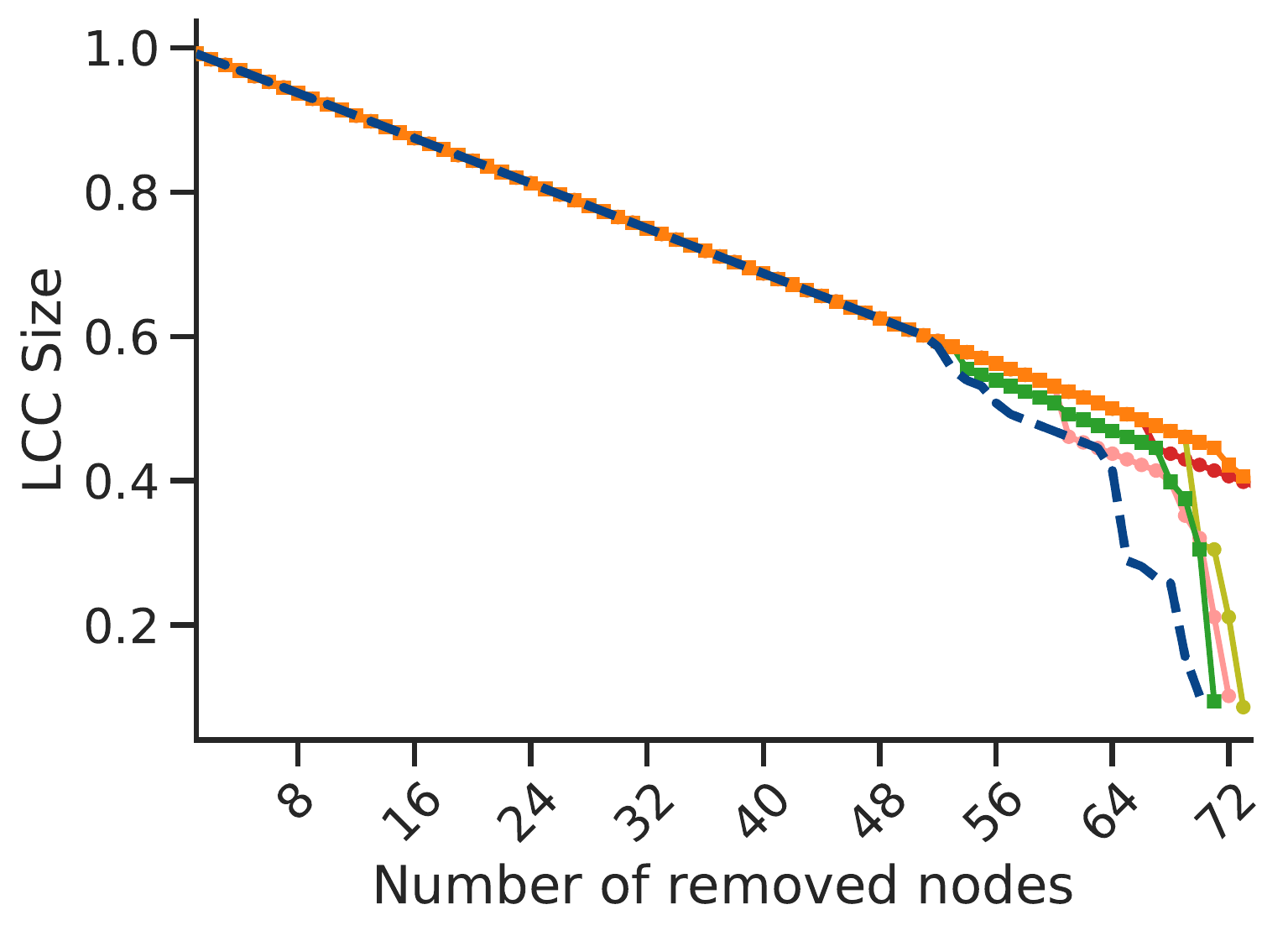}
		\caption{foodweb-baywet}
		\label{f:foodweb-baywet_reinserted_dismantling}
	\end{subfigure}%
	\begin{subfigure}{0.5\textwidth}
		\centering
		\includegraphics[width=\textwidth]{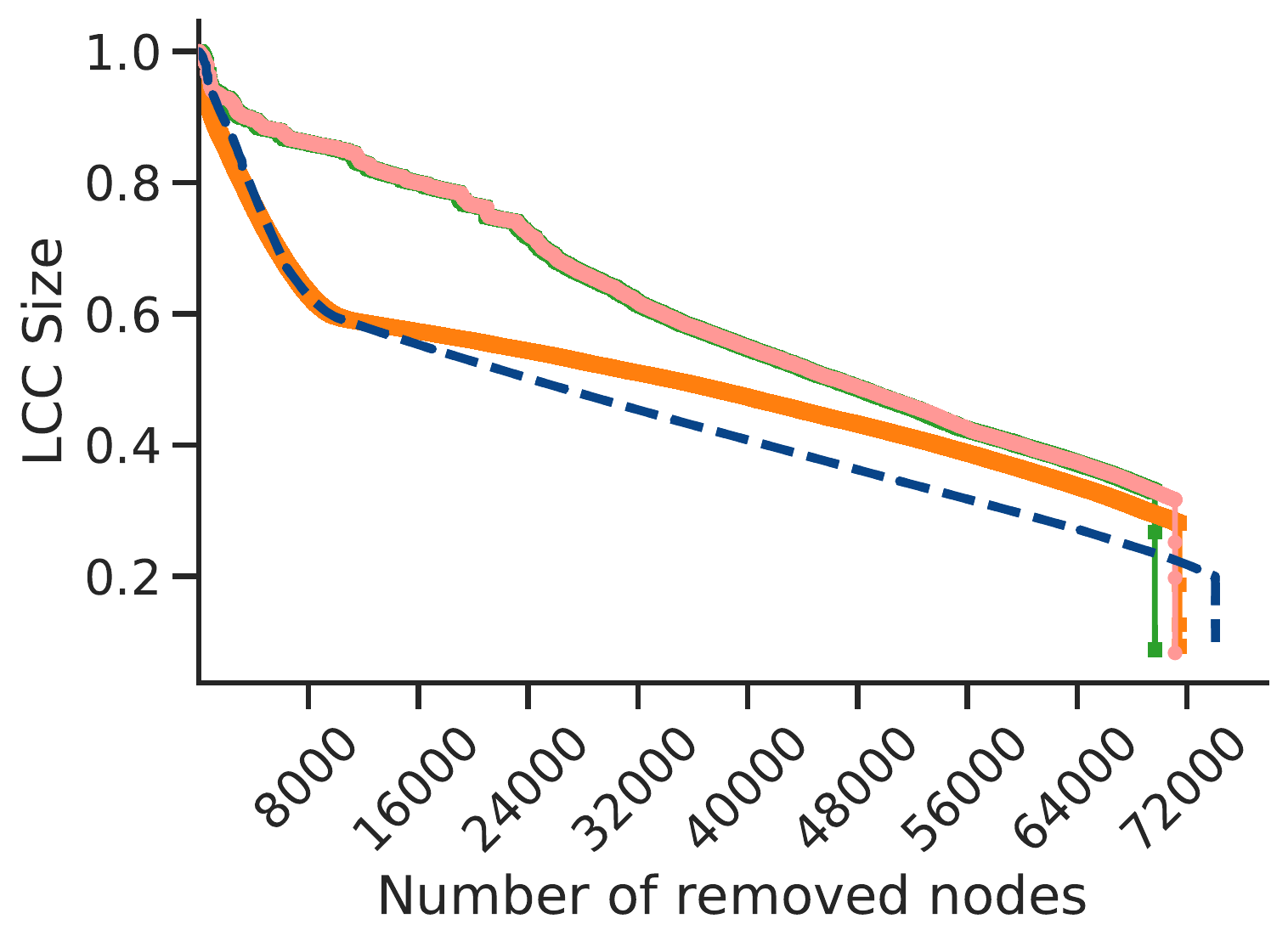}
		\caption{hyves}
		\label{f:hyves_reinserted_dismantling}
	\end{subfigure}%
\end{figure}
\begin{figure}[!ht]\ContinuedFloat
	\centering
	\begin{subfigure}{0.5\textwidth}
		\centering
		\includegraphics[width=\textwidth]{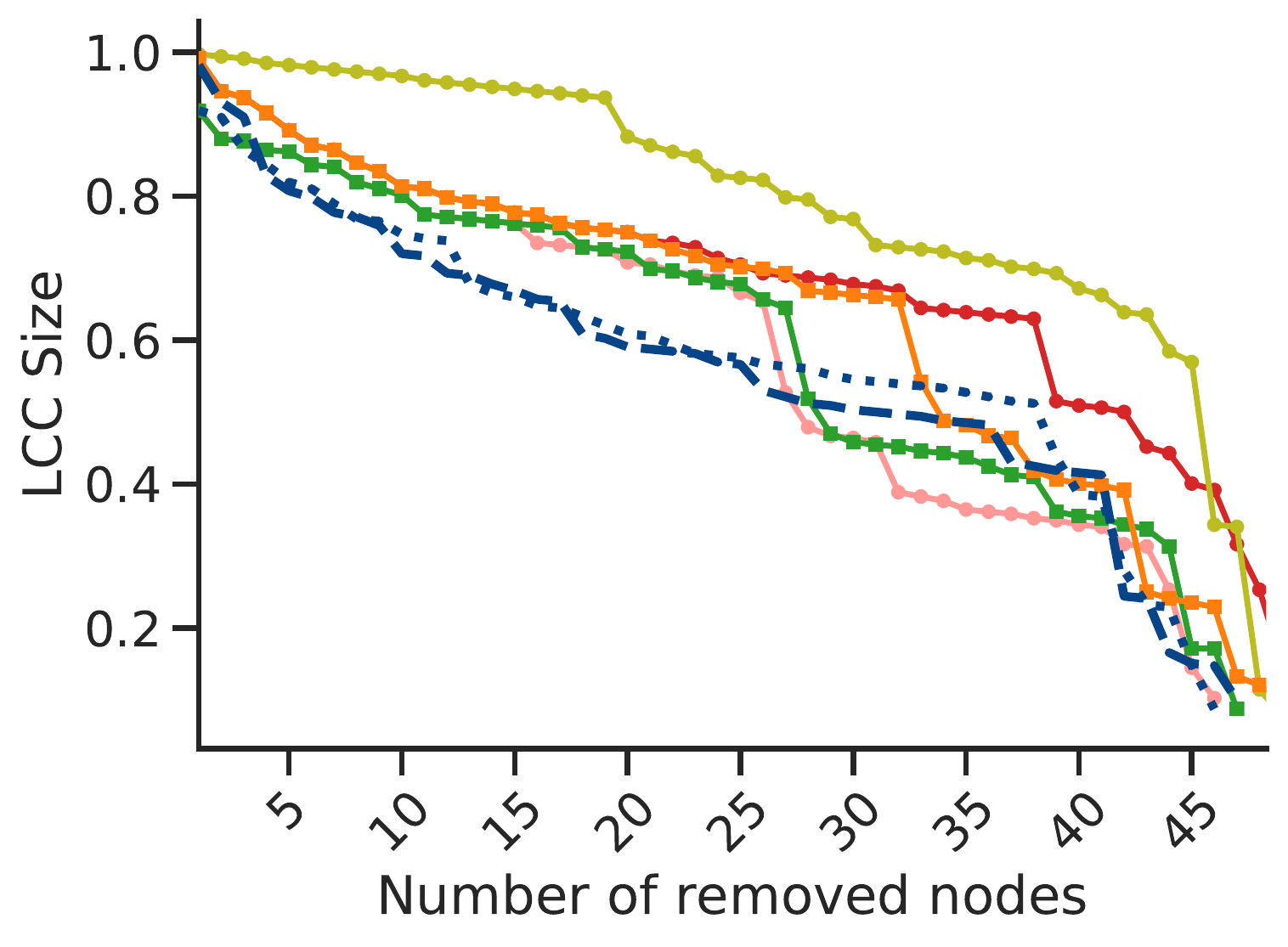}
		\caption{inf-USAir97}
		\label{f:inf-USAir97_reinserted_dismantling}
	\end{subfigure}%
	\begin{subfigure}{0.5\textwidth}
	 \centering
	 \includegraphics[width=\textwidth]{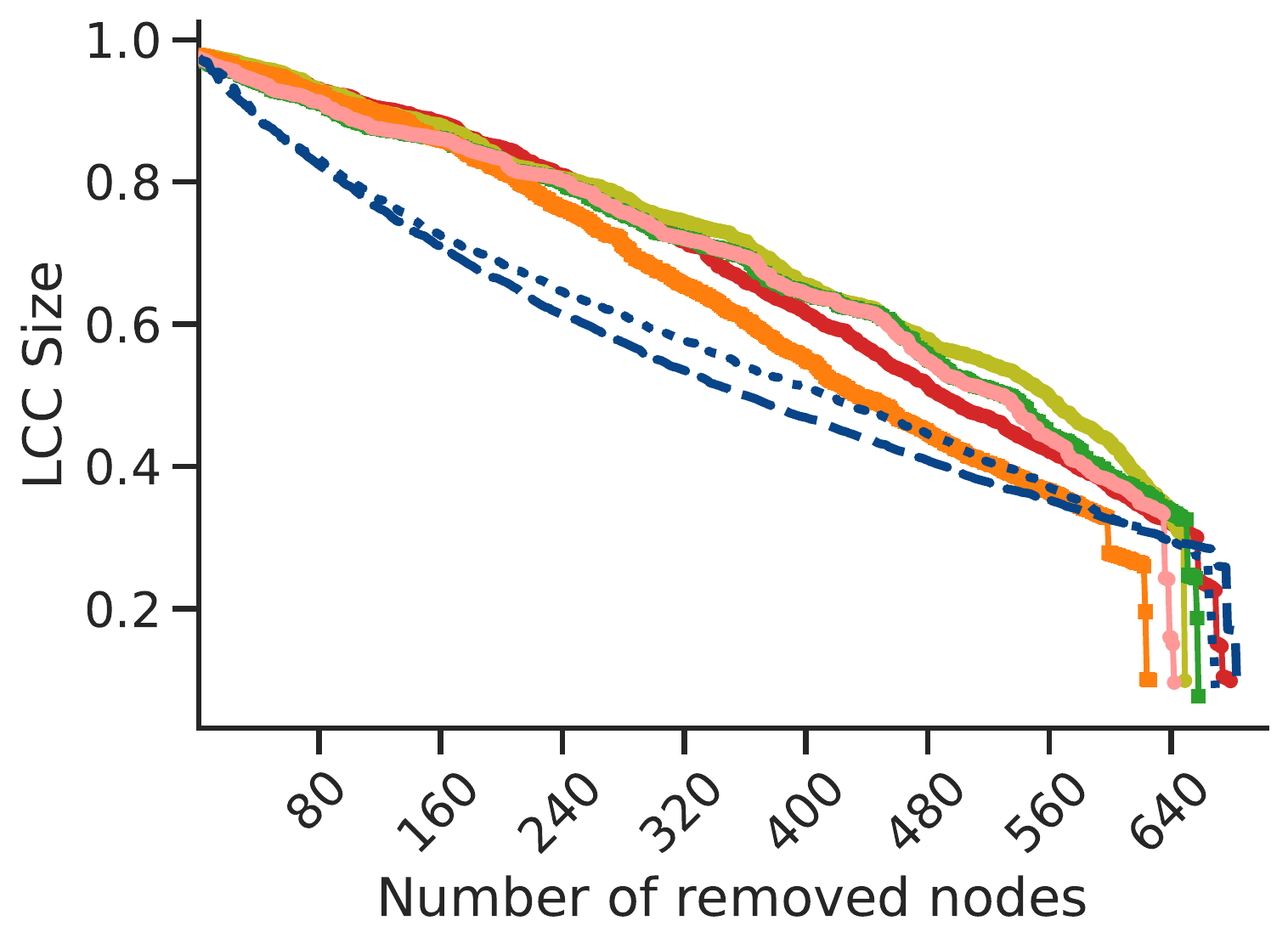}
	 \caption{librec-ciaodvd-trust}
	 \label{f:librec-ciaodvd-trust_reinserted_dismantling}
	 \end{subfigure}%
\hfill
	\begin{subfigure}{0.5\textwidth}
		\centering
		\includegraphics[width=\textwidth]{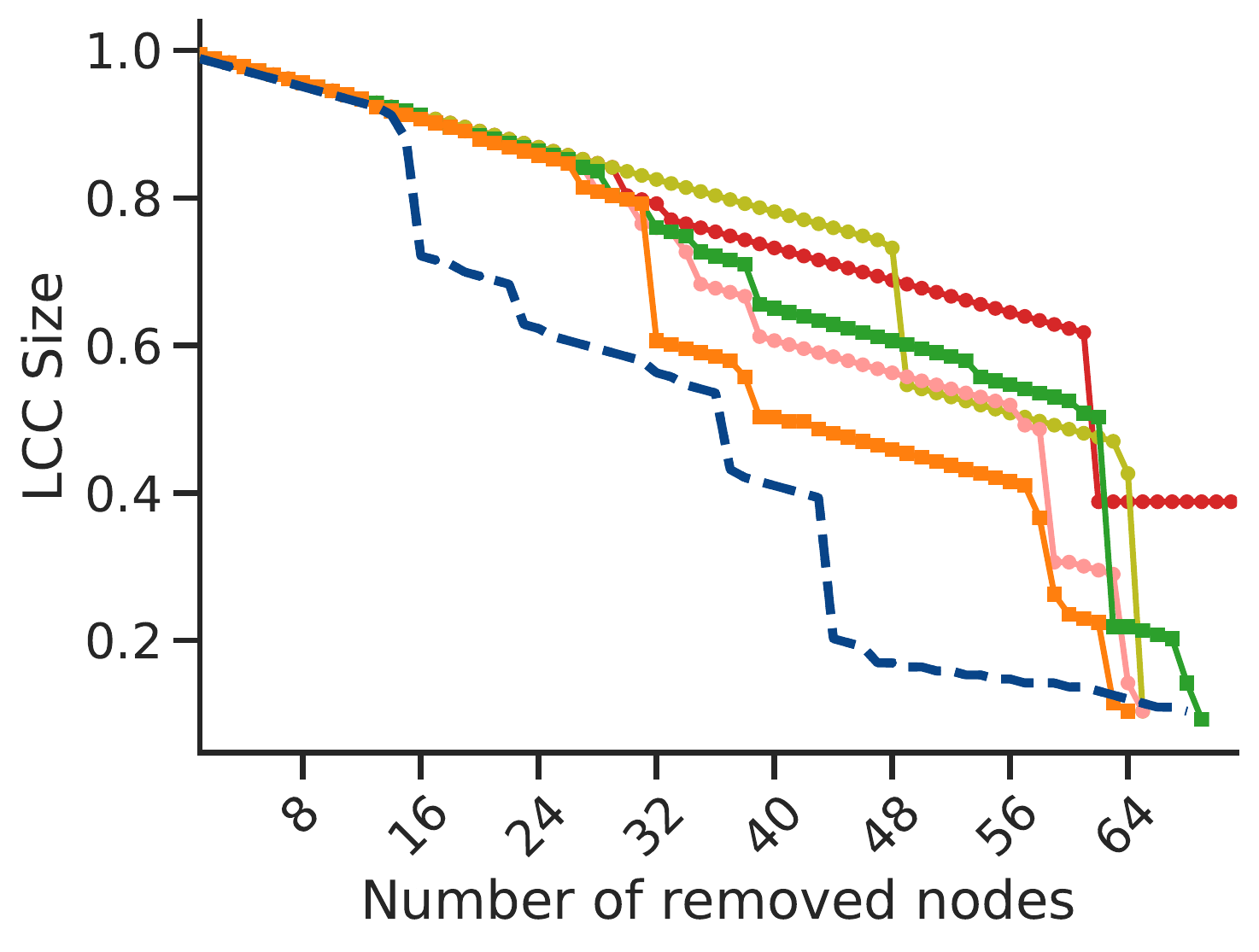}
		\caption{maayan-foodweb}
		\label{f:maayan-foodweb_reinserted_dismantling}
	\end{subfigure}%
	\begin{subfigure}{0.5\textwidth}
		\centering
		\includegraphics[width=\textwidth]{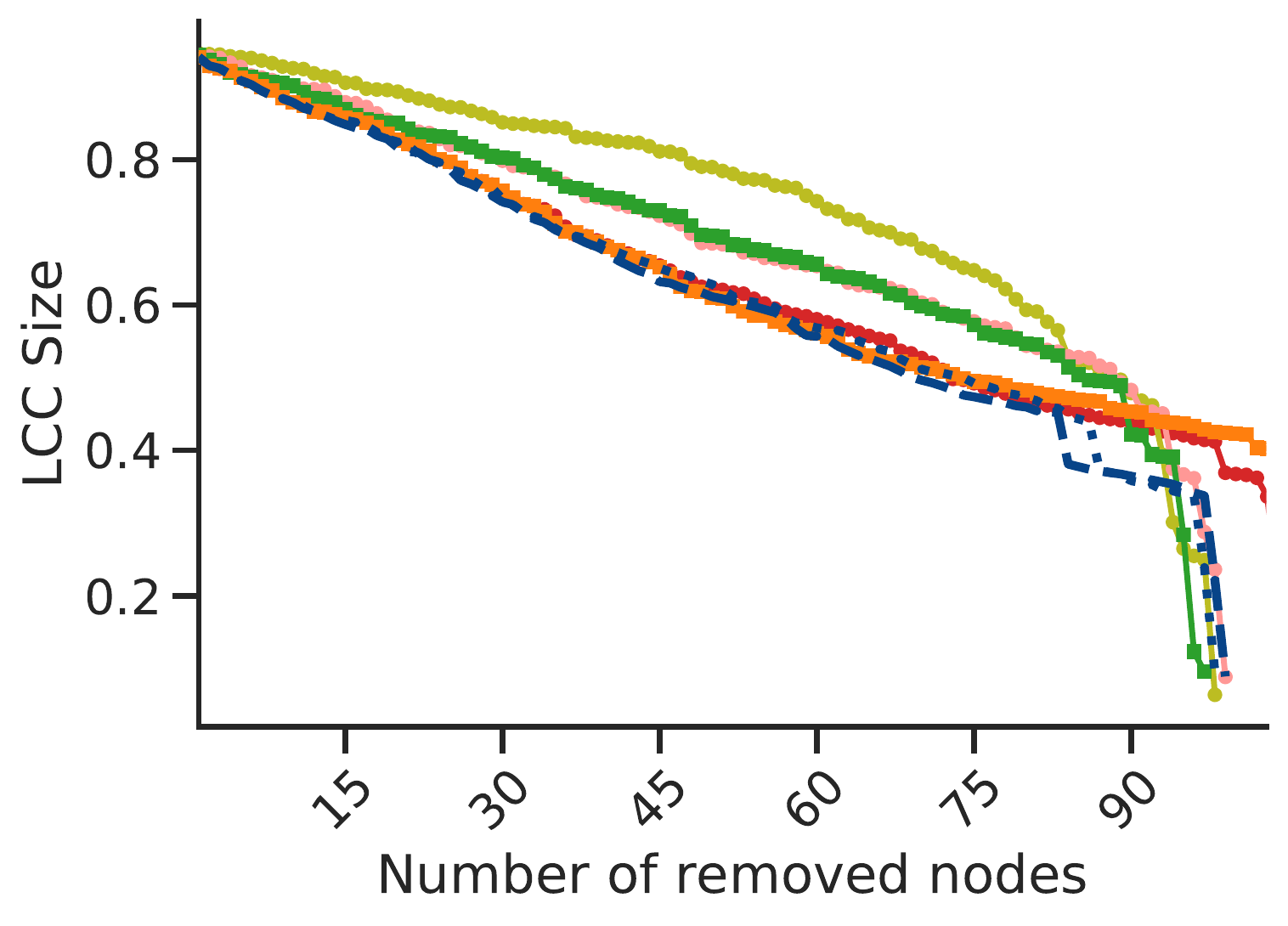}
		\caption{maayan-Stelzl}
		\label{f:maayan-Stelzl_reinserted_dismantling}
	\end{subfigure}%

	\begin{subfigure}{0.5\textwidth}
		\centering
		\includegraphics[width=\textwidth]{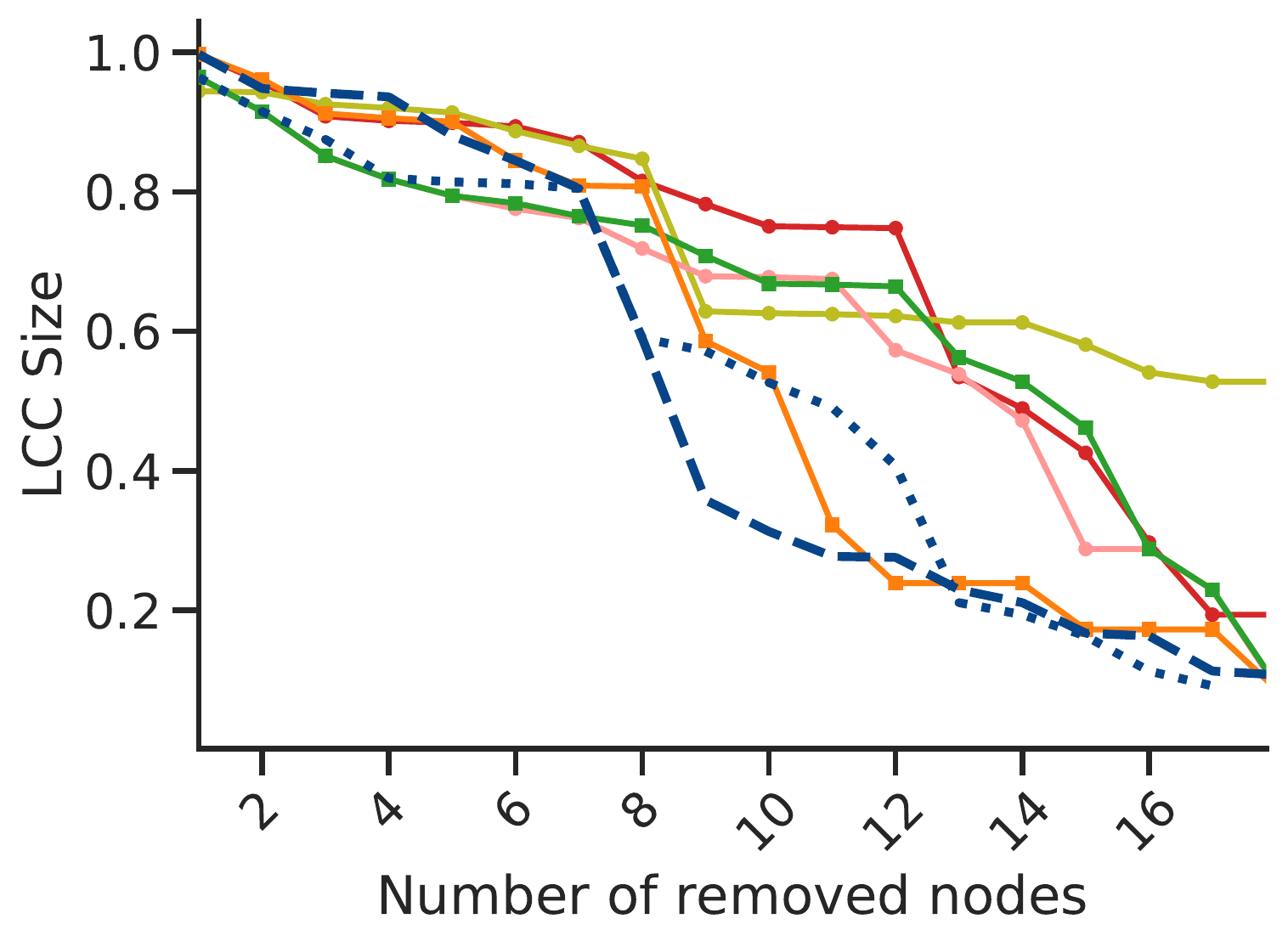}
		\caption{moreno-crime-projected}
		\label{f:moreno_crime_projected_reinserted_dismantling}
	\end{subfigure}%
	\begin{subfigure}{0.5\textwidth}
		\centering
		\includegraphics[width=\textwidth]{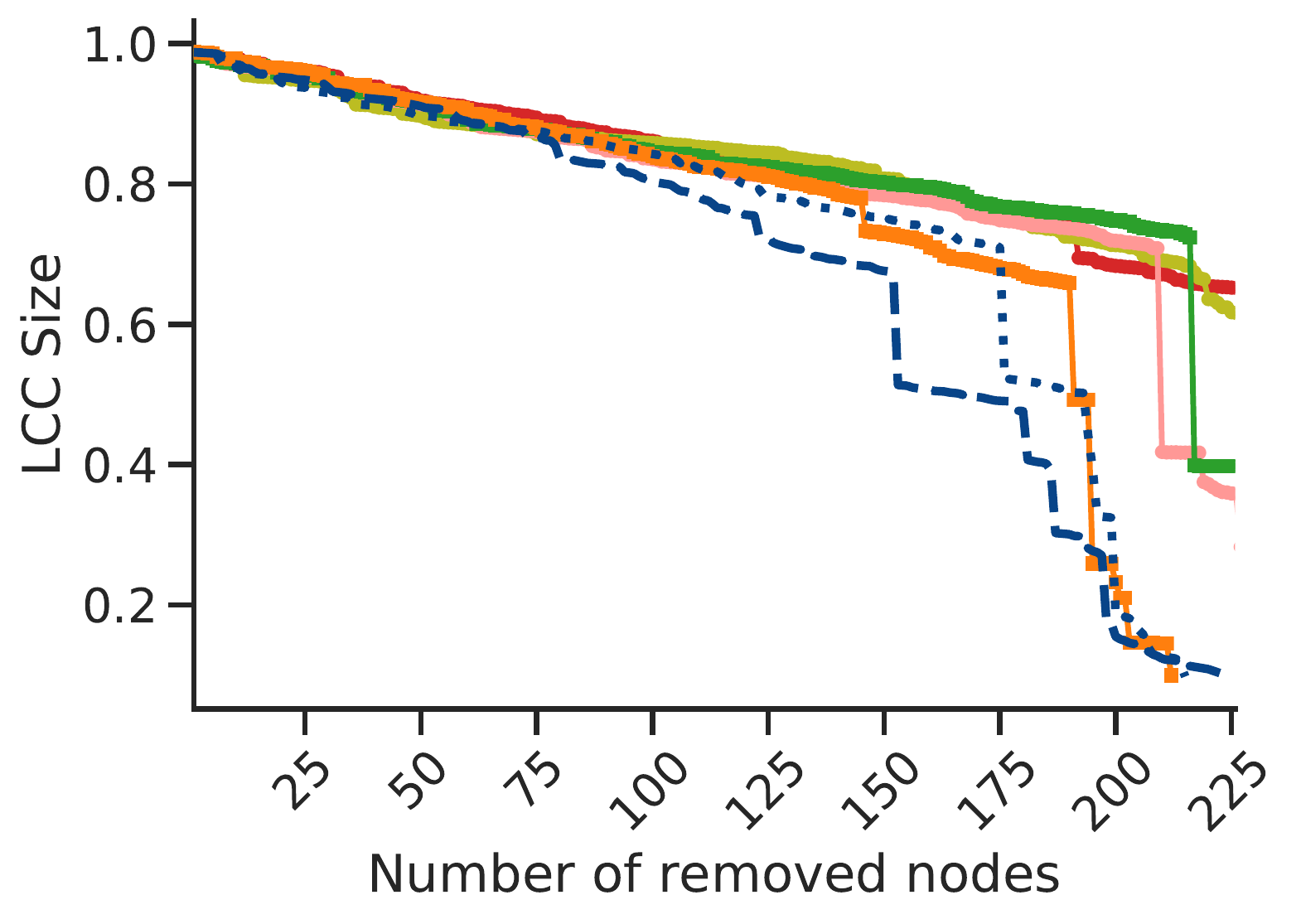}
		\caption{opsahl-openflights}
		\label{f:opsahl-openflights_reinserted_dismantling}
	\end{subfigure}%
\end{figure}

\begin{figure}[!ht]\ContinuedFloat
	\centering
	\begin{subfigure}{0.5\textwidth}
		\centering
		\includegraphics[width=\textwidth]{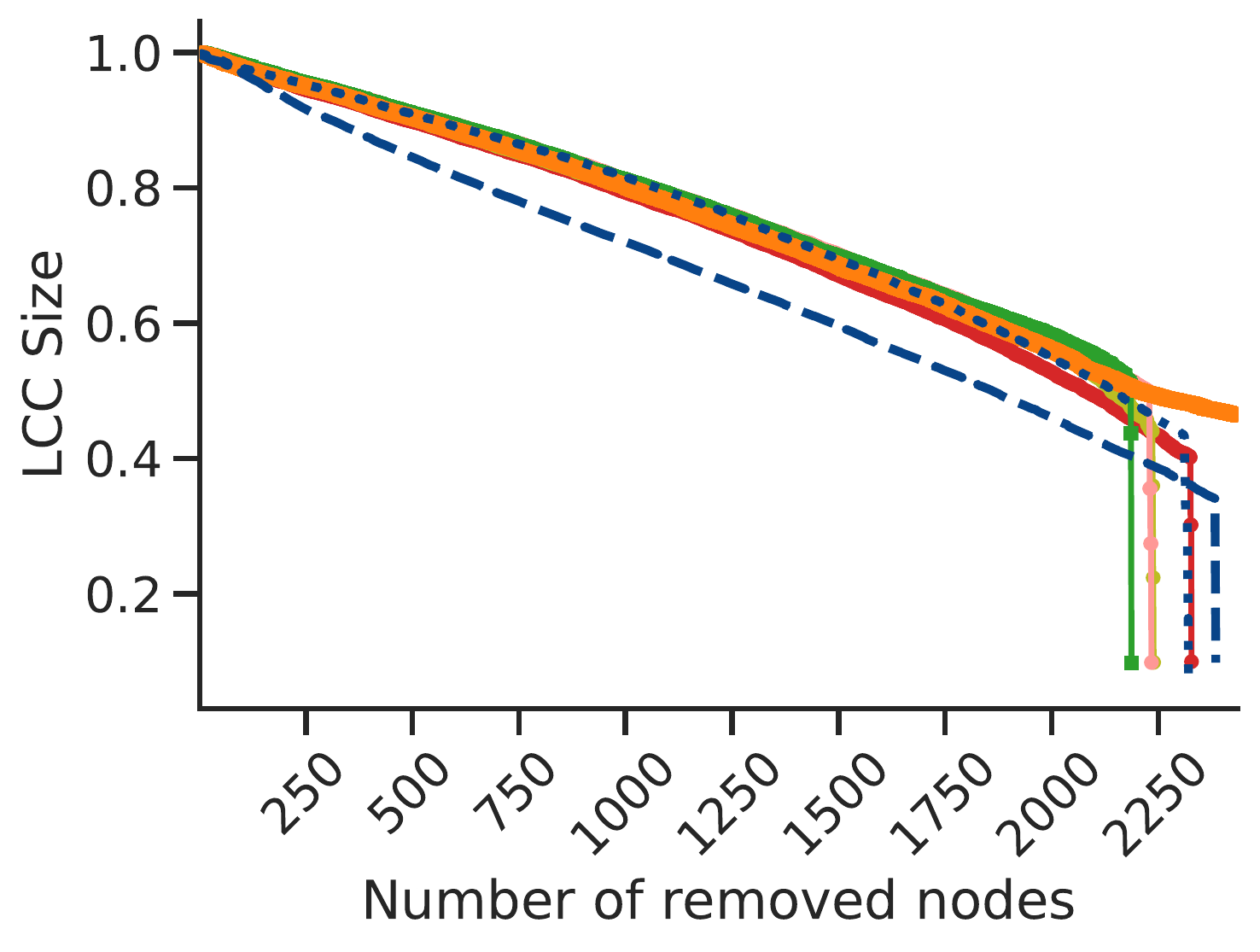}
		\caption{p2p-Gnutella06}
		\label{f:p2p-Gnutella06_reinserted_dismantling}
	\end{subfigure}%
	\begin{subfigure}{0.5\textwidth}
		\centering
		\includegraphics[width=\textwidth]{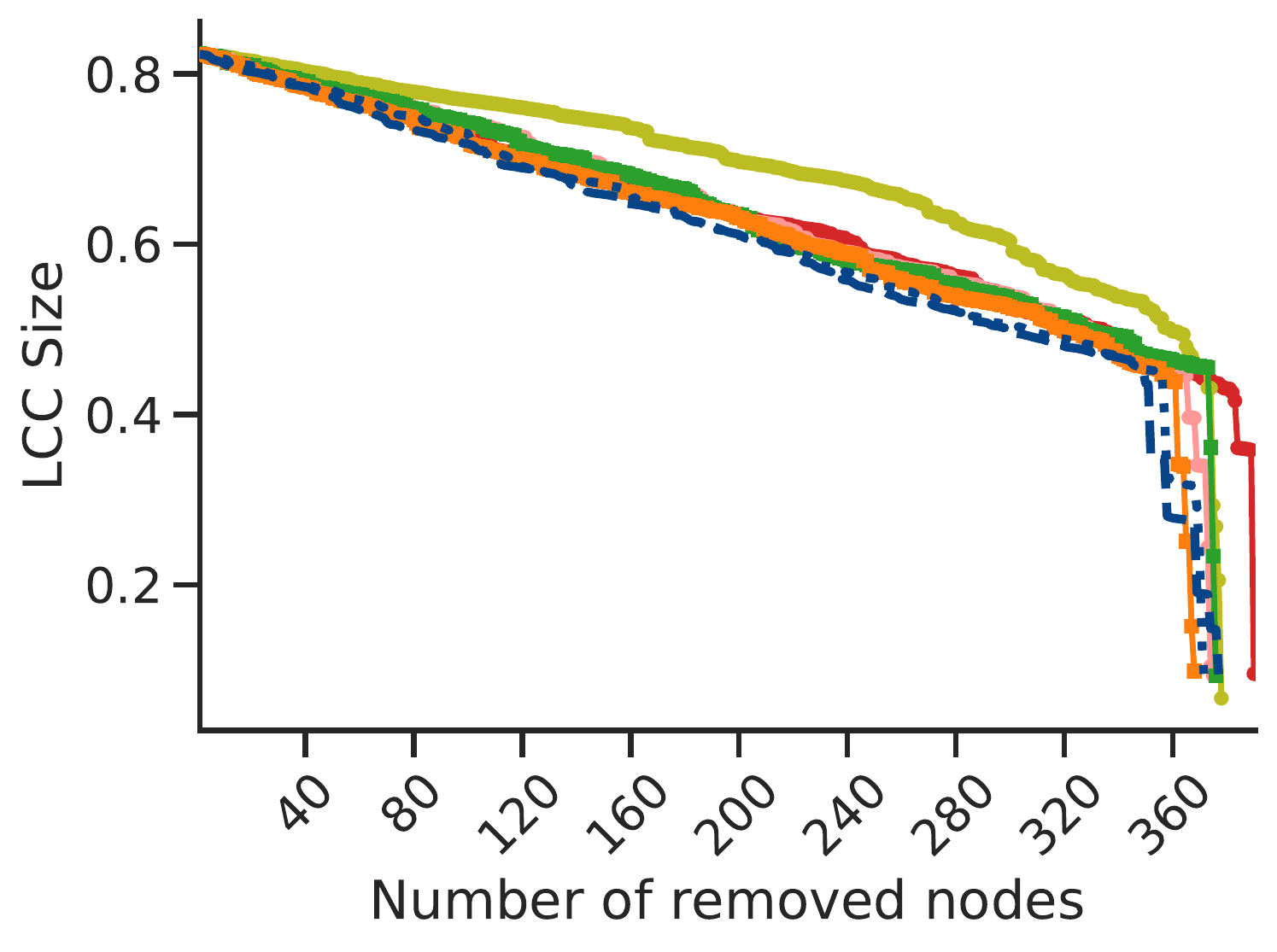}
		\caption{petster-hamster}
		\label{f:petster-hamster_reinserted_dismantling}
	\end{subfigure}%
\hfill
	\begin{subfigure}{0.5\textwidth}
		\centering
		\includegraphics[width=\textwidth]{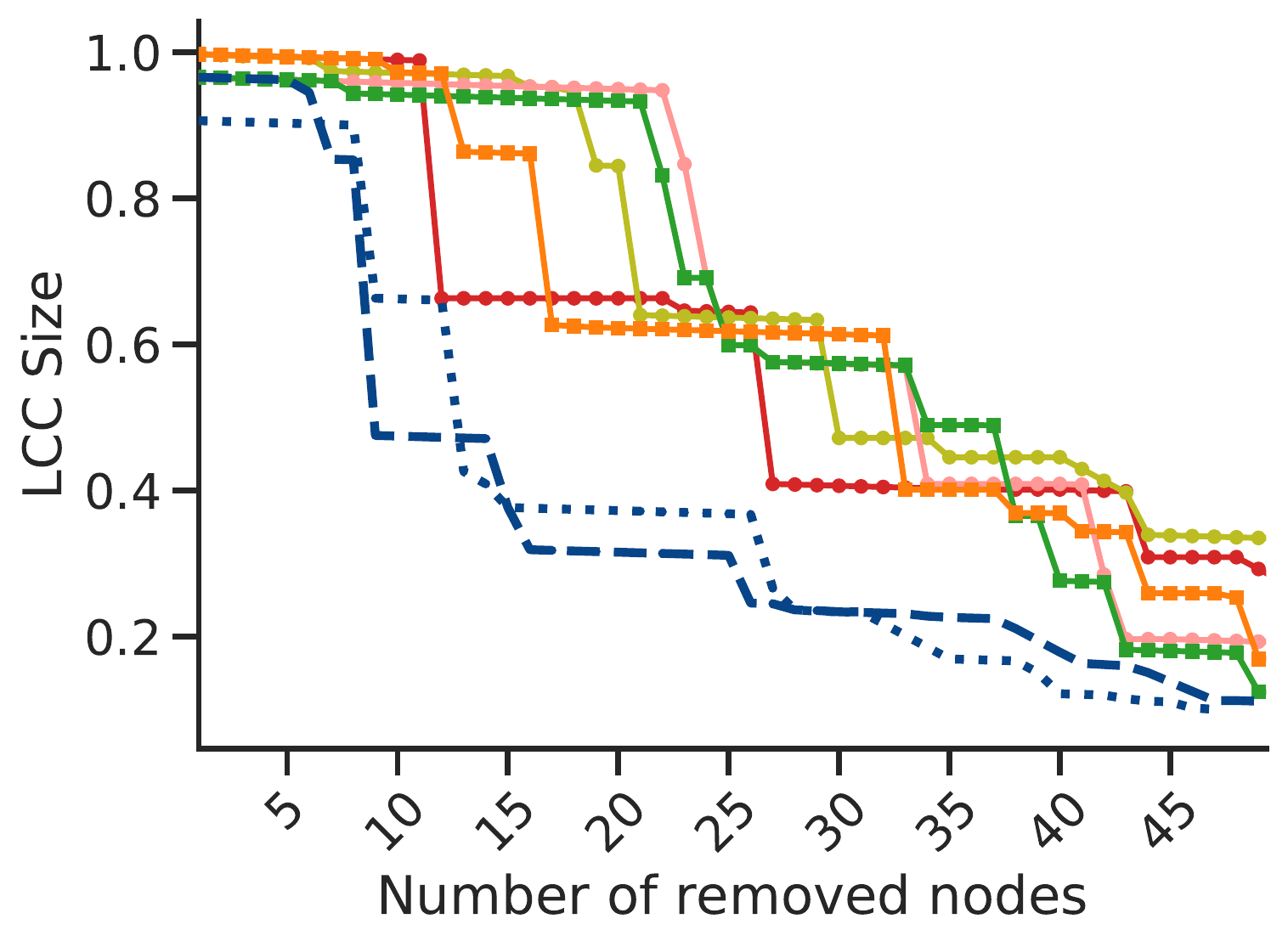}
		\caption{power-eris1176}
		\label{f:power-eris1176_reinserted_dismantling}
	\end{subfigure}%
	\begin{subfigure}{0.5\textwidth}
		\centering
		\includegraphics[width=\textwidth]{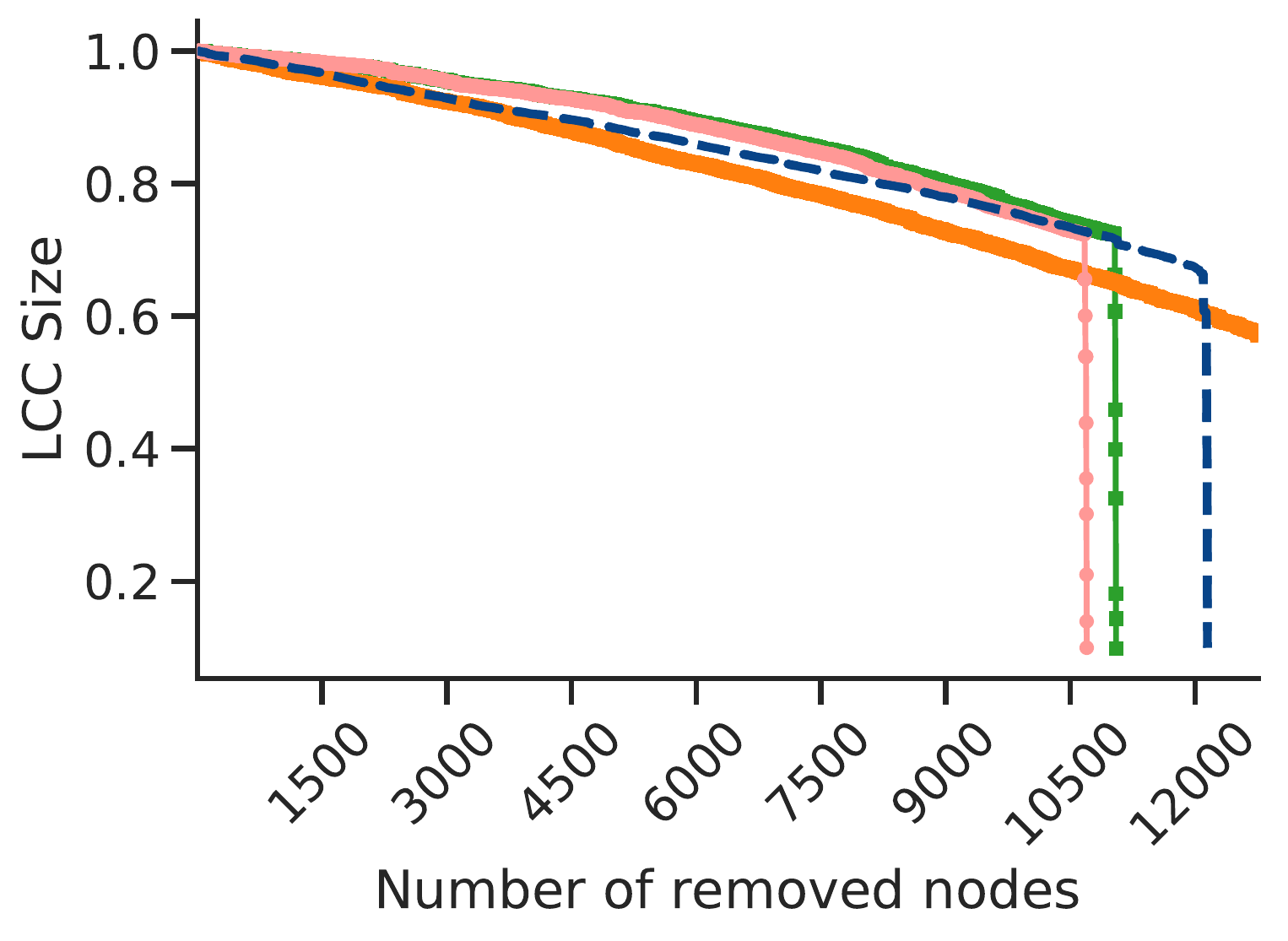}
		\caption{tech-RL-caida}
		\label{f:tech-RL-caida_reinserted_dismantling}
	\end{subfigure}%
\hfill
	\begin{subfigure}{0.5\textwidth}
		\centering
		\includegraphics[width=\textwidth]{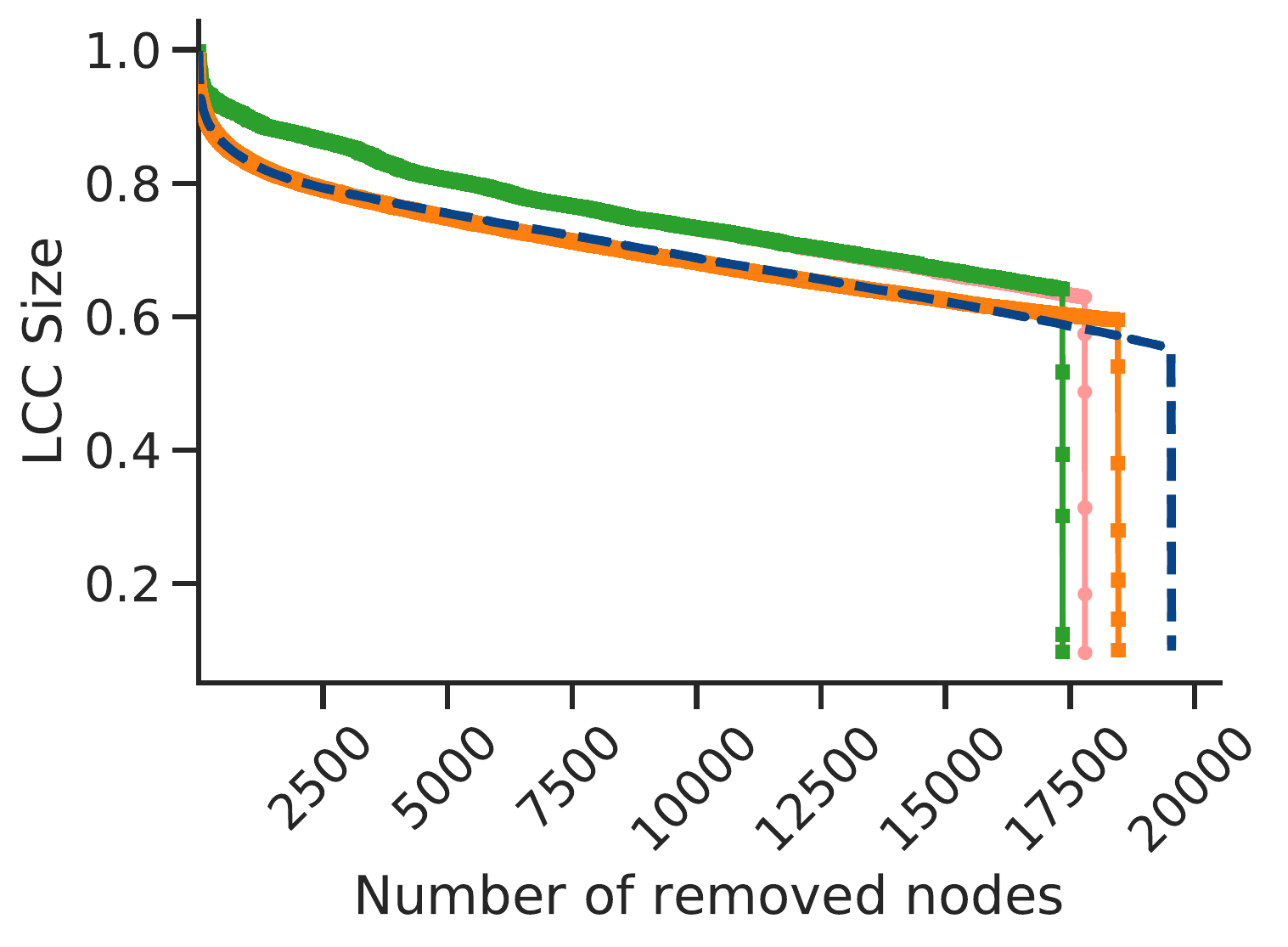}
		\caption{twitter\_LCC}
		\label{f:twitter_LCC_reinserted_dismantling}
	\end{subfigure}%
	\begin{subfigure}{0.5\textwidth}
		\centering
		\includegraphics[width=\textwidth]{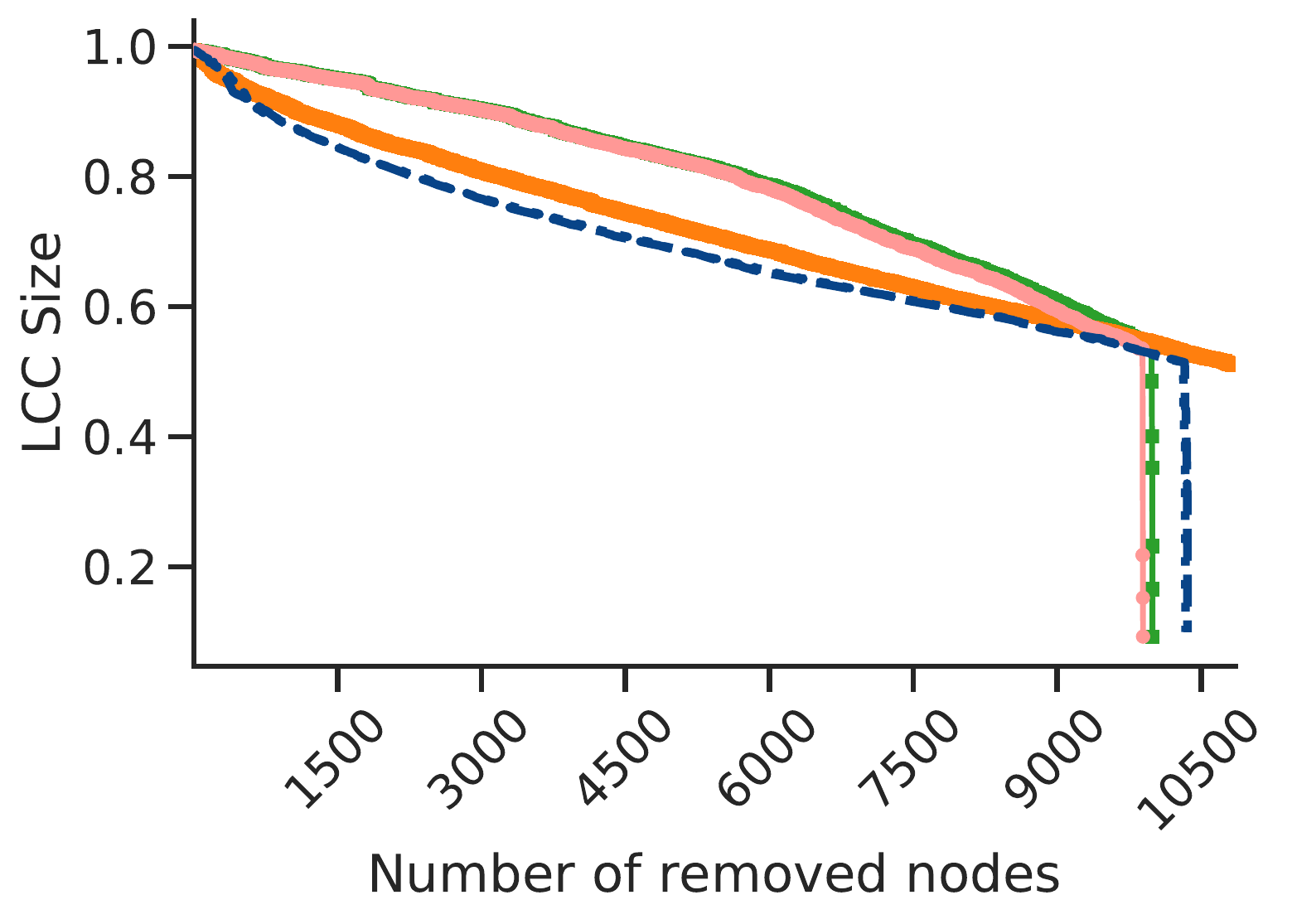}
		\caption{wordnet-words}
		\label{f:wordnet-words_reinserted_dismantling}
	\end{subfigure}%
	\hfill
	\caption{Dismantling of some networks in our test set. We compare against the algorithms with reinsertion phase in Tables~\ref{t:full_test_network_table} and~\ref{t:large_test_networks} and show both the models with lower area under the curve (GDM +R AUC) and with lower number of removals (GDM +R \#Removals), which may overlap for some networks.}
   	\label{f:dismantling_curves_reinserted}
\end{figure}

\paragraph*{More Early Warning $\Omega$ examples}

\begin{figure}[htbp!]
	\centering
	\begin{subfigure}{\textwidth}
		\centering
		\includegraphics[width=0.5\textwidth]{figures/EarlyWarning/legend_horizontal}
	\end{subfigure}%
	\hfill
	\begin{subfigure}{0.5\textwidth}
		\centering
		\includegraphics[width=\textwidth]{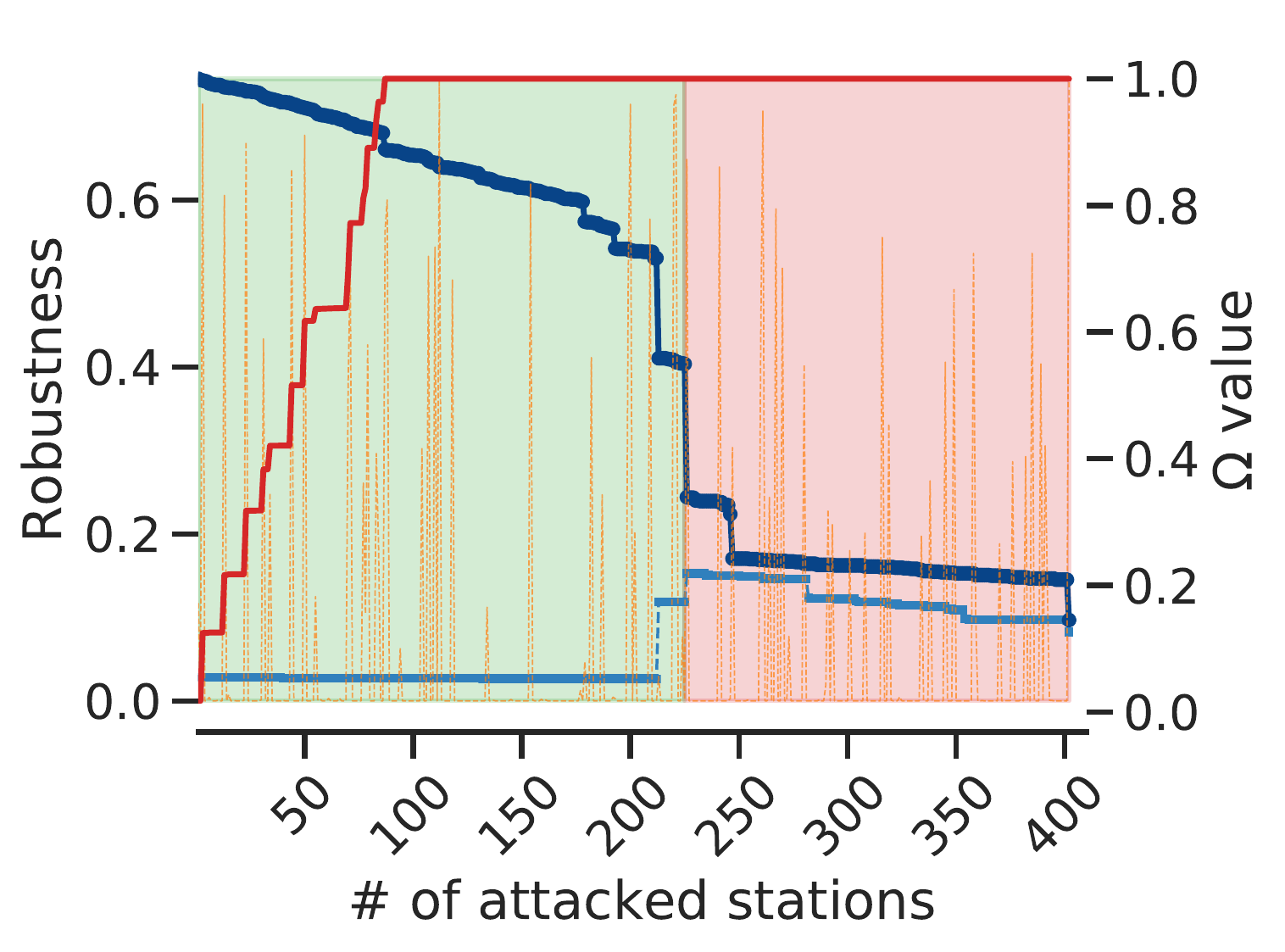}
		\caption{Random}
		\label{f:eu_powergrid_ew_random}
	\end{subfigure}%
	\begin{subfigure}{0.5\textwidth}
		\centering
		\includegraphics[width=\textwidth]{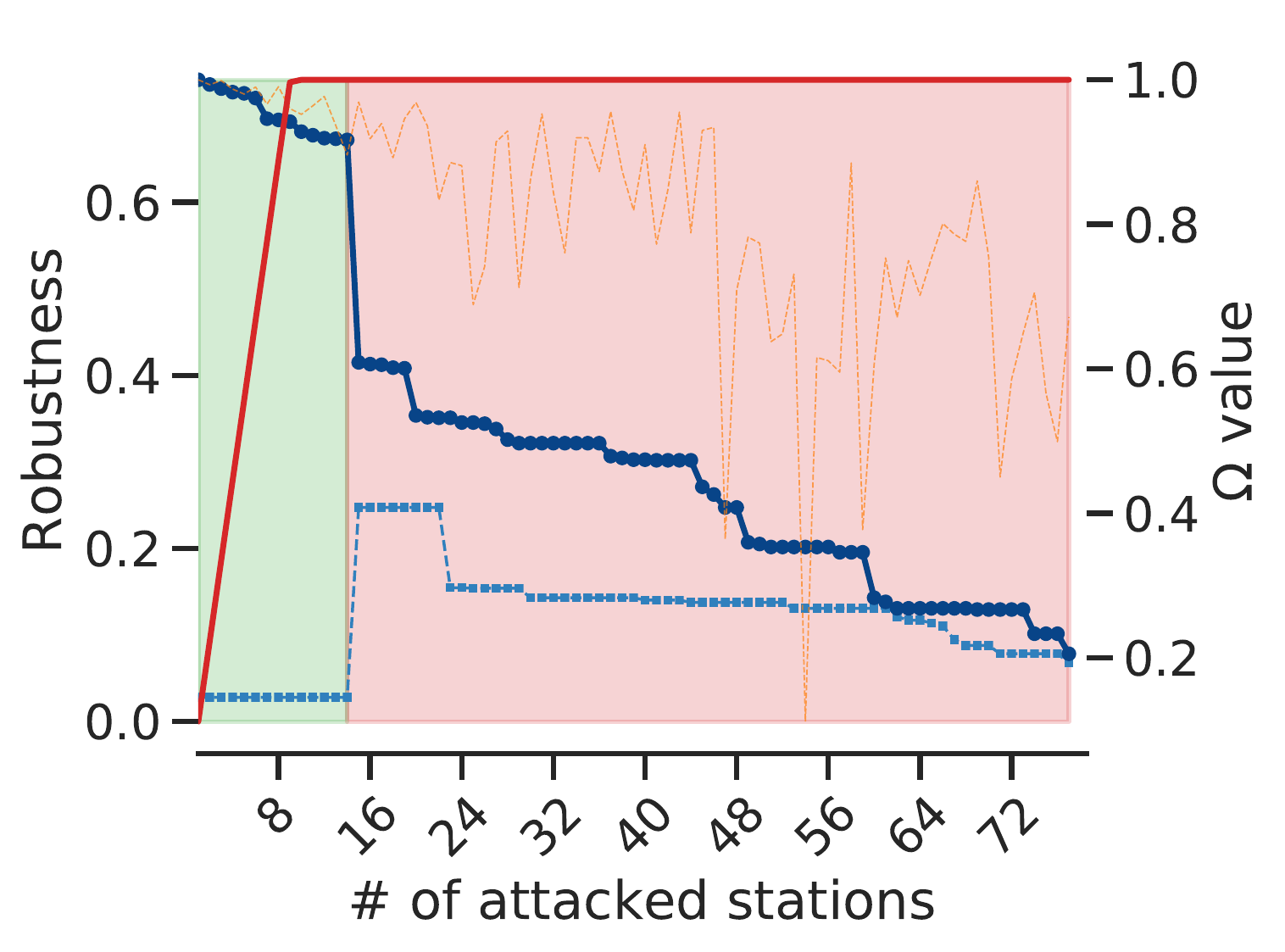}
    \caption{Degree}
    \label{f:eu_powergrid_ew_degree}
	\end{subfigure}%
	\hfill
	\begin{subfigure}{0.5\textwidth}
		\centering
		\includegraphics[width=\textwidth]{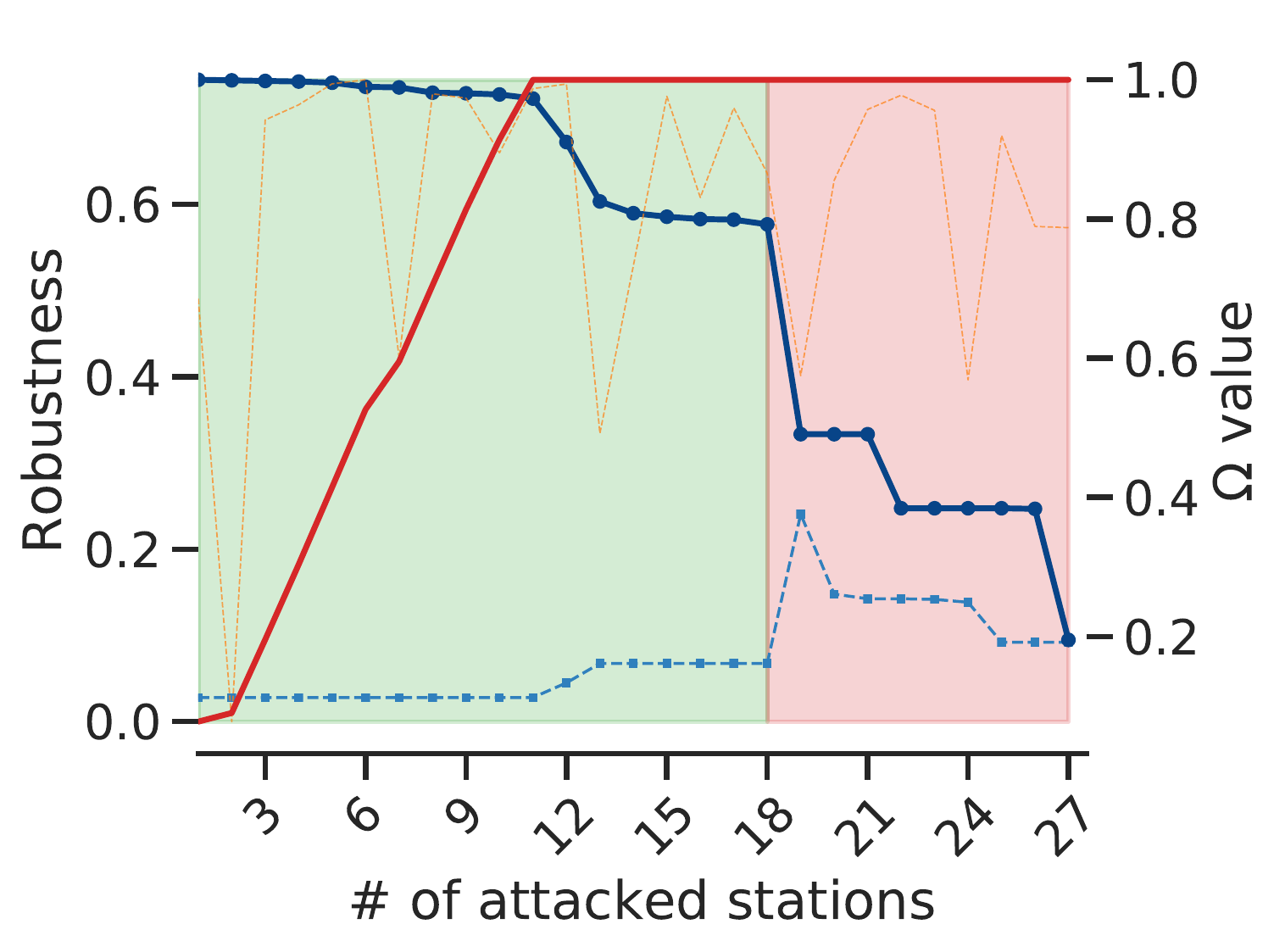}
    \caption{MS + R}
    \label{f:eu_powergrid_ew_msr}
	\end{subfigure}%
	\hfill
	\caption{Early Warning values for the SciKit European powergrid under random failures and targeted attacks.}
	\label{f:eu_powergrid_ew}
\end{figure}

\begin{figure}[htbp!]
	\centering
	\begin{subfigure}{\textwidth}
		\centering
		\includegraphics[width=0.5\textwidth]{figures/EarlyWarning/legend_horizontal}
	\end{subfigure}%
	\hfill
	\begin{subfigure}{0.5\textwidth}
		\centering
		\includegraphics[width=\textwidth]{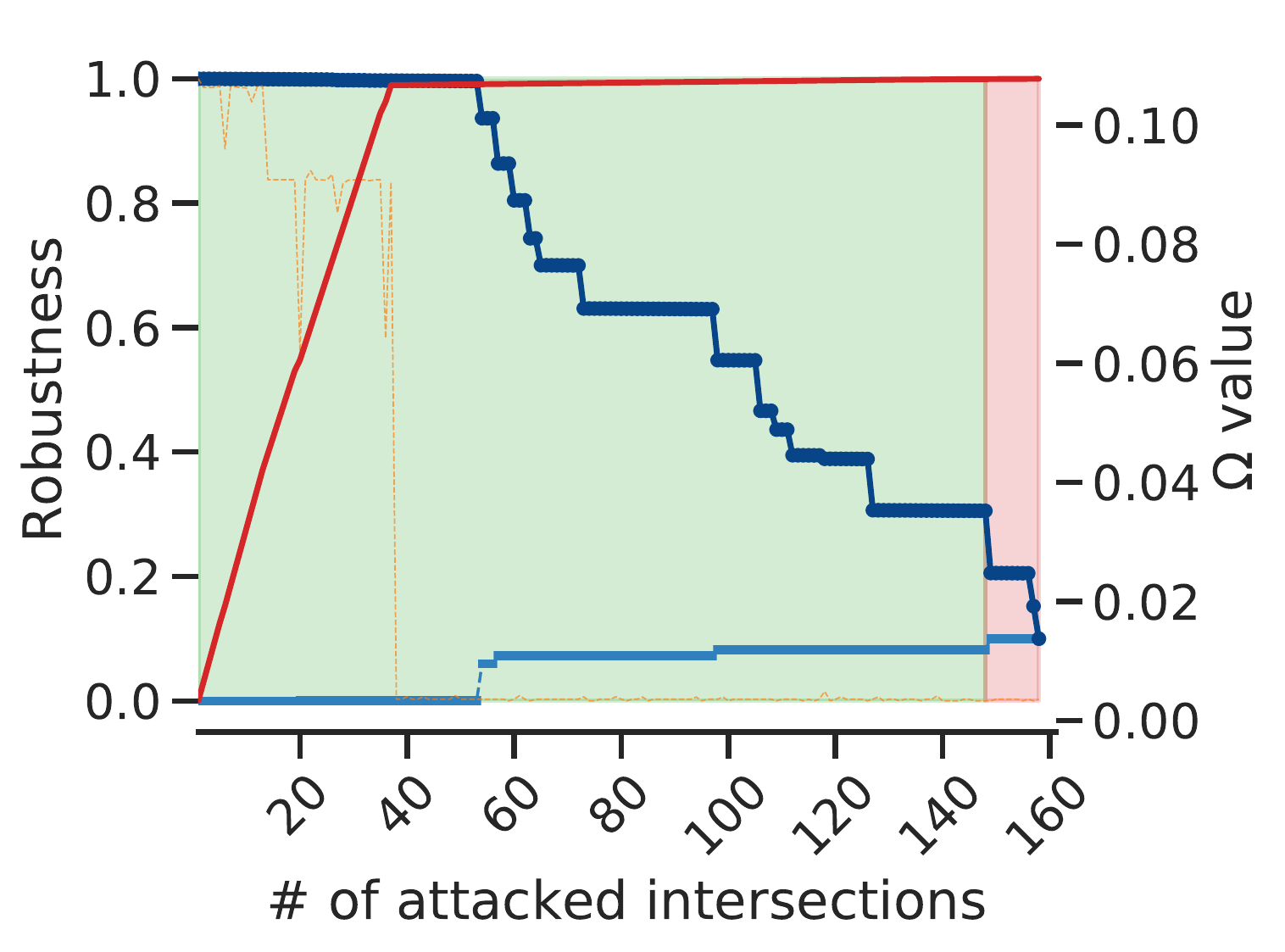}
		\caption{California Roads}
		\label{f:roads-california_gndr}
	\end{subfigure}%
	\begin{subfigure}{0.5\textwidth}
		\centering
		\includegraphics[width=\textwidth]{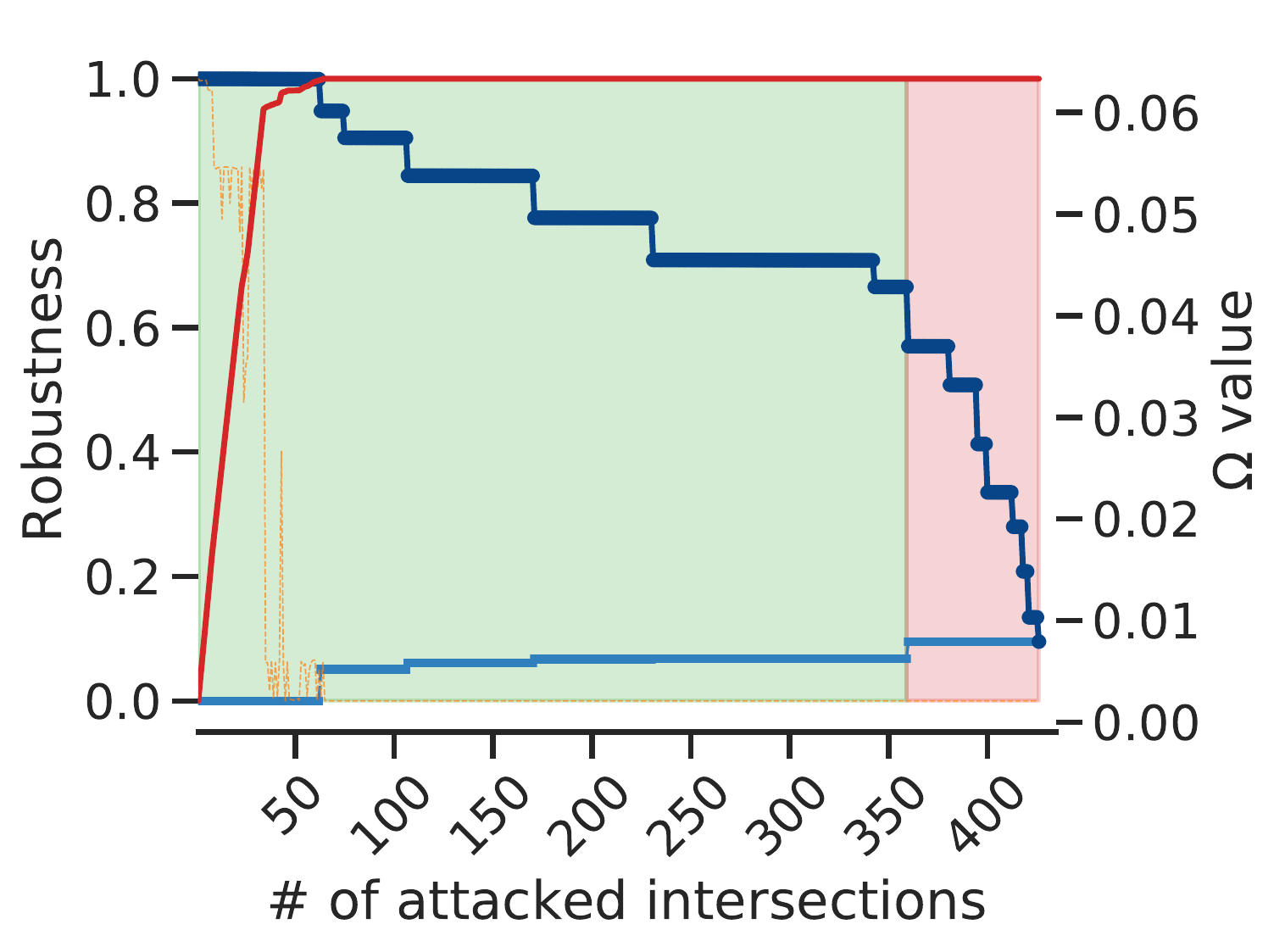}
    \caption{North-America roads}
    \label{f:roads-northamerica_gndr}
	\end{subfigure}%
	\hfill
	\begin{subfigure}{0.5\textwidth}
		\centering
		\includegraphics[width=\textwidth]{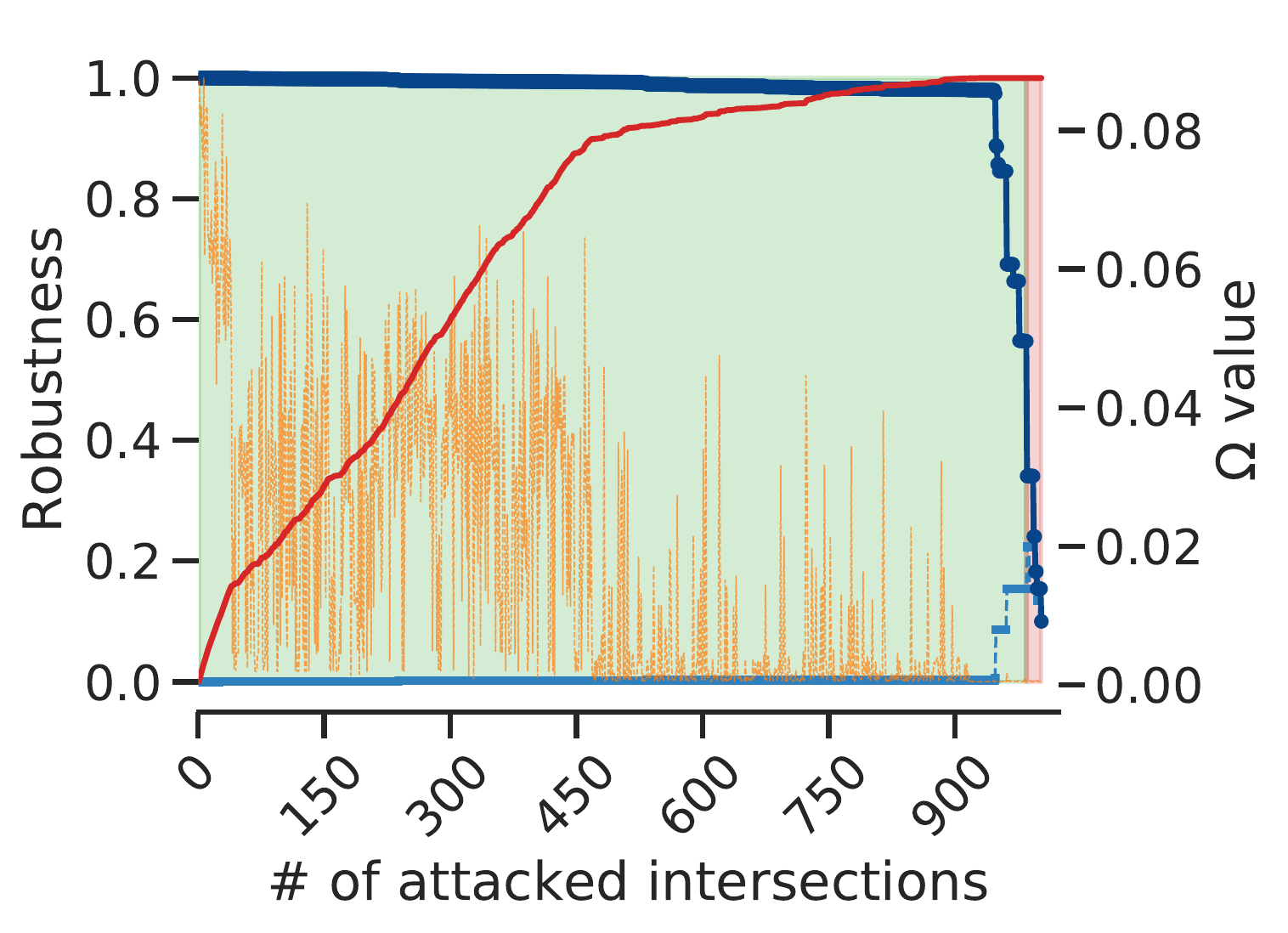}
    \caption{San Francisco roads}
    \label{f:roads-sanfrancisco_gndr}
	\end{subfigure}%
	\hfill
	\caption{$\Omega$ values for three different American road networks under \emph{GND} +R attacks (with cost matrix $\mathbf{W} = \mathbf{I}$).}
	\label{f:early_warning_gndr_roads}
\end{figure}

In addition to the example applications of $\Omega$ illustrated in the main paper, we also test if it can detect the collapse of other systems.
In particular, we show the SciKit European powergrid (eu-powergrid) under random failures, degree or Min-Sum + Reinsertion phase attacks in
Figure~\ref{f:eu_powergrid_ew},
and also various American roads under Generalized Network Dismantling + Reinsertion phase attacks in
Figure~\ref{f:early_warning_gndr_roads}.
In all these scenarios, $\Omega$ is able to detect the system damage and reaches warning levels before the system collapse actually happens, even in case of multiple large connected components detaching from the larger one as the attack goes on.

\begin{figure}[htbp!]
	\centering
	\begin{subfigure}{\textwidth}
		\centering
		\includegraphics[width=0.5\textwidth]{figures/EarlyWarning/legend_horizontal}
	\end{subfigure}%
	\hfill
	\begin{subfigure}{0.5\textwidth}
		\centering
		\includegraphics[width=\textwidth]{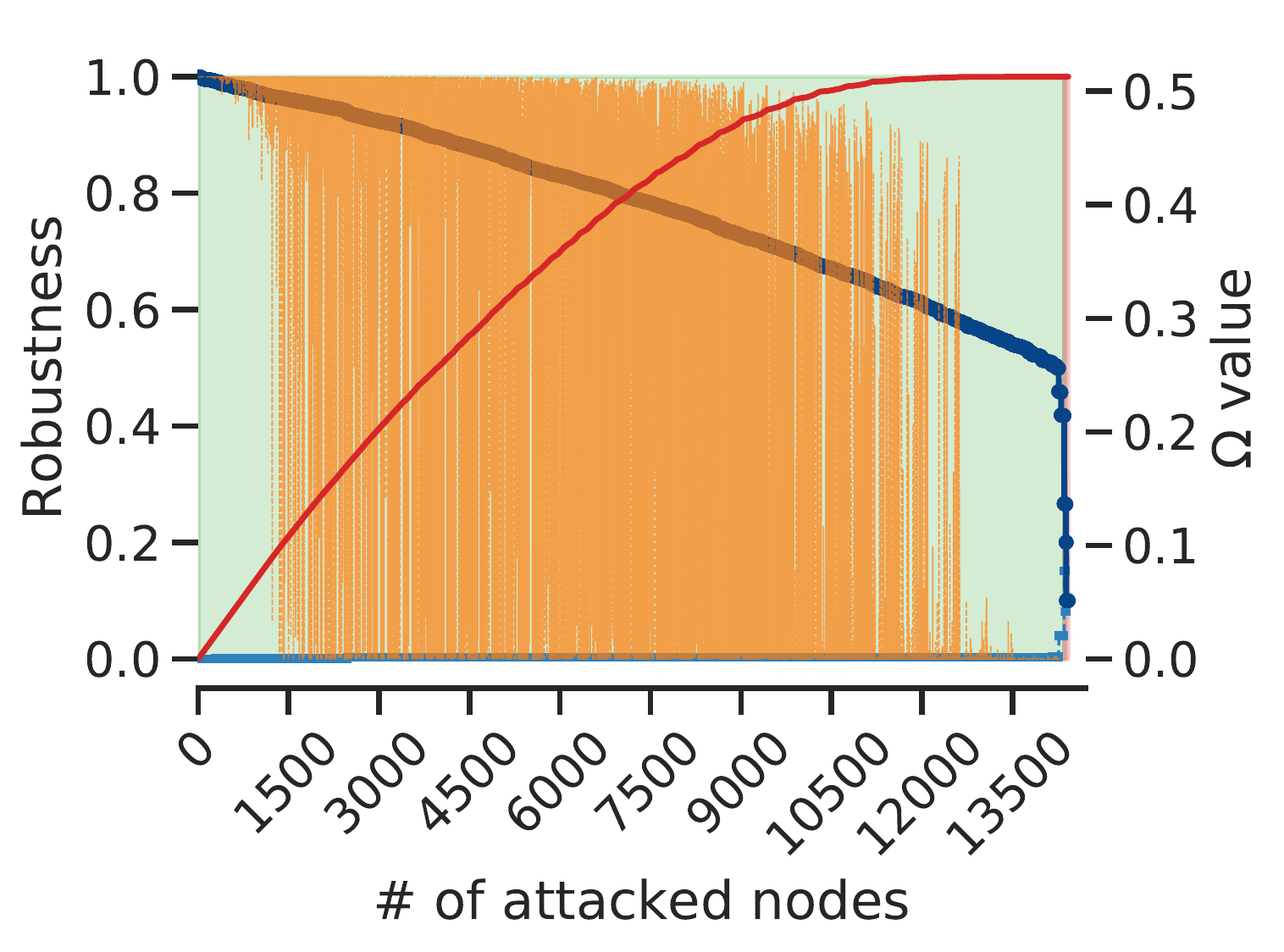}
		\caption{Internet topology (tech-RL-caida)}
		\label{f:tech-RL-caida_gndr}
	\end{subfigure}%
	\hfill
	\begin{subfigure}{0.5\textwidth}
		\centering
		\includegraphics[width=\textwidth]{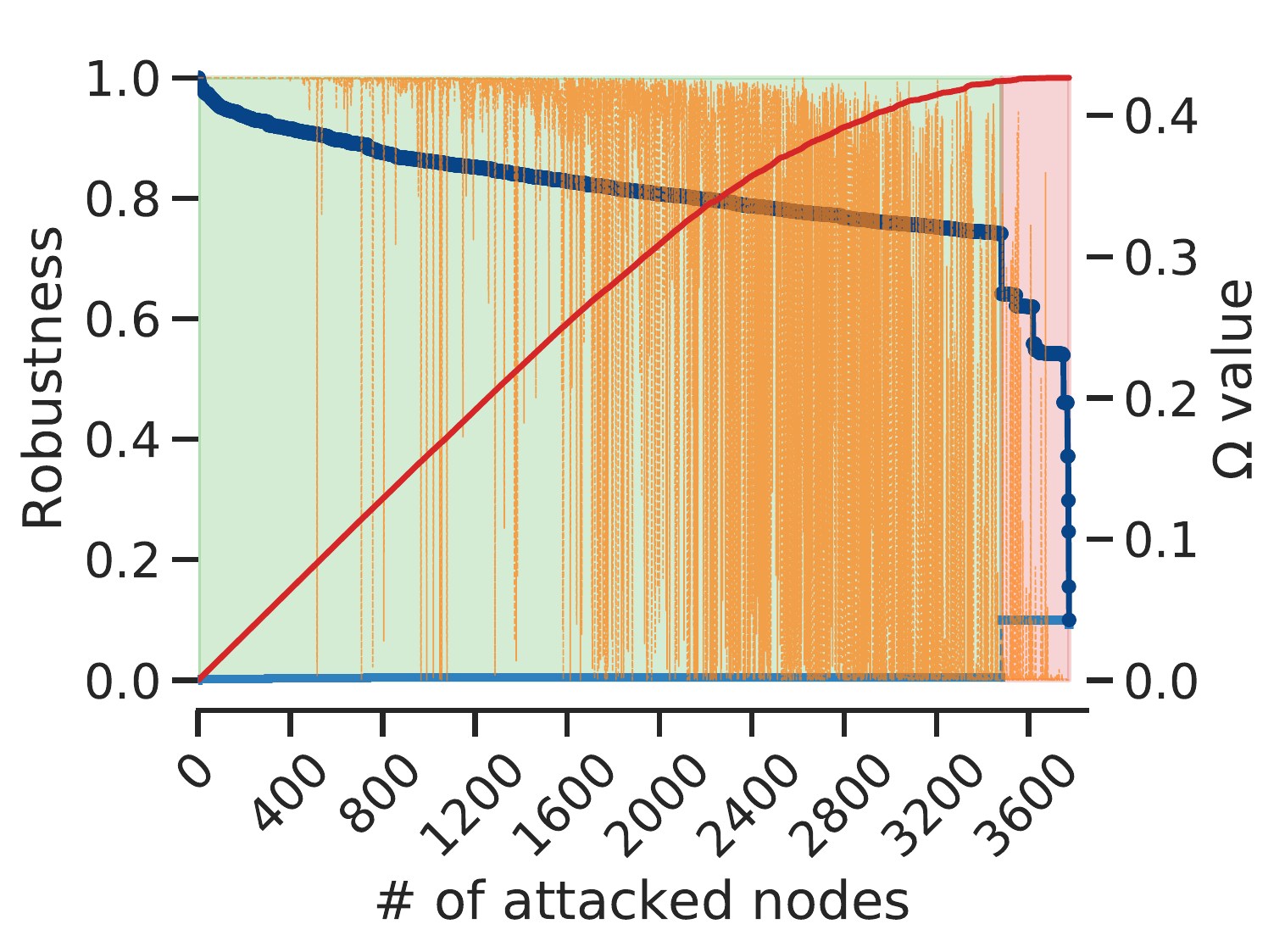}
    \caption{University of Notre Dame website hyperlinks network}
    \label{f:web-NotreDame_gndr}
	\end{subfigure}%
	\hfill
	\begin{subfigure}{0.5\textwidth}
		\centering
		\includegraphics[width=\textwidth]{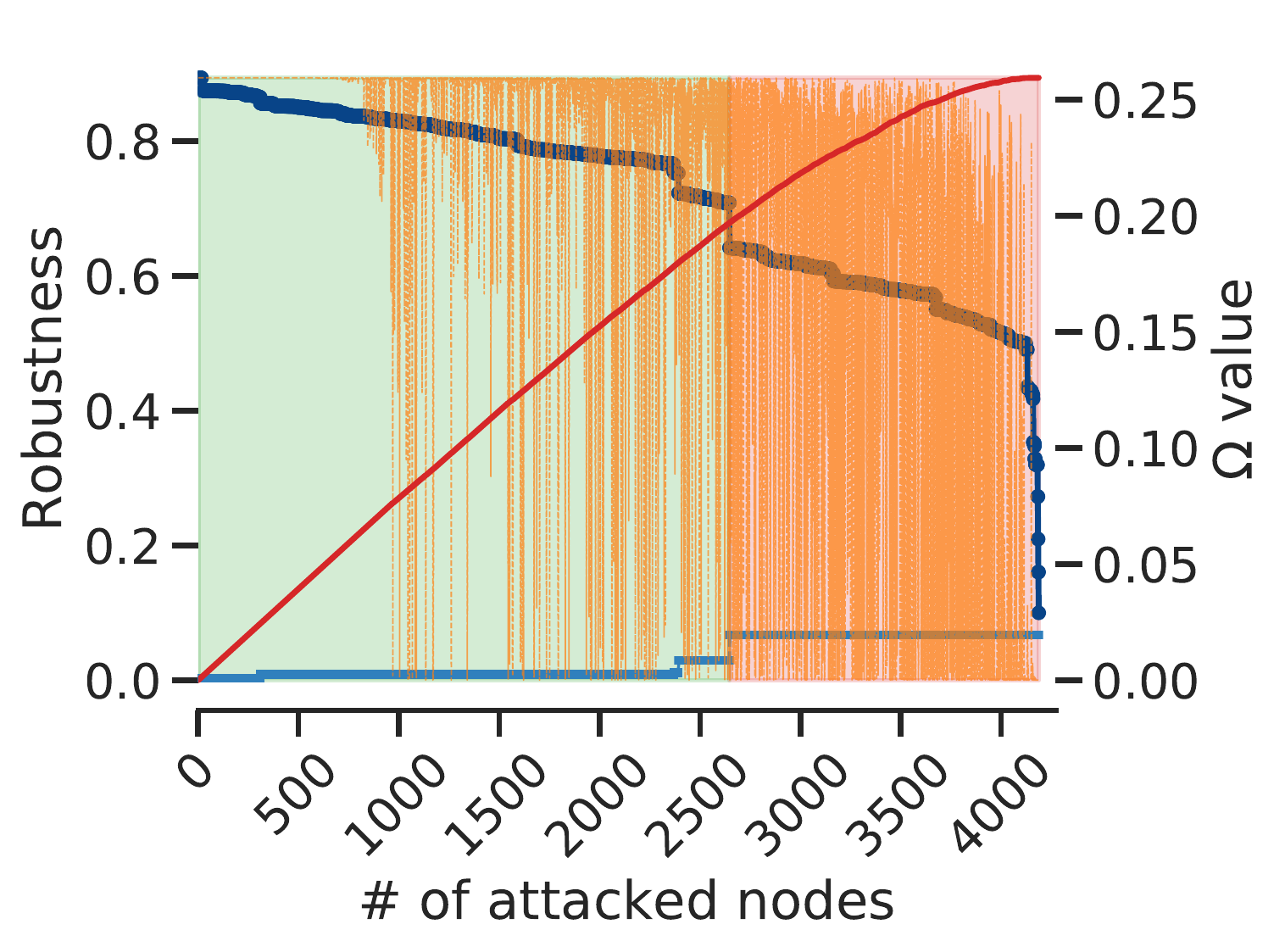}
    \caption{Stanford University website hyperlinks network}
    \label{f:web-Stanfordo_gndr}
	\end{subfigure}%
	\hfill
	\caption{$\Omega$ values for three different internet networks under GND +R attacks (with cost matrix $\mathbf{W} = \mathbf{I}$).}
	\label{f:early_warning_gndr_internet}
\end{figure}

\subsection*{Dataset}
In Table~\ref{t:test_networks} we list the test networks used in our experiments with their category and size (number of nodes and edges).
Those networks model systems from various domains (e.g., biological, infrastructure and social data and so on), and range from a few hundred of nodes to more than one million.
For more about each network, we refer the reader to the original source.

\begin{table}[ht]
  \begin{adjustbox}{width=\textwidth}
    \begin{tabular}{ | l | c | c | r | r | r | }
	    \hline
	    Network                            & Name                                     & Category       & $|N|$      & $|E|$      & References \\
			\hline
      ARK201012\_LCC                     & CAIDA ARK (Dec 2010) (LCC)               & Infrastructure & 29.3K      & 78.1K      & \cite{CAIDA_ARK}                                                                     \\
      advogato                           & Advogato trust network                   & Social         & 6.5K       & 43.3K      & \cite{konect:2017:advogato,konect:massa09}                                            \\
      arenas-meta                        & C. elegans                               & Metabolic      & 453        & 2.0K       & \cite{konect:2017:arenas-meta,konect:duch05}                                          \\
      cfinder-google                     & Google.com internal                      & Hyperlink      & 15.8K      & 149.5K     & \cite{konect:2017:cfinder-google,konect:palla07}                                      \\
      citeseer                           & CiteSeer                                 & Citation       & 384.4K     & 1.7M       & \cite{konect:2017:citeseer,b524}                                                      \\
      com-dblp                           & DBLP co-authorship                       & Coauthorship   & 317.1K     & 1.0M       & \cite{konect:2017:com-dblp,konect:leskovec2012}                                       \\
      corruption                         & Corruption Scandals                      & Social         & 309        & 3.3K       & \cite{10.1093/comnet/cny002}                                                          \\
      dblp-cite                          & DBLP citation                            & Citation       & 12.6K      & 49.6K      & \cite{konect:2017:dblp-cite,konect:DBLP}                                              \\
      digg-friends                       & Digg friends                             & Social         & 279.6K     & 1.5M       & \cite{konect:2017:digg-friends,konect:digg}                                           \\
      dimacs10-celegansneural            & C. elegans (neural)                      & Neural         & 297        & 2.1K       & \cite{konect:2018:dimacs10-celegansneural,konect:duncan98,konect:celegansneural1}     \\
      dimacs10-polblogs                  & Political blogs (LCC)                    & Hyperlink      & 1.2K       & 16.7K      & \cite{konect:2018:dimacs10-polblogs,konect:adamic2005}                                \\
      douban                             & Douban social network                    & Social         & 154.9K     & 327.2K     & \cite{konect:2017:douban,konect:socialcomputing}                                      \\
      econ-wm1                           & Economic network WM1                     & Economic       & 260        & 2.6K       & \cite{nr}                                                                             \\
      ego-twitter                        & Twitter lists                            & Social         & 23.4K      & 32.8K      & \cite{konect:2017:ego-twitter,konect:McAuley2012}                                     \\
      email-EuAll                        & EU institution email                     & Communication  & 265.2K     & 365.6K     & \cite{konect:2017:email-EuAll,konect:leskovec107}                                     \\
      eu-powergrid                       & SciGRID Power Europe                     & Power          & 1.5K       & 1.8K       & \cite{SciGRIDv0.2}                                                                    \\
      foodweb-baydry                     & Florida ecosystem dry                    & Trophic        & 128        & 2.1K       & \cite{konect:2017:foodweb-baydry,konect:foodweb}                                      \\
      foodweb-baywet                     & Florida ecosystem wet                    & Trophic        & 128        & 2.1K       & \cite{konect:2017:foodweb-baywet,konect:foodweb}                                      \\
      gridkit-eupowergrid                & GridKit Power Europe                     & Power          & 13.8K      & 17.3K      & \cite{wiegmans_2016}                                                                 \\
      gridkit-north\_america             & GridKit Power North-America              & Power          & 16.2K      & 20.2K      & \cite{wiegmans_2016}                                                                 \\
      hyves                              & Hyves social network                     & Social         & 1.4M       & 2.8M       & \cite{konect:2017:hyves,konect:socialcomputing}                                       \\
      inf-USAir97                        & US Air lines (1997)                      & Infrastructure & 332        & 2.1K       & \cite{nr,pajek_repo}                                                                 \\
      internet-topology                  & Internet (AS) topology                   & Infrastructure & 34.8K      & 107.7K     & \cite{konect:2017:topology,konect:zhang05}                                            \\
      librec-ciaodvd-trust               & CiaoDVD trust network                    & Social         & 4.7K       & 33.1K      & \cite{konect:2018:librec-ciaodvd-trust,konect:ciaodvd}                                \\
      librec-filmtrust-trust             & FilmTrust trust network                  & Social         & 874        & 1.3K       & \cite{konect:2018:librec-filmtrust-trust,konect:filmtrust}                            \\
      linux                              & Linux source code files                  & Software       & 30.8K      & 213.7K     & \cite{konect:2017:linux}                                                              \\
      loc-brightkite                     & Brightkite friendships                   & Social         & 58.2K      & 214.1K     & \cite{konect:2017:loc-brightkite_edges,konect:cho2011}                               \\
      loc-gowalla                        & Gowalla friendships                      & Social         & 196.6K     & 950.3K     & \cite{konect:2017:loc-gowalla_edges,konect:cho2011}                                  \\
      london\_transport\_multiplex\_aggr & Aggregated London Transportation network & Transport      & 369        & 430        & \cite{DeDomenico8351}                                                                 \\
      maayan-Stelzl                      & Human protein (Stelzl)                   & Metabolic      & 1.7K       & 3.2K       & \cite{konect:2017:maayan-Stelzl,konect:stelzl}                                        \\
      maayan-figeys                      & Human protein (Figeys)                   & Metabolic      & 2.2K       & 6.4K       & \cite{konect:2017:maayan-figeys,konect:figeys}                                        \\
      maayan-foodweb                     & Little Rock Lake food web                & Trophic        & 183        & 2.5K       & \cite{konect:2017:maayan-foodweb,konect:little-rock-lake}                             \\
      maayan-vidal                       & Human protein (Vidal)                    & Metabolic      & 3.1K       & 6.7K       & \cite{konect:2017:maayan-vidal,konect:proteome}                                       \\
      moreno\_crime\_projected           & Crime (projection)                       & Social         & 754        & 2.1K       & \cite{konect:2017:moreno_crime}                                                      \\
      moreno\_propro                     & Protein                                  & Metabolic      & 1.9K       & 2.3K       & \cite{konect:2017:moreno_propro,konect:coulomb2005,konect:han2005,konect:stumpf2005} \\
      moreno\_train                      & Train bombing terrorist contacts         & Human contact  & 64         & 243        & \cite{konect:2017:moreno_train,konect:hayes}                                         \\
      munmun\_digg\_reply\_LCC           & Digg social network replies (LCC)        & Communication  & 29.7K      & 84.8K      & \cite{konect:2017:munmun_digg_reply,konect:choudhury09}                             \\
      munmun\_twitter\_social            & Twitter follows (ICWSM)                  & Social         & 465.0K     & 833.5K     & \cite{konect:2017:munmun_twitter_social,konect:choudhury10}                         \\
      opsahl-openflights                 & OpenFlights                              & Infrastructure & 2.9K       & 15.7K      & \cite{konect:2017:opsahl-openflights,konect:opsahl2010b}                              \\
      opsahl-powergrid                   & US power grid                            & Infrastructure & 4.9K       & 6.6K       & \cite{konect:2017:opsahl-powergrid,konect:duncan98}                                   \\
      opsahl-ucsocial                    & UC Irvine messages                       & Communication  & 1.9K       & 13.8K      & \cite{konect:2017:opsahl-ucsocial,konect:opsahl09}                                   \\
      oregon2\_010526                    & Autonomous systems Oregon-2              & Infrastructure & 11.5K      & 32.7K      & \cite{10.1145/1081870.1081893}                                                        \\
      p2p-Gnutella06                     & Gnutella P2P, August 8 2002              & Computer       & 8.7K       & 31.5K      & \cite{konect:ripeanu02,snapnets}                                                      \\
      p2p-Gnutella31                     & Gnutella P2P, August 31 2002             & Computer       & 62.6K      & 147.9K     & \cite{konect:2017:p2p-Gnutella31,konect:ripeanu02}                                    \\
      pajek-erdos                        & Erdős co-authorship network              & Coauthorship   & 6.9K       & 11.8K      & \cite{konect:2018:pajek-erdos,pajek_repo}                                             \\
      petster-catdog-household           & Catster/Dogster familylinks (LCC)        & Social         & 324.9K     & 2.6M       & \cite{konect:2017:petster-carnivore}                                                  \\
      petster-hamster                    & Hamsterster full                         & Social         & 2.4K       & 16.6K      & \cite{konect:2017:petster-hamster}                                                    \\
      power-eris1176                     & Power network problem                    & Power          & 1.2K       & 9.9K       & \cite{nr}                                                                             \\
      roads-california                   & California Road Network                  & Infrastructure & 21.0K      & 21.7K      & \cite{10.1007/11535331_16}                                                           \\
      roads-northamerica                 & North-America Road Network               & Infrastructure & 175.8K     & 179.1K     & \cite{dcws}                                                                           \\
      roads-sanfrancisco                 & San Francisco Road Network               & Infrastructure & 175.0K     & 221.8K     & \cite{brinkhoff2002framework}                                                         \\
      route-views                        & Autonomous systems AS-733                & Infrastructure & 6.5K       & 13.9K      & \cite{konect:2017:as20000102,konect:leskovec107}                                      \\
      slashdot-threads                   & Slashdot threads                         & Communication  & 51.1K      & 117.4K     & \cite{konect:2017:slashdot-threads,konect:slashdot-threads}                           \\
      slashdot-zoo                       & Slashdot Zoo                             & Social         & 79.1K      & 467.7K     & \cite{konect:2017:slashdot-zoo,kunegis:slashdot-zoo}                                 \\
      subelj\_jdk                        & JDK dependency network                   & Software       & 6.4K       & 53.7K      & \cite{konect:2016:subelj_jdk}                                                        \\
      subelj\_jung-j                     & JUNG and Javax dependency network        & Software       & 6.1K       & 50.3K      & \cite{konect:2017:subelj_jung-j,konect:dependency1}         \\
      tech-RL-caida                      & Internet router network                  & Infrastructure & 190.9K     & 607.6K     & \cite{nr}                                                                             \\
      twitter\_LCC                       & Twitter users (LCC)                      & Social         & 532.3K     & 694.6K     & \cite{morone2015influence}                                                            \\
      web-EPA                            & Pages linking to epa.gov                 & Hyperlink      & 4.3K       & 8.9K       & \cite{nr}                                                                             \\
      web-NotreDame                      & Notre Dame web pages                     & Hyperlink      & 325.7K     & 1.1M       & \cite{konect:2017:web-NotreDame,konect:albert1999}                                   \\
      web-Stanford                       & Stanford University web pages            & Hyperlink      & 281.9K     & 2M         & \cite{konect:2017:web-Stanford,konect:snap}                                          \\
      web-webbase-2001                   & Web network                              & Hyperlink      & 16.1K      & 25.6K      & \cite{nr}                                                                             \\
      wikipedia\_link\_kn                & Wikipedia links (KN)                     & Hyperlink      & 29.5K      & 278.7K     & \cite{konect:2018:wikipedia_link_li}                                                 \\
      wikipedia\_link\_li                & Wikipedia links (LI)                     & Hyperlink      & 49.1K      & 294.3K     & \cite{konect:2018:wikipedia_link_kn}                                                 \\
      wordnet-words                      & WordNet lexical network                  & Lexical        & 146.0K     & 657.0K     & \cite{konect:2017:wordnet-words,konect:fellbaum98}          \\
	    \hline
    \end{tabular}%
    \end{adjustbox}
	\caption{The networks used to evaluate our approach. For each network, we report the name, the number of nodes and edges, the category it belongs to and some references.}
	\label{t:test_networks}
\end{table}%

\subsection*{Test environment}

Here we detail the environment where our experiments were performed and the tools used.

All experiments ran on a shared machine equipped with two Intel Xenon E5-2620 CPUs, 128GB RAM and a two core nVidia Tesla K80 (with 12GB VRAM each). More details about the drivers used and the full package dependency list of our code can be found in the code package.

Concerning the other algorithms used in our comparison (i.e., \emph{GND}, \emph{EGND} \emph{MS}, \emph{CoreHD} and \emph{EI}), we use authors' official code with default parameters. Specifically, we use identity weight input matrix for both \emph{GND} and \emph{EGND} (and the relative fine-tuning algorithm), $1K$ trials for the \emph{EGND}.


\bibliographystyle{plain}
\bibliography{scibib}

%
%

